\documentclass[10pt,journal,compsoc]{IEEEtran}

\usepackage{color}

\ifCLASSOPTIONcompsoc
  \usepackage[nocompress]{cite}
\else
  \usepackage{cite}
\fi

\ifCLASSINFOpdf
   \usepackage[pdftex]{graphicx}
   \graphicspath{{../pdf/}{../jpeg/}{./figs/}{./figures}}
\else
\fi

\usepackage{graphicx}
\ifCLASSINFOpdf

\else

\fi

\usepackage[cmex10]{amsmath}
\usepackage{array}
\usepackage{multirow}
\usepackage{lineno}
\usepackage{url}
\usepackage{amsmath}

\ifCLASSOPTIONcompsoc
  \usepackage[nocompress]{cite}
\else
  \usepackage{cite}
\fi

\def\x{{\bf x}}

\def\0{{\bf 0}}
\def\1{{\bf 1}}

\def\eg{{\em e.g.}}
\def\ie{{\em i.e.}}

\usepackage{amsmath}
\usepackage{amssymb}
\usepackage{multirow}
\usepackage{array}
\usepackage{booktabs}
\usepackage{multirow}
\usepackage{bm}
\usepackage{graphicx}
\usepackage{epstopdf}
\usepackage{subfigure}
\usepackage{makecell}
\usepackage[pagebackref=false,breaklinks=true,letterpaper=true,colorlinks,bookmarks=false]{hyperref}
\usepackage{cite}
\usepackage{multirow}
\usepackage{color}
\usepackage{CJK}
\usepackage{soul}
\usepackage[numbers,sort&compress]{natbib}
\usepackage{hyperref}

\usepackage[colorinlistoftodos]{todonotes}

\def\eg{{\em e.g.}}
\def\ie{{\em i.e.}}

\begin{document}
\title{Functional Imaging Constrained Diffusion for Brain PET Synthesis from Structural MRI}

\author{Minhui~Yu, 
Mengqi~Wu,
Ling~Yue,
Andrea~Bozoki,
Mingxia~Liu,~\IEEEmembership{Senior Member,~IEEE}

\IEEEcompsocitemizethanks{
\IEEEcompsocthanksitem M.~Yu, M.~Wu and M.~Liu are with the Department of Radiology and Biomedical Research Imaging Center (BRIC), University of North Carolina at Chapel Hill, Chapel Hill, NC 27599, USA. 
M.~Yu and M.~Wu are also with the Joint Department of Biomedical Engineering, University of North Carolina at Chapel Hill and North Carolina State University, Chapel Hill, NC 27599, USA. 
L.~Yue is with the Department of Geriatric Psychiatry, Shanghai Mental Health Center, Shanghai Jiao Tong University School of Medicine, Shanghai 200240, China. 
A.~Bozoki is with the Department of Neurology, University of North Carolina at Chapel Hill, Chapel Hill, NC 27599, USA. 
\IEEEcompsocthanksitem 
Corresponding author: M.~Liu (Email: mxliu@med.unc.edu). 
\protect\\
}
}

\IEEEtitleabstractindextext{%
\begin{abstract}

Magnetic resonance imaging (MRI) and positron emission tomography (PET) are increasingly used in multimodal analysis of neurodegenerative disorders. 
While MRI is broadly utilized in clinical settings, PET is less accessible. 
Many studies have attempted to use deep generative models to synthesize PET from MRI scans. 
However, they often suffer from unstable training and inadequately preserve brain functional information conveyed by PET. 
To this end, we propose a functional imaging constrained diffusion (FICD) framework for 3D brain PET image synthesis with paired structural MRI as input condition, through a new constrained diffusion model (CDM). 
The FICD introduces noise to PET and then progressively removes it with CDM, ensuring high output fidelity throughout a stable training phase. 
The CDM learns to predict denoised PET with a functional imaging constraint introduced to ensure voxel-wise alignment between each denoised PET and its ground truth. 
Quantitative and qualitative analyses conducted on 293
subjects with paired T1-weighted MRI and $^{18}$F-fluorodeoxyglucose (FDG)-PET scans suggest that FICD achieves superior performance in generating FDG-PET data compared to state-of-the-art methods. 
We further validate the effectiveness of the proposed FICD on data from a total of 1,262 subjects through three downstream tasks, with experimental results suggesting its utility and generalizability.
\end{abstract}

\begin{IEEEkeywords}
Brain, Image Synthesis, PET, Structural MRI, Diffusion Model. 
\end{IEEEkeywords}}

\maketitle

\IEEEdisplaynontitleabstractindextext

\IEEEpeerreviewmaketitle

\IEEEraisesectionheading{\section{Introduction}\label{S1}}


\IEEEPARstart{B}{rain} magnetic resonance imaging (MRI) and positron emission tomography (PET) offer a synergistic diagnosis of the brain. 
MRI, with its detailed structural imaging capabilities, is one of the main processes in clinical diagnostics of neurodegenerative disorders~\cite{ridha2006tracking,dickerson2009cortical,hohenfeld2018resting}. 
PET provides unique insights into the brain's metabolic patterns and neuronal activity through specific radioactive tracers~\cite{nestor2018clinical,young2020imaging,dominguez2023review}. 
Conducting a multimodal study with these two imaging techniques proves to be especially advantageous in the exploration of neurodegenerative disorders, due to the intricate interplay between brain anatomy and its biochemical processes~\cite{hinrichs2011predictive,nasrallah2014multimodality}. 
Despite their combined value, the acquisition of brain PET lags behind MRI, primarily due to PET's higher operational costs and complexities in handling radioactive materials~\cite{wittenberg2019economic}. 
The frequent unavailability of brain PET scans poses challenges to the progress of multimodal studies~\cite{sharma2022comprehensive,pan2021disease}.

To address this modality-missing issue, numerous efforts have explored employing deep generative models to synthesize PET images from MRI scans. 
Generative adversarial networks (GANs) have been widely used due to their capability of high-fidelity image generation~\cite{hu2021bidirectional,pan2021disease,zhang2022bpgan,vega2024image,zhou2021synthesizing,wei2020predicting}.
These approaches frequently encounter issues with training instability attributed to their training nature and also suffer from limited diversity in the generated outputs due to mode collapse~\cite{goodfellow2016nips}. 
Variational autoencoder (VAE) models, including hybrids like VAEGAN, are also popular in medical image synthesis due to their ability to handle complex data structures effectively~\cite{li2021new,yang2021synthesizing}.
However, VAE models usually suffer from image blurriness and face posterior collapse, where a subset of the latent space becomes redundant and does not contribute to the data generation process~\cite{lucas2019don}.

\begin{figure*}[!t]
\setlength{\abovecaptionskip}{0pt}
\setlength{\belowcaptionskip}{0pt}
\setlength{\abovedisplayskip}{0pt}
\setlength\belowdisplayskip{0pt}
\setlength{\abovecaptionskip}{0pt}
\centering
\includegraphics[width=1\textwidth]{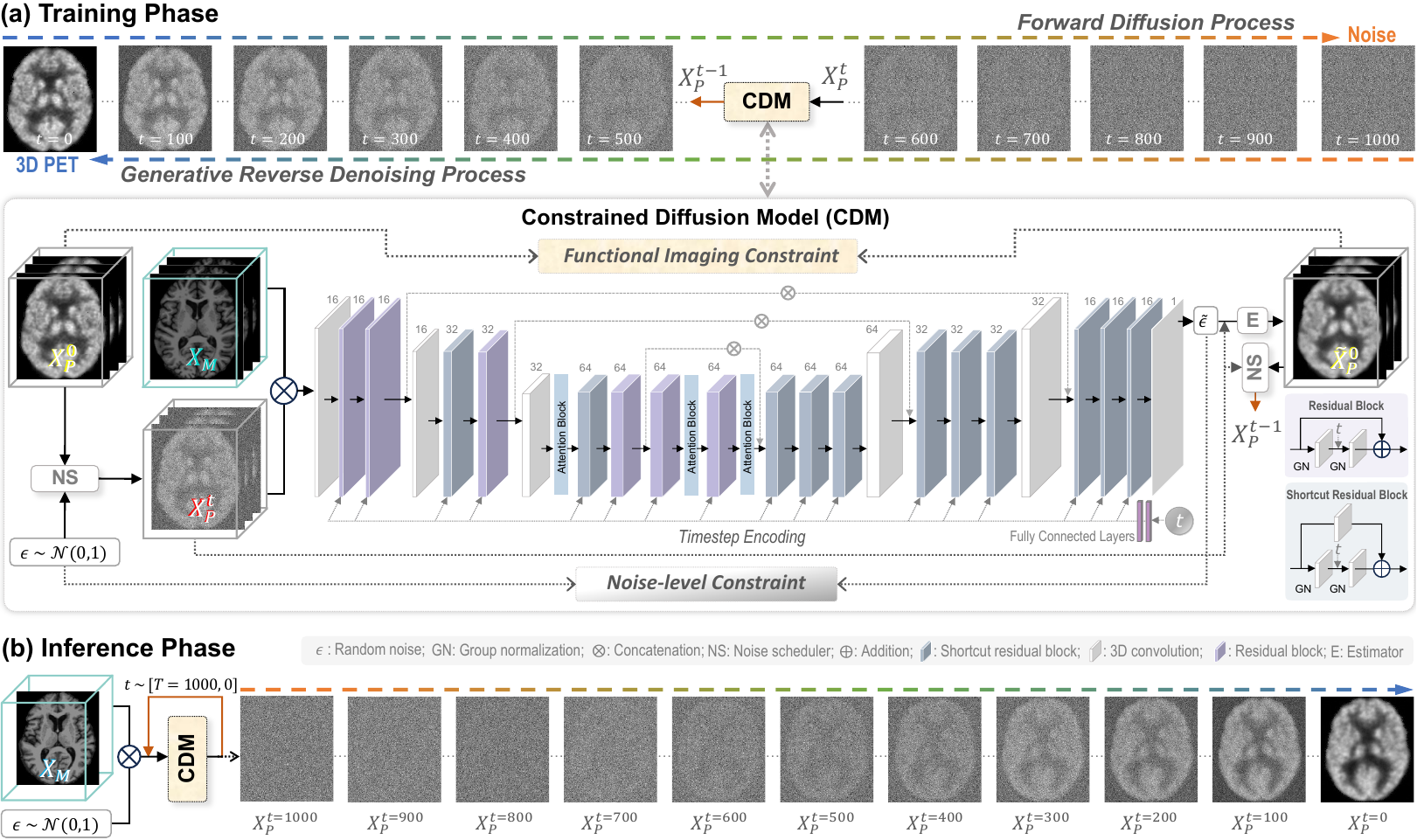}
\caption{Illustration of the proposed functional imaging constrained diffusion (FICD) framework for 3D brain PET image synthesis with paired structural MRI as the condition. 
(a) The training phase consists of a \emph{forward diffusion process} that incrementally adds noise to an input PET image, and a \emph{generative reverse denoising process} that gradually removes the noise. 
The MRI condition and noise-corrupted PET are input to the proposed constrained diffusion model (CDM) to predict the noise on PET, then the noise is
further used to estimate the denoised PET in the previous
timestep and in the final timestep. The outputs of CDM are
optimized by a unique functional imaging constraint and a
noise-level constraint, respectively. 
(b) In the inference phase, an MRI and a pure noise are input to the CDM, which progressively removes the noise to generate a synthetic PET image.}
\label{framework}
\end{figure*}

Diffusion models have gained significant attention as a robust type of deep generative model. 
By initially introducing noise to data and then progressively denoising it through iterative processes~\cite{sohl2015deep,ho2020denoising}, diffusion models have been applied in a wide range of fields and produce better outcomes than other types of generative models~\cite{dhariwal2021diffusion,cao2024survey}.
Most recently, it has gained increasing interest in medical research for cross-modality image translation~\cite{kazerouni2023diffusion,xie2023synthesizing}. 
However, the optimization of diffusion models only relies on the constraint of noise, yielding an indirect estimation of image outputs, which is inadequate to generate unique outcomes. 
This outcome uniqueness is particularly critical in 3D PET synthesis, since voxel intensities of PET images convey essential brain functional information (\eg, measurement of regional glucose consumption in $^{18}$F-fluorodeoxyglucose (FDG)-PET and detection of amyloid plaques in amyloid PET) that is crucial for brain disorder analysis~\cite{chetelat2020amyloid}. 
Intuitively, it is meaningful to introduce an image-level constraint to ensure the one-to-one voxel correspondence between each synthetic PET image and its ground truth, helping improve the output uniqueness of diffusion models.

To this end, we propose a functional imaging constrained diffusion (FICD) framework for 3D PET image synthesis from structural MRI, through a new constrained diffusion model (CDM). 
The FICD framework is designed to generate samples with the Markov chain transition of a denoising diffusion model, to synthesize PET images with the MRI as a condition.  
As illustrated in Fig.~\ref{framework}~(a), it consists of a \emph{forward diffusion process} that incrementally adds noise to an input PET image, and a \emph{generative reverse denoising process} that gradually removes the noise with CDM. 
Specifically, the condition (\ie, MRI) and noise-corrupted PET are input to the CDM to predict the noise on PET, then the noise is further used to estimate the denoised PET in the previous timestep and in the final timestep. 
The outputs of CDM are optimized by a new functional imaging constraint and a noise-level constraint, respectively. 
Extensive experiments are performed on T1-weighted MRI and PET scans in both image synthesis and three downstream tasks (\ie, forecasting the progression of brain diseases, predicting future cognitive functions, and generating amyloid PET images), with experimental results indicating the superiority of our method over several state-of-the-art methods. 
The source code and our trained model can be accessed \href{https://github.com/minhuiyu0418/FICD}{online}.

The major contributions of this work can be summarized as follows. 
\begin{itemize}
\vspace{-1mm}
\item A new generative framework called functional imaging constrained diffusion (FICD) is developed to synthesize 3D PET from structural MRI, helping alleviate the modality-missing issue in multimodal studies. 
Quantitative and qualitative results show that FICD is capable of producing high-quality PET images.

\item A functional imaging constraint is designed to 
encourage voxel-wise alignment between an estimated PET and its ground truth, further constraining the denoising process. 
This is different from traditional diffusion models that only rely on noise constraints for optimization.  
Experimental results suggest that this constraint improves the fidelity of synthesized images and significantly reduces output variability.

\item Extensive experiments are conducted to evaluate the utility of synthesized PET images in two downstream tasks: forecasting the progression of preclinical Alzheimer's disease (AD) 
and predicting cognitive function at future time points. 
The experimental results demonstrate the improvements achieved by 
our method in multimodal analysis of preclinical AD.

\item 
We further adapt the FICD model, initially trained on FDG-PET, to accommodate PET with three common amyloid radiotracers (\ie, Pittsburgh Compound-B, $^{18}$F-flutemetamol, and Florbetapir), by fine-tuning this model to generate PET scans with these tracers.
This adaptation highlights the model's flexibility and generalizability across various problem settings.
\end{itemize}

The remainder of this paper is organized as follows. 
We review the most relevant studies in Section~\ref{S2}. 
In Section~\ref{S3}, we introduce the details of the proposed method and the implementation details. 
In Section~\ref{S4}, we present data involved in this work, competing methods, experimental settings, and results in forecasting disease stages and clinical scores. 
We further study the influence of several key components in Section~\ref{S5}. 
This paper is finally concluded in Section~\ref{S6}.

\section{Related Work}
\label{S2}
\subsection{Multimodal Medical Image Analysis}
Neuroimaging biomarkers provided by MRI and PET are among the most pivotal 
tools for monitoring Alzheimer's disease (AD) progression and understanding its underlying pathology~\cite{weiner2013alzheimer}. 
While structural MRI helps unveil brain atrophy, $^{18}$F-fluorodeoxyglucose (FDG)-PET can capture regional hypometabolism, and amyloid PET helps detect amyloid plaque accumulation in the brain~\cite{perrin2009multimodal}. 
Since each modality offers unique insights into brain structure and function, the integration of these modalities significantly improves disease progression prediction~\cite{walhovd2010multi,jack200811c,sharma2022comprehensive}. 
Walhovd et al.~\cite{walhovd2010multi} report that MRI and FDG-PET complement each other due to their varied sensitivity to memory performance across healthy aging and various stages of cognitive decline.  
Jack et al.~\cite{jack200811c} observe that using amyloid PET and MRI together can enhance clinical diagnosis compared to using a single modality. 
Sharma et al.~\cite{sharma2022comprehensive} highlight that missing data or incomplete modality are common challenges in multimodal analyses, impacting the comprehensive integration and interpretation of multimodal imaging data. 
In particular, while MRI is widely used in clinical settings, PET scans are less readily available due to their high cost and requirement for radioactive tracers~\cite{wittenberg2019economic,chetelat2020amyloid}.

Some previous studies attempt to solve the modality-missing issue by imputing those missing neuroimage features. 
Abdelaziz et al.\cite{abdelaziz2021alzheimer} employ linear interpolation to address missing modalities. 
Ritter et al.~\cite{ritter2015multimodal} utilize a hybrid approach combining mean imputation and the Expectation-Maximization (EM) algorithm to impute features for missing modalities. 
Some studies~\cite{yuan2012multi,xiang2014bi} employ multi-view learning methods to make use of samples with incomplete modalities, without directly imputing those missing feature values. 
However, these methods generally focus on predefined features that are specifically engineered by experts to represent neuroimages. 
Considering these human-engineered features may not be sufficient to represent the rich information in brain functional and structural images, direct synthesis of missing neuroimages could provide a more effective solution to this problem~\cite{hu2021bidirectional,zhang2022bpgan,vega2024image,xie2023synthesizing}.

\subsection{Cross-Modality Medical Image Translation}
Recent initiatives have utilized deep generative models, particularly generative adversarial networks (GANs), for medical image synthesis to address missing modalities~\cite{goodfellow2020generative}. 
A typical GAN model consists of two main elements: a generator and a discriminator. The generator's goal is to replicate the data distribution of real data, while the discriminator acts as a binary classifier, distinguishing between real and synthetic samples via ongoing optimization. These components engage in a competitive but synergistic relationship. 
Many studies have utilized GANs to synthesize PET images from structural MRI scans. 
Hu et al.~\cite{hu2021bidirectional} introduce a bidirectional mapping GAN  that incorporates semantic information derived from PET images into the latent space learned by GAN, to utilize image contexts and latent vectors. 
Zhang et al.~\cite{zhang2022bpgan} develop a brain PET GAN, a 3D end-to-end framework that employs a multiple convolution U-Net~\cite{ronneberger2015u} generator architecture, augmented with gradient profile loss and structural similarity index measure loss.
Vega et al.~\cite{vega2024image} employ a conditional GAN architecture that inputs a pair of MRI and synthetic PET scans into the discriminator to determine if it is synthetic or real. 
Wei et al.~\cite{wei2020predicting} propose a conditional flexible self-attention GAN (CF-SAGAN) model to predict the parametric map of PET scans from multisequence MRI. 
Zhou et al.~\cite{zhou2021synthesizing} develop a 3D unified anatomy-aware cyclic
adversarial network (UCAN) for translating multi-tracer PET scans with a
generative model, where MRI with anatomical information is incorporated. 
Hu et al.~\cite{hu2021bidirectional} develop a bidirectional mapping GAN method for brain MR-to-PET synthesis, where image contexts and latent vectors are used and jointly optimized for image synthesis. 
A critical challenge in training GAN models is their inherent instability, where the generator and discriminator engage in a complex dynamic that can be difficult to balance effectively. 
This often manifests in issues such as mode collapse, where the generator produces a limited variety of samples~\cite{goodfellow2016nips}.

Variational autoencoders (VAEs) are another type of generative model. 
They employ deep neural networks to first encode input data into a latent space with predefined distributions, optimized by maximizing the evidence lower bound between the encoded distributions and a prior, and then decode from this space to reconstruct the input~\cite{kingma2013auto}. 
The integration of VAE with GANs, known as VAEGANs, can further enhance its performance and has gained widespread utilization~\cite{larsen2016autoencoding}.
Yang et al.~\cite{yang2021synthesizing} leverage VAEGAN to create cross-domain and multi-contrast MR images from CT scans. 
Similarly, Li et al.~\cite{li2021new} utilize VAEGAN to generate arterial spin labeling (ASL) images from structural MRI and demonstrate an enhancement in diagnosis accuracy facilitated by synthesized ASL images. 
While achieving notable success, VAEs also present distinct limitations, since they are prone to producing outputs lack of image sharpness~\cite{dai2018diagnosing}. 
Additionally, these models tend to suffer from posterior collapse, where a subset of the latent space becomes redundant and does not contribute to the data generation process~\cite{bowman2015generating}.

Diffusion models have recently gained increasing attention in image synthesis~\cite{sohl2015deep,ho2020denoising,song2020score}. 
They are distinguished by their high training stability and capability to generate images of exceptional quality, offering solutions to the challenges commonly faced by GANs and VAEs~\cite{dhariwal2021diffusion,cao2024survey}. 
A few studies have attempted to use them to synthesize missing medical images~\cite{lyu2022conversion,xie2023synthesizing}. 
Lyu et al.~\cite{lyu2022conversion} utilize the Denoising Diffusion Probabilistic Model (DDPM) and score-based diffusion model to translate MR images into computed tomography (CT) scans, tackling output uncertainty through Monte-Carlo sampling~\cite{lyu2022conversion}. 
Their findings demonstrate that diffusion models outperform traditional convolutional neural networks and GANs in this modality conversion task.
Xie et al.~\cite{xie2023synthesizing} employ a joint diffusion attention model (JDAM) to generate synthetic PET images from high-field and ultra-high-field MRI scans (with MRI as a condition), by focusing on learning information about the joint probability distribution between MRI and noise PET. 
However, diffusion models tend to produce diverse outputs due to the randomness of input noise~\cite{lyu2022conversion}. 
This output variability may not be advantageous for cross-modality medical image translation where consistency and accuracy are crucial. 
In particular, the outcome uniqueness is critical in brain PET synthesis since image voxel intensities convey functional measures of brain activity, such as regional glucose consumption in FDG-PET and amyloid plaques in amyloid PET images~\cite{chetelat2020amyloid}. 
In this work, we introduce a functional imaging constraint to guide the diffusion model toward generating more consistent outputs.

\section{Methodology}
\label{S3}
\subsection{Problem Formulation} 
This work focuses on synthesizing 3D brain PET images based on structural MRI scans. 
We denote  $X_M$ as a 3D MR image and $X_P$ as its corresponding 3D PET image. 
We aim to train a model that can map a structural MRI to its real PET image, denoted as $\bm{f}$: $\{\x\in X_M\}$$\to$$ \{\x' \in X_P\}$. 
In the inference phase, an MRI scan can be input to the trained model to generate a PET image, thus helping alleviate the modality-missing issue. 
Considering the diffusion model's high training stability and the capability to generate high-quality images through learning Markov chain transitions for sample generation, we propose to use the diffusion model to synthesize 3D PET images with the guidance of structural MRI condition, as depicted in Fig.~\ref{framework}.

\subsection{Proposed Methodology}
\subsubsection{Overview}
As shown at the top of Fig.~\ref{framework}~(a), our functional imaging constrained diffusion (FICD) framework consists of a \emph{forward diffusion process} that incrementally adds Gaussian noise to the PET image, and a \emph{generative reverse denoising process} that progressively removes the noise. 
Given $T$ timesteps, the forward diffusion process is a Markov chain, defined as:
\begin{equation}
q(X_P^{1},\cdots,X_P^{T} | X_P^0) := \prod\nolimits_{t=1}^{T} q(X_P^t | X_P^{t-1}), 
\end{equation}
where $q$ is the posterior distribution, $X_P^0$ is the real PET image and $\{X_P^{1},\cdots,X_P^{T}\}$ is the disturbed samples from timestep 1 to \(T\).
The distribution of disturbed samples at timestep $t$ can be 
formulated with Gaussian transitions: 
\begin{equation}
q(X_P^t | X_P^0) = \mathcal{N}\left(X_P^t; \sqrt{\bar{\alpha}^t} \cdot X_P^0, (1 - \bar{\alpha}^t) \cdot \mathbf{I}\right),
\label{scheduler}
\end{equation}
where \(\mathcal{N}(\mu, \sigma)\) is Gaussian distribution with mean \(\mu\) and variance \(\sigma\), while $\bar{\alpha}^t := \prod_{s=1}^{t} \alpha^s$
is a time-dependent hyperparameter and $\mathbf{I}$ is the identity matrix indicating isotropic variance. 
In the generative reverse denoising process, the proposed constrained diffusion model (CDM) predicts noise on each sample on timestep $t$ and removes it to predict $X^{t-1}_P$ from $X^{t}_P$ iteratively, where $t\sim [T,0]$ and $X^t_P$ is the synthesized PET at timestep $t$. 

The bottom part of Fig.~\ref{framework}~(a) shows the architecture of CDM, which contains three steps.  
1) Input processing: At timestep $t$, a noise scheduler $\mathbf{NS}$ introduces random noise $\epsilon\sim N(0,1)$ into the PET image $X^0_P$ to create the input noisy image $X^t_P$, following the distribution in Eq.~\eqref{scheduler}. 
This image is then concatenated with a condition image $X_M$, and the model is also fed with the timepoint $t$ to determine the state of the noise scheduler.
2) Neural network learning: The input is first downsampled, and then upsampled to produce the output noise $\tilde \epsilon$, with the downsampled features integrated into the upsampling, and enhanced with attention blocks.
3) Output processing: With the predicted noise $\tilde \epsilon$, denoised $X^{t-1}_P$ and $\tilde X_P^0$ is estimated through an estimator $\mathbf{E}$ and a new $\mathbf{NS}$.
Additionally, Fig.~\ref{framework}~(b) depicts the inference phase where the trained model generates a synthetic PET image starting from pure noise (with an MRI as a condition).

\textbf{Condition}.  
To guide the synthesizing process toward a specific output, a condition can be introduced alongside the input noisy image.
In cross-modality image translation, the source modality can serve as a condition to facilitate the synthesis of the target modality. 
In FICD, we achieve this by concatenating the paired MRI $X_M$ with the noisy PET image, both having the same image dimension, as the input.

An alternative approach could be using the gradient image of MRI as the condition.
As proposed in~\cite{gong2023gradient}, when the diffusion denoising model is used to predict gradient, it can yield a higher training efficiency.
Given that PET images typically lack distinct gradient information due to their functional imaging nature, one can alternatively use the gradient of MRI $\nabla X_M$ that conveys rich anatomical information about the brain as the condition. 
A comparative analysis of this gradient-based approach with the direct use of MR images as the condition will be detailed in Section~\ref{S5_condition}.

\textbf{Training Phase}. 
In the \emph{training phase}, instead of iterating through every timestep, an intermediate timestep $t$ is randomly selected for each input subject to learn the denoising from $t$ to $t-1$.
To achieve this, a random Gaussian noise $\epsilon\sim N(0,1)$ is employed to corrupt the data point $X^0_P$, \ie, the PET scan, with the noise scheduler $\mathbf{NS}$. 
By reparameterizing $\alpha^t$ in Eq.~\eqref{scheduler}~\cite{ho2020denoising}, this corruption process can be mathematically expressed as:
\begin{equation}
X^t_P = \sqrt{\bar{\alpha}^t} X^0_P + \epsilon \sqrt{1 - \bar{\alpha}^t}.
\label{x_t}
\end{equation}
The corrupted data $X^t_P$ is then concatenated with the corresponding condition, \ie, the MRI scan, to form the input for the neural network. With the timestep $t$ embedded, the neural network outputs the predicted noise $\tilde{\epsilon}$ at $t$. 
A noise-level constraint using mean squared error loss is incorporated to encourage the estimated noise to be close to the real noise $\epsilon$: 
\begin{equation}
L_{N} = \frac{1}{n} \sum\nolimits_{i=1}^{n} \left( \epsilon - \tilde{\epsilon} \left( (X_M)^i, 
(X^t_P)^i,
t \right) \right)^2,
\label{eq_noiseLoss}
\end{equation}
where $n$ is the training sample size, while $(X_M)^i$ and $ (X^t_P)^i$ denote the MRI and noisy PET data for the $i$-th subject.

\begin{figure*}[!t]
\setlength{\belowdisplayskip}{-2pt}
\setlength{\abovedisplayskip}{-1pt}
\setlength{\abovecaptionskip}{-4pt}
\setlength{\belowcaptionskip}{-2pt}
\centering
\includegraphics[width=1\textwidth]{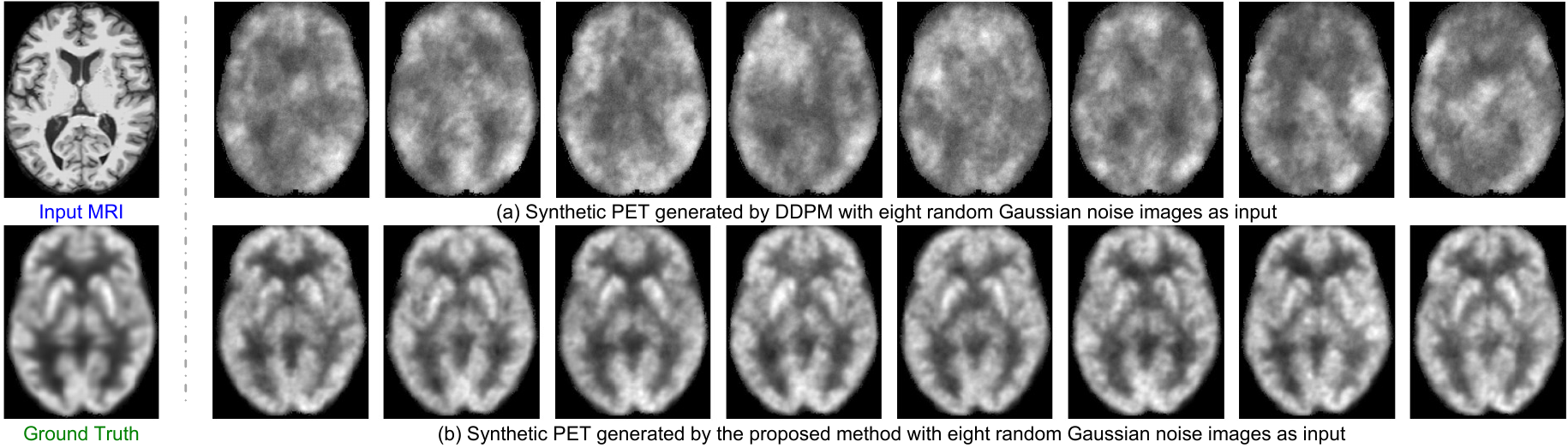}
\caption{Synthetic PET images of the same subject (ID: 002\_S\_4270) from ADNI~\cite{jack2008alzheimer} generated through eight separate sampling iterations by (a) DDPM~\cite{ho2020denoising} and (b) our FICD, with each iteration introducing new random noise.
The input MRI and the ground-truth PET are displayed on the left.}
\label{fig_variance}
\end{figure*}

\subsubsection{Functional Imaging Constraint}
In a vanilla denoising diffusion model~\cite{ho2020denoising}, while the noisy input is composed of the original image and input noise $\epsilon$, only the predicted noise $\tilde{\epsilon}$ is constrained in optimization. 
Therefore, the model only receives indirect information about the images it is tasked to synthesize, which may not be sufficient to ensure the production of images with voxel-level correspondence to the ground truth. 
Moreover, relying solely on noise for constraints introduces uncertainty throughout the entire framework, potentially affecting its performance. 
On the other hand, as a vital functional imaging tool, PET is particularly effective for assessing metabolic processes in the brain. 
The voxel intensities in PET images represent functional metrics of brain activity, such as regional glucose consumption in FDG-PET and amyloid plaques in amyloid PET. 
Motivated by this fact, we propose to use a \emph{functional imaging constraint} (FIC) to guide the model training, ensuring the similarity between synthetic and real PET images at each timestep. 
Specifically, we propose to generate a synthetic PET at the timestep $t$ through an image estimator $\mathbf{E}$, by reformulating Eq.~\eqref{x_t} as follows: 
\begin{equation}
\small
\tilde{X}^0_P = \frac{X^t_P  - \sqrt{1 - \bar{\alpha}^t} \, \tilde\epsilon(X_M, X^t_P, t)}{\sqrt{\bar{\alpha}^t}}. 
\label{x_0}
\end{equation} 
With the estimated PET image $\tilde{X}^0_P$, we employ an $l_1$ loss in this \emph{functional imaging constraint}, defined as:
\begin{equation}
L_{I} = \frac{1}{n} \sum\nolimits_{i=1}^{n} | X^0_{P} - \tilde X^0_P |,
\label{eq_immagLoss}
\end{equation}
which aims to minimize the voxel-wise discrepancy between the estimated image $\tilde X^0_P$ and the real PET $X^0_P$. 
This encourages the 
PET images generated in the intermediate process of the diffusion model to be close to the real one, thus helping preserve the functional information conveyed by voxel intensities in PET scans. 
This differs from the noise-level constraint defined in Eq.~\eqref{eq_noiseLoss} which focuses on accurately modeling the noise removal process at each diffusion step.   
It is worth noting that other types of functional constraints can also be utilized and the proposed FIC is readily adaptable to different methods like the latent diffusion model~\cite{rombach2022high}, which will be discussed in Section~\ref{S5}.

\subsubsection{Constrained Diffusion Model}
By incorporating the above-mentioned functional imaging constraint, we develop a constrained diffusion model (CDM), used in each timestep of our FICD. 
At timestep $t$, the input of CDM contains the perturbed PET image  $X^t_P$ and the paired 3D MRI $X_M$ that is integrated as a conditional variable. 
During training, the output of CDM is fed into an image estimator $\mathbf{E}$ to generate a PET $\tilde{X}_{P}^{0}$ (with noise removed) to facilitate the computation of functional imagining constraint. 
During inference (see Section~\ref{S3_inference}), the $\tilde{X}_{P}^{0}$ is fed into a noise scheduler with $X^t_P$ to generate a denoised PET $X_P^{t-1}$ (with noise removed according to the timestep).  
With Eq.~\eqref{eq_noiseLoss} and Eq.~\eqref{eq_immagLoss}, the CDM will be optimized by minimizing the following hybrid loss:
\begin{equation}
L=L_N+L_I. 
\label{eq_overallLoss}
\end{equation}

The bottom part of Fig.~\ref{framework}~(a) illustrates the network architecture of the CDM, consisting of three downsampling blocks and three upsampling blocks.  
Each of the three downsampling blocks contains a convolutional layer to reduce spatial dimensions and two residual sub-blocks that use skip connections. 
Each residual sub-block includes two convolutional layers with each of them going through a group normalization, a timestep embedding, and a shortcut connection~\cite{ho2020denoising}. 
The timestep embedding component projects the encoded timestep features to align with the number of feature channels, thereby integrating the timestep information into feature learning. 
The number of feature channels increases throughout the downsampling blocks, from 16 to 32, then to 64.
To accommodate the increment in feature channel numbers, the architecture includes two variations of residual blocks. 
When the number of input feature channels matches the output, the input and output are summed for a skipping connection. 
If the output feature channels exceed the number of input channels, a convolutional layer is introduced to the shortcut connection to align the dimensions accordingly. 
The three upsampling blocks work in reverse to downsampling, enlarging image dimensions while decreasing the number of feature channels. 
The output from each downsampling block is concatenated with the corresponding upsampling block to preserve image details~\cite{ronneberger2015u}. 
This architecture is enhanced by incorporating attention blocks that perform self-attention on the learned intermediate features, thereby emphasizing the important features for improved model performance.
Each attention block has 8 attention heads with 64 channels~\cite{vaswani2017attention}.

\subsection{Inference Phase}
\label{S3_inference}
As shown in Fig.~\ref{framework}~(b), to obtain synthesized PET from MRI, CDM is applied to remove noise from the input through generative reverse denoising process, 
starting with Gaussian noise $\epsilon\sim \mathcal{N}(0,1)$ concatenated with MRI of matching dimensions as the input. 
With the input noisy image $X_P^t$ (pure noise $\epsilon$ when $t=T$) and the noise scheduler $\mathbf{NS}$ at timestep $t$, 
and the estimated output $\tilde{X}_P^0$ in Eq.~\eqref{x_0}, the noise distribution~\cite{ho2020denoising} at timestep $t-1$ can be written as: 
\begin{equation}
q(X_P^{t-1} | X_P^t,\tilde{X}^0_P) = \mathcal{N}\left(X_P^{t-1}; \tilde{\mu}^t(X_P^t,\tilde{X}^0_P) 
, \tilde \alpha^t \mathbf{I}\right),
\label{noise_t-1}
\end{equation}
where 
\begin{equation}
\small
\tilde\mu^t (X_P^t, \tilde{X}^0_P) := \frac{\sqrt{\bar{\alpha}^{t-1}}(1-\alpha^t)}{1 - \bar\alpha^t}\tilde{X}^0_P + \frac{\sqrt{\alpha^t}(1 - \bar\alpha^{t-1})}{1 - \bar\alpha^t}X_P^t,
\label{noise_t-1_u}
\end{equation}
\begin{equation}
\small
\tilde \alpha^t = \frac{1 - \bar{\alpha}^{t-1}}{1 - \bar\alpha^t} (1-\alpha^t) \cdot 
\label{noise_t-1_alpha}
\end{equation}
The sample result of $X_P^{t-1}$, serving as the noisy PET input for the subsequent iteration, can therefore be formulated as:
\begin{equation}
\small
X_P^{t-1} = \frac{1}{\sqrt{\alpha^t}} \left( X_P^t - \frac{1 - \alpha^t}{\sqrt{1 - \bar\alpha^t}} \tilde\epsilon(X_M,X^t_P,t) \right) + \sigma^t z,
\label{x_t-1}
\end{equation}
where $z\sim N(0,1)$ and $\sigma^t$ denotes the variance~\cite{ho2020denoising}. 
Through $T=1,000$ inference timesteps, the model gradually removes noise and ultimately yields the PET image.

During inference, the randomness of noise would introduce uncertainty to the output. 
This is a common challenge in denoising diffusion models for medical
image synthesis~\cite{lyu2022conversion}. 
Figure~\ref{fig_variance} shows eight outputs for the same subject, each image synthesized by   DDPM~\cite{ho2020denoising} and the proposed FICD using different instances of Gaussian random noise. 
From Fig.~\ref{fig_variance}~(a), one can observe that each of the synthesized images generated by DDPM is different. 
Figure~\ref{fig_variance}~(b) suggests that the synthesized images generated by FICD have an overall resemblance, 
which could be attributed to the proposed functional imaging constraint to reduce output uncertainty to some extent. 
In addition, a common post-processing strategy to address this issue is to employ the Monte-Carlo (MC) sampling method~\cite{lyu2022conversion}, by repeating the denoising process multiple times to generate multiple synthetic outputs and using the averaged results as the final output. 
This helps to further mitigate the effects of randomness in image synthesis.  
We will discuss the influence of MC sampling strategy in \emph{Supplementary Materials}.

\subsection{Implementation Details}
The FICD is implemented using the PyTorch-based MONAI framework [45] and operates within two environments: Python 3.11.5 on an RTX 3090 GPU with 24 GB of memory, and Python 3.12.7 on a GPU cluster with four H100 GPUs (each with 80 GB of memory). 
The training of FICD involves 50 epochs with a batch size of 2. 
The Adam optimizer is employed with a learning rate established at $5\times10^{-5}$.
In line with~\cite{ho2020denoising}, we empirically set the timestep $T$ as $1,000$ and linearly increase scheduler $1-\alpha^t$ from 0.0005 to 0.0195.
All 3D MRI and PET scans used for training image synthesis models in this work are paired for each subject. 
With FICD, the inference time for generating a synthesized 3D PET image is approximately 30 seconds on the GPU cluster.

\section{Experiment}
\label{S4}

\subsection{Subjects and Multimodal Image Preprocessing}

Three cohorts are used for performance evaluation: 1) the Alzheimer's Disease Neuroimaging Initiative (ADNI)~\cite{jack2008alzheimer}, 
2) the Chinese Longitudinal Aging Study (CLAS)~\cite{xiao2016china}, and 
3) the Australian Imaging, Biomarkers and Lifestyle (AIBL) study~\cite{ellis2009australian}. 
This work uses baseline imaging data from the three cohorts. 
Detailed information on the studied subjects 
is given in Table~S4 of \emph{Supplementary Materials}.

1) \textbf{ADNI}. 
A total of $856$ subjects from ADNI are utilized in this study, including $359$ individuals diagnosed with Alzheimer's disease (AD), $436$ cognitively normal (CN) individuals, and $61$ subjects identified as clinically normal but exhibiting significant memory concerns (SMC), which may indicate preclinical AD~\cite{jessen2014conceptual}.
Among the SMC subjects, based on 5-year follow-up results, $19$ are identified as having progressive significant memory concern (pSMC), and $42$ subjects are classified as stable significant memory concern (sSMC). 
Specifically, as all of these subjects exhibit SMC at baseline time, those who maintain SMC over a 5-year follow-up are categorized as sSMC, while those who progress to mild cognitive impairment (MCI) or AD during the same period are classified as pSMC. 
Among all of those subjects, $293$ CN subjects, $240$ AD subjects, and all SMC subjects have both baseline T1-weighted structural MRI and $^{18}$F-fluorodeoxyglucose (FDG)-PET scans available. 
The remaining subjects only have MRI data. 

2) \textbf{CLAS}. 
The CLAS is a de-identified database, designed to provide information about the cognitive, mental, and psychosocial health of older people in China.  
This is a joint effort of 15 research centers located in the eastern, middle, and western parts of China, and samples are randomly selected from all permanent residents aged over 60 in the 2010 national census.  
It comprises 75 preclinical AD subjects who self-report having significant cognitive decline (SCD) at baseline. 
To be noted, SCD generally refers to the same disease stage as SMC, with slight differences in diagnostic standards~\cite{abdulrab2008subjective,aisen2015alzheimer}.
Based on a 7-year longitudinal follow-up, 51 subjects retained SCD status, and thus, are classified as stable SCD (sSCD). 
The remaining 24 subjects progress to MCI and are identified as progressive SCD (pSCD). 
The CLAS dataset exclusively contains baseline T1-weighted MRI data but lacks PET data.

3) \textbf{AIBL}. 
The AIBL provides both MRI and PET images from multiple sites. 
With the development of amyloid PET radiotracers, the study updates the radiotracers multiple times throughout their longitudinal study, providing PET images of three types of radiotracers, namely Pittsburgh Compound-B (PiB), $^{18}$F-flutemetamol (FLUTE), and Florbetapir (AV45). 
In this work, we employ the paired baseline amyloid PET and T1-weighted MRI from 331 CN subjects.
These subjects are divided into three groups based on PET radiotracers, including 119 subjects with PiB-PET, 143 subjects with FLUTE-PET, and 69 subjects with AV45-PET.  
We manually exclude scans that fail during preprocessing or exhibit significantly low image quality.

\textbf{Image Processing}.
We utilize a standardized processing pipeline to unify images across all datasets. 
For 3D MRI scans, we apply skull stripping to each brain, normalize image intensity, and then register the brain to the MNI space. 
For 3D PET scans, we first apply skull stripping, then linearly align each PET to its paired MRI, and finally register the linearly aligned PET to MNI space with the deformation matrix created by MRI registration.
All MRI and PET scans are uniformly cropped from the original dimensions  (\ie, $181 \times 217 \times 181$ with voxel dimensions of $1 \times 1 \times 1$) to the dimensions of $160 \times 180 \times 160$, to discard the uninformative background while still keep the whole brain.
The image intensity is normalized to $[-1,1]$ in the training phase, and $[0,1]$ for computing the quantitative results. 

\subsection{Competing Methods}
We compare the proposed FICD with six state-of-the-art 3D deep generative methods, including two GAN-based models, two VAE-based models, and two diffusion models.

1) \textbf{GAN}: 
The GAN contains a generator and a discriminator. 
The generator has a 3-layer encoder (with convolutional channel numbers of 16, 32, and 64), six residual blocks, and a 2-layer decoder (with deconvolutional channel numbers of 32 and 16), followed by a convolutional layer. 
Filter sizes are $7\times7\times7$ for the first and last layers, and $3\times3\times3$ for all others. 
The discriminator includes five convolutional layers with channels increasing from 32 to 256, ending with a single output channel.

2) \textbf{CycleGAN}: 
It includes two generators and two adversarial discriminators that form two branches for MRI-to-PET and PET-to-MRI prediction. 
The generator has a 3-layer encoder (with convolutional channel numbers of 8, 16, and 32), six residual blocks, and a 2-layer decoder (with deconvolutional channel numbers of 16 and 8), followed by a final convolutional layer. 
The generator's first and last layers use a $7\times7\times7$ filter size, while all other layers employ a $3\times3\times3$ filter size.
The discriminator consists of five convolutional layers with channel counts rising from 16 to 128, culminating in a single output channel.  
We leverage the public pre-trained CycleGAN~\cite{pan2018synthesizing} in the experiments. 

3) \textbf{VAE}: 
It consists of an encoder and a decoder. 
The encoder is built from five convolutional blocks, with the channel numbers increasing from 16 to 256, and with the feature map size progressively decreasing. 
Following each convolutional block is a residual block. 
The encoder's output is then directed through two separate linear layers for computing the mean and variance. 
These are utilized for calculating the Kullback-Leibler divergence and shaping the distribution that serves as input to the decoder. 
Mirroring the encoder, the decoder features a symmetrical architecture but in reverse order, reconstructing the input data from the latent space representations generated by the encoder.

4) \textbf{VAEGAN}: 
The VAEGAN~\cite{larsen2016autoencoding} integrates a VAE as the generator, alongside a discriminator for adversarial training.
Due to computational power limitation, the channel numbers for this model are half of that in VAE, \ie, 8, 16, 32, 64, and 128.
The discriminator has four convolutional layers, with channel numbers 8, 16, 32, and 1 for label prediction.

5) \textbf{DDPM}: 
The DDPM~\cite{ho2020denoising} contains a forward diffusion process based on a Markov chain and learns the reverse denoising process.
Similar to our method, MRI is used as the condition in this model. 
The DDPM can be treated as a vanilla framework of our FICD. 
It shares a similar foundational structure with FICD while only the noise-level constraint is utilized for  training. 
For a fair comparison, we maintain identical parameters, model structures, and Monte-Carlo sampling times in the diffusion-based models as those used in FICD's training throughout all  experiments.

6) \textbf{LDM}: This method is based on the latent diffusion model (LDM) introduced in~\cite{rombach2022high}, which includes an encoder for image feature extraction, a decoder for image reconstruction, and a denoising diffusion model for modality translation within the latent space (dimension: $10 \times12\times 10$). 
Initially, the encoder and decoder are trained jointly across 100 epochs to reconstruct PET images, utilizing the entire set of PET images in the training dataset. 
We then use the locked weights of the encoder to extract features from PET and MRI. 
Afterward, we utilize the denoising diffusion model to convert MRI latent features into PET latent features, and the decoder, with its weights fixed as well, is applied to reconstruct the synthesized PET images.
The Monte-Carlo sampling time is the same as FICD and DDPM.

\begin{figure*}[!t]
\setlength{\abovecaptionskip}{0pt}
\setlength{\belowcaptionskip}{0pt}
\setlength\belowdisplayskip{0pt}
\setlength{\abovecaptionskip}{-4pt}
\centering
\includegraphics[width=1\textwidth]{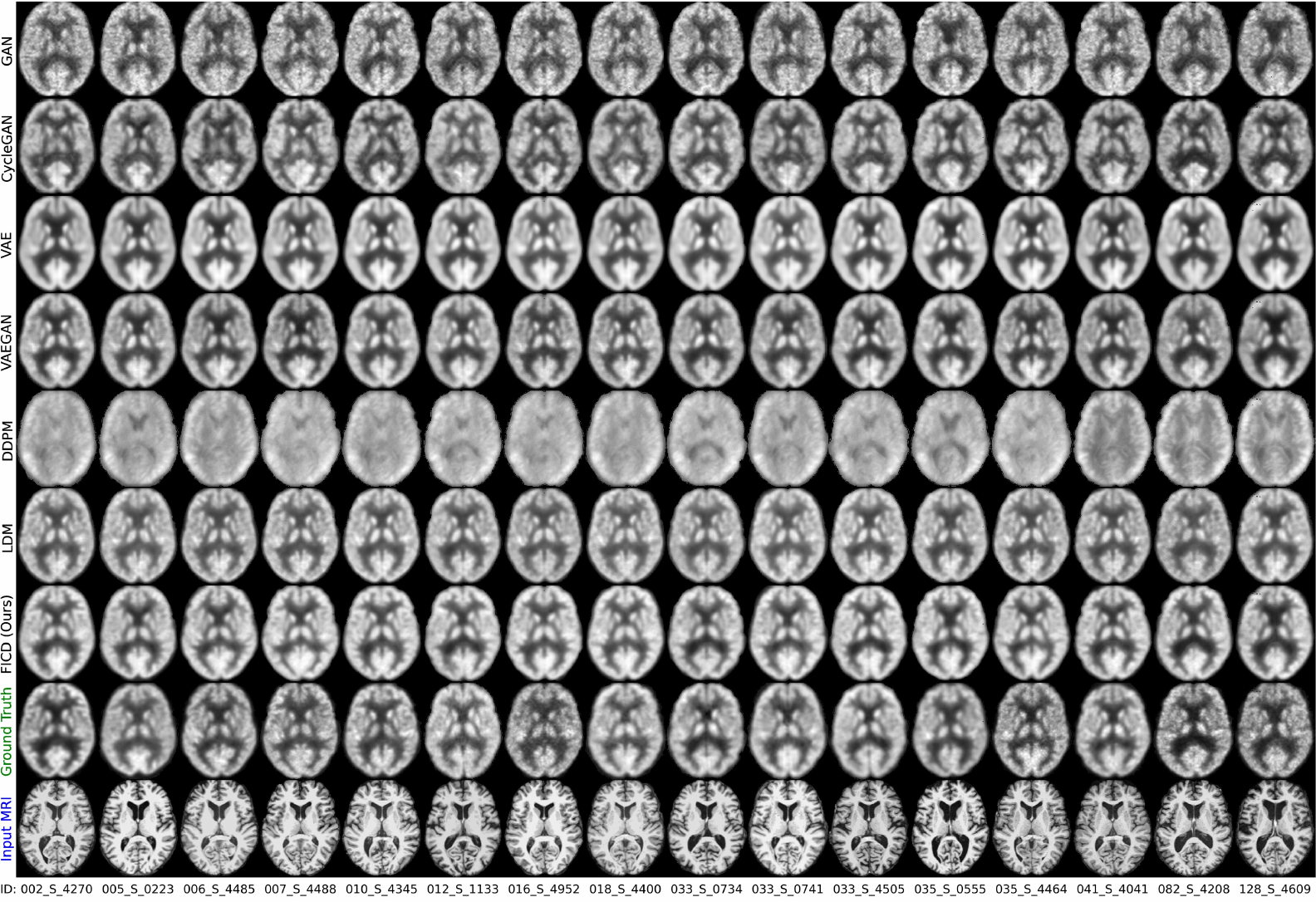}
\caption{Visualization of PET images synthesized by the proposed FICD and six competing methods on cognitively normal subjects from the test set in ADNI~\cite{jack2008alzheimer}. The ground-truth PET images and input MRI are displayed at the bottom with the corresponding subject ID.}
\label{qualitative}
\end{figure*}

\subsection{Experimental Settings}
Four tasks are performed in the experiments: 
1) image synthesis, 
2) disease progression forecasting,  
3) prediction of future cognitive function, and 
4) generalization evaluation. 
For Task 1, we train the proposed FICD and competing methods on ADNI CN subjects, and evaluate their performance in generating FDG-PET images with ground-truth scans available. 
For Task 2 and Task 3, we first use the models trained in Task 1 to synthesize missing PET data for ADNI and CLAS, and then perform downstream prediction. 
For Task 4, we use the AIBL data, where amyloid PET scans with three radiotracers (\ie, PiB, FLUTE, and AV45) are available. 
Subject IDs can be found in Tables S5-S7 of \emph{Supplementary Materials} to facilitate reproducible research.

\textbf{Task 1: Image Synthesis}. 
To assess the quality of the synthesized image, a total of $293$ CN subjects with paired T1-weighted MRI and FDG-PET from ADNI are used in this task. 
Specifically, $263$ subjects are used for model training, while the rest $30$ subjects are used for testing. 
For diffusion models (\ie, DDPM, LDM, and FICD), the outputs are the average of 10 times using MC sampling strategy.  
We use four evaluation metrics to assess the image quality at the 3D volume level, including 1) peak signal-to-noise ratio (PSNR), 2) structural similarity index measure (SSIM), 3) mean absolute error (MAE), and 4) normalized mutual information (NMI).
For evaluation, the synthesized images are padded to their original size.

\textbf{Task 2: Disease Progression Forecasting}.
This task aims to assess how effectively synthesized PET scans from a generative model can aid in training a downstream model for predicting the progression of preclinical AD. 
We deploy the trained FICD and each competing method from Task 1 to synthesize the missing PET for $119$ AD subjects and $143$ CN subjects from the ADNI dataset, as well as for all the subjects from the CLAS dataset.
Given the time-consuming nature of the process, we apply Monte-Carlo with a sampling time of five for the diffusion-based models (\ie, DDPM, LDM, and FICD) for this task and all subsequent ones.
Following this, a classification model is trained on all available MRI and PET pairs, including both real and synthesized PET, to identify AD patients from CN.
The classification model incorporates a dual-branch convolutional neural network (CNN) architecture with two branches of identical encoders for each modality. 
The features extracted from these encoders are then concatenated and subsequently fused in a transformer self-attention module~\cite{vaswani2017attention} for final result prediction.
Driven by the challenge of limited sample sizes in preclinical AD-related research~\cite{jack2008alzheimer,xiao2016china}, 
the model (trained for AD vs. CN classification) is then used to classify between progressive and stable preclinical AD in a transfer learning manner. 
Two preclinical AD datasets are used: 1) SMC subjects from ADNI (denoted as ADNI-SMC), who have both modalities data, and 2) SCD subjects from CLAS (denoted as CLAS-SCD), which only has MRI and has PET synthesized by each method. 
We repeat the training phase for this classification model five times with different random parameter initializations and record the mean and standard deviation results. 
Six evaluation metrics are used:  area under the ROC curve (AUC), accuracy (ACC), sensitivity (SEN), specificity (SPE), balanced accuracy (BAC), and F1-Score (F1s).

\textbf{Task 3: Prediction of Future Cognitive Function}.
Mini-mental state examination (MMSE) and clinical dementia rating (CDR) 
are pivotal tools for cognitive function evaluation and are instrumental in diagnosing and tracking progression of neurodegenerative diseases, including Alzheimer's disease and other dementia~\cite{aisen2010clinical,xiao2016china,morris1997clinical}. 
Therefore, we include the task of predicting the future MMSE and CDR scores for preclinical AD subjects based on baseline MRI and PET. 
For the ADNI-SMC cohort, we focus on predicting cognitive scores at 2-year follow-up, with $58$ subjects having MMSE scores available (18 classified as pSMC and $40$ classified as sSMC) and $17$ subjects have CDR scores available ($11$ classified as pSMC and $6$ classified as sSMC).  
For the CLAS-SCD cohort, all the subjects have MMSE scores available and have no CDR scores. 
We employ the same model structure as Task 2 with the output layer modified for linear regression to predict each type of cognitive score. 
The training phase for this prediction model is repeated five times with different random initializations. 
The outcomes of these repeated training are then averaged, and we assess the model's predictive accuracy by calculating the correlation coefficient (CC), root mean squared error (RMSE), and mean absolute error (MAE) between the predicted and real scores.

\begin{table}[!t]
\setlength{\abovecaptionskip}{0pt}
\setlength{\belowcaptionskip}{0pt}
\setlength{\abovedisplayskip}{0pt}
\setlength\belowdisplayskip{-0pt}
\setlength{\tabcolsep}{3.5pt}
\scriptsize
\renewcommand{\arraystretch}{0.9}
\tiny
\centering
\caption{Quantitative results achieved by FICD and six competing methods for FDG-PET generation on ADNI, with best results shown in bold.}
\label{sites}
\begin{tabular}{l|cccc}
\toprule
Method & PSNR$\uparrow$ & SSIM$\uparrow$ & MAE$\downarrow$ & NMI$\uparrow$\\ 
\midrule
GAN & $24.3918$$\pm$$0.4501$ & $0.8279$$\pm$$0.0001$& $0.0252$$\pm$$0.0022$& $0.7313$$\pm$$0.0208$\\
CycleGAN & $25.0409$$\pm$$1.0553$ & $0.8691$$\pm$$0.0005$& $0.0236$$\pm$$0.0031$& $0.7490$$\pm$$0.0278$\\
VAE & $26.6605$$\pm$$0.8572$ & $0.8838$$\pm$$0.0007$ & $0.0195$$\pm$$0.0022$ & $0.7944$$\pm$$0.0329$\\
VAEGAN & $26.8577$$\pm$$0.7049$ & $0.8821$$\pm$$0.0005$ & $0.0191$$\pm$$0.0021$ & $0.8044$$\pm$$0.0243$\\
LDM & $26.3088$$\pm$$0.9892$ & $0.8770$$\pm$$0.0215$ & $0.0206$$\pm$$0.0026$ & $0.8008$$\pm$$0.0287$\\
DDPM & $21.6955$$\pm$$0.6776$ & $0.8319$$\pm$$0.0290$ & $0.0367$$\pm$$0.0030$ & $0.6929$$\pm$$0.0218$\\
FICD~(Ours) & $\mathbf{27.8847}$$\pm$$1.1676$ & $\mathbf{0.9124}$$\pm$$0.0239$ & $\mathbf{0.0173}$$\pm$$0.0026$ & $\mathbf{0.8603}$$\pm$$0.0355$\\
\bottomrule
\end{tabular}
\label{quantitative}
\end{table}

\begin{figure*}[!t]
\setlength{\abovecaptionskip}{0pt}
\setlength{\belowcaptionskip}{0pt}
\setlength\belowdisplayskip{0pt}
\setlength{\abovecaptionskip}{-2pt}
\centering
\includegraphics[width=0.98\textwidth]{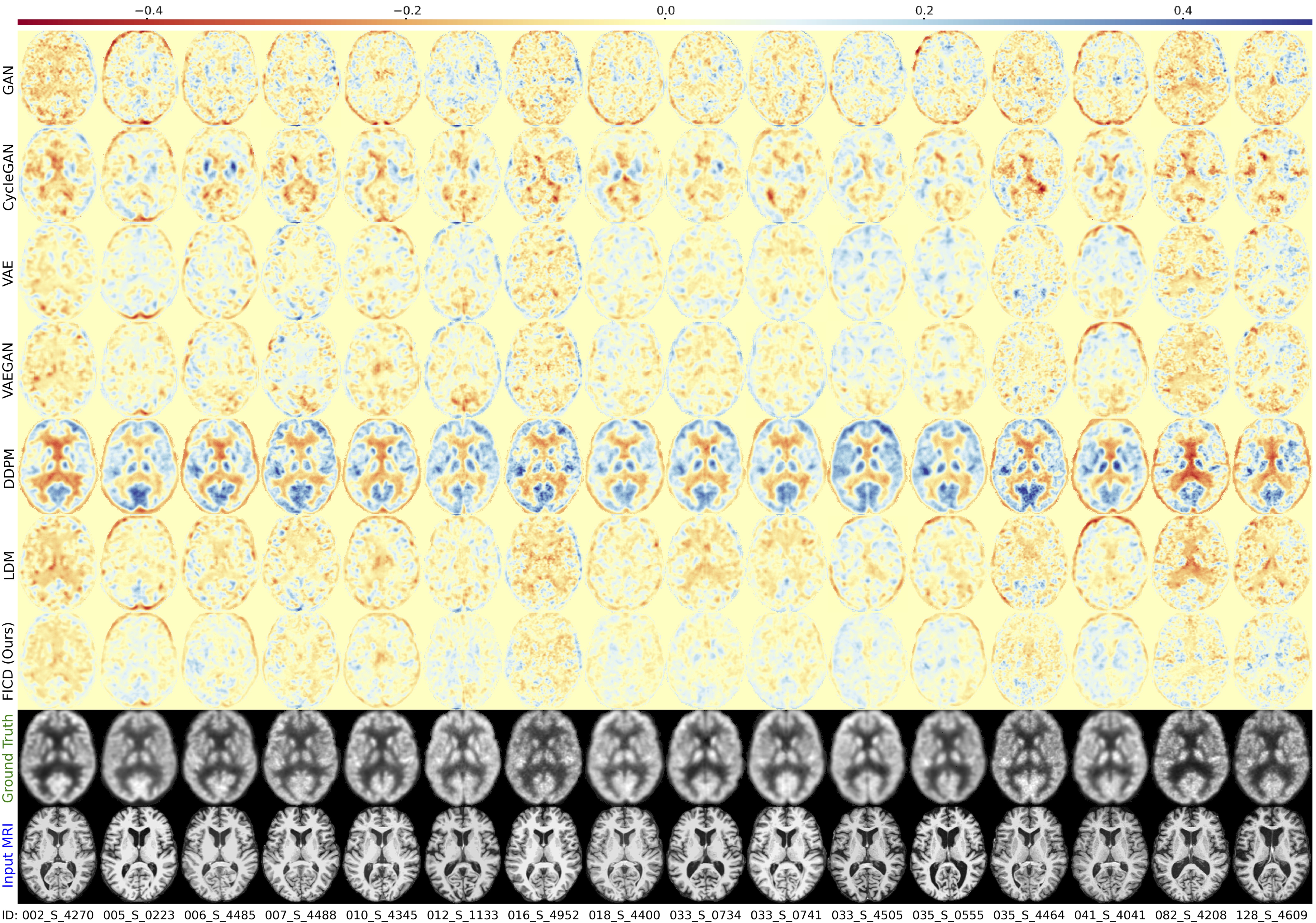}
\caption{Difference maps for synthetic PET images generated by the proposed FICD and six competing methods on cognitively normal subjects from the test set in ADNI~\cite{jack2008alzheimer}. The ground-truth PET images and input MRI are displayed at the bottom with the corresponding subject ID.}
\label{error_map}
\end{figure*}

\begin{table*}[!t]
\setlength{\abovecaptionskip}{0pt}
\setlength{\belowcaptionskip}{-3pt}
\setlength{\abovedisplayskip}{0pt}
\setlength\belowdisplayskip{0pt}
\renewcommand{\arraystretch}{0.86}
\scriptsize
\centering
\caption{Results (mean$\pm$standard deviation) of different methods in the tasks of CLAS-SCD progression prediction (\ie, pSCD vs. sSCD classification) and ADNI-SMC progression prediction (\ie, pSMC vs. sSMC classification) with the input of PET and MRI scans.}%
\setlength\tabcolsep{2pt}
\begin{tabular*}{1\textwidth}{@{\extracolsep{\fill}}l|cc cc cc c|c cc cc cc}
\toprule
\multirow{2}{*}{~Method} &
\multicolumn{6}{c}{pSCD vs. sSCD classification on CLAS-SCD} &&
\multicolumn{6}{c}{pSMC vs. sSMC classification on ADNI-SMC} \\
\cmidrule{2-14}
&AUC (\%) $\uparrow$&ACC (\%)$\uparrow$ &SEN (\%)$\uparrow$ &SPE (\%)$\uparrow$ &BAC (\%)$\uparrow$ &F1s (\%)$\uparrow$ &&AUC (\%)$\uparrow$ &ACC (\%)$\uparrow$ &SEN (\%)$\uparrow$ &SPE (\%)$\uparrow$ &BAC (\%)$\uparrow$ &F1s (\%)$\uparrow$  \\ 

\midrule
~GAN 
&60.89±7.62&58.13±4.59&61.67±7.17&56.47±3.37&59.07±5.27&48.52±5.64
&&67.12±7.33&60.33±4.91&65.26±7.88&58.10±3.56&61.68±5.72&50.61±6.11\\  
~CycleGAN 
&57.11±3.27&55.47±2.00&57.50±3.12&54.51±1.47&56.00±2.29&45.25±2.45
&&66.89±2.30&58.36±2.45&62.11±3.94&56.67±1.78&59.39±2.86&48.16±3.05 \\  
~VAE
&56.62±5.00&52.80±3.11&53.33±4.86&52.55±2.29&52.94±3.57&41.97±3.82
&&63.66±7.69&60.33±5.33&65.26±8.55&58.10±3.87&61.68±6.21&50.61± 6.63\\  
~VAEGAN
&57.22±5.93&54.40±2.72&55.83±4.25&53.73±2.00&54.78±3.12&43.93±3.34
&&67.89±4.57&59.02±2.07&63.16±3.33&57.14±1.51&60.15±2.42&48.98±2.58 \\ 
~DDPM
&53.64±5.64&51.73±3.62&51.67±5.65&51.76±2.66&51.72±4.16&40.66±4.45
&&65.99±5.33&59.67±3.21&64.21±5.16&57.62±2.33&60.91±3.74&49.80±4.00 \\ 
~LDM 
&57.17±4.32&56.00±4.46&58.33±6.97&54.90±3.28&56.62±5.13&45.90±5.49
&&68.80±5.52&61.64±6.36&67.37±10.21&59.05±4.62&63.21±7.41&52.24±7.91 \\ 
~FICD~(Ours) &\textbf{63.77}±2.72&\textbf{58.67}±2.39&\textbf{62.50}±3.73&\textbf{56.86}±1.75&\textbf{59.68}±2.74&\textbf{49.18}±2.93
&&\textbf{72.78}±2.83&\textbf{63.61}±2.62&\textbf{70.53}±4.21&\textbf{60.48}±1.90&\textbf{65.50}±3.06&\textbf{54.69}±3.27\\
\bottomrule
\end{tabular*}
\label{image_input_classification}
\end{table*}

\begin{figure*}[!tbp]
\setlength{\belowdisplayskip}{-0pt}
\setlength{\abovedisplayskip}{-0pt}
\setlength{\abovecaptionskip}{-2pt}
\setlength{\belowcaptionskip}{-0pt}
\centering
\includegraphics[width=0.98\textwidth]{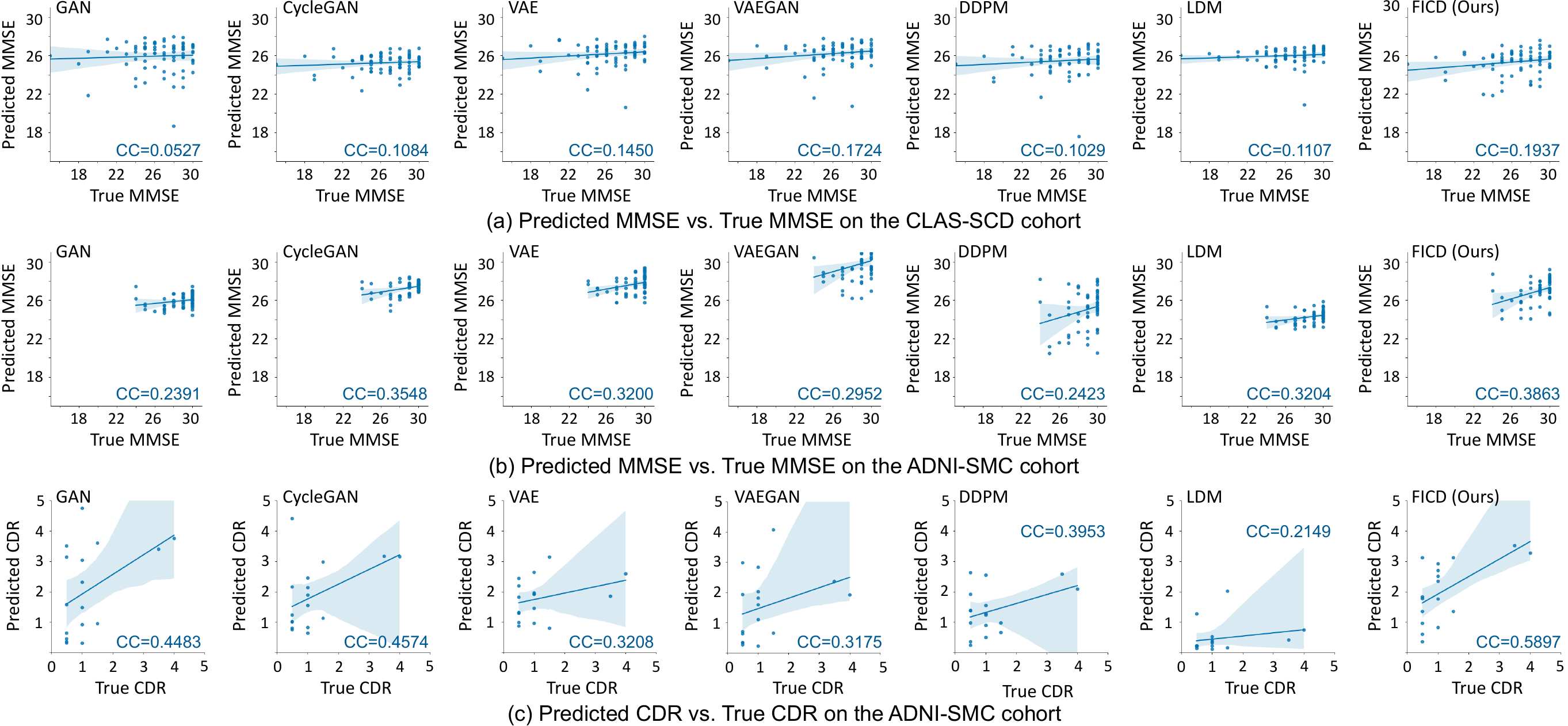}
\caption{Scatter diagrams of cognitive scores (\ie, MMSE and CDR) predicted by different methods for CLAS-SCD at 7-year follow-up and ADNI-SMC at 2-year follow-up, with baseline images as input. CC: correlation coefficient; MMSE: mini-mental state examination; CDR: clinical dementia rating.}
\label{fig_regression}
\end{figure*}

\textbf{Task 4: Generalization Evaluation}. 
Amyloid PET which shows brain amyloid deposition is another pivotal molecular imaging biomarker for the investigation of dementia, providing complementary information for FDG-PET that measures regional glucose consumption~\cite{chetelat2020amyloid}. 
Therefore, we extend the application of FICD to encompass the synthesis of amyloid PET images associated with several commonly employed radiotracers (\ie, PiB, FLUTE, and AV45). 
Specifically, for each type of 
radiotracer, we fine-tune FICD model initially pre-trained on FDG-PET data from the ADNI for 7 epochs, using MRI inputs to generate amyloid PET images.
Same as the settings in Task 1, the AIBL data is partitioned into 90\% for fine-tuning and 10\% for test, with the detailed subject numbers reported in Table~S4 and subject IDs in Table~S7 of \emph{Supplementary Materials}.

\subsection{Results of Image Synthesis}
\subsubsection{Quantitative Results}
Table~\ref{quantitative} shows the quantitative results of our proposed FICD with six other competing methods.
The results illustrate that FICD outperforms all other methods across every evaluation metric. 
Notably, compared to DDPM, our method enhances the PSNR by nearly 30\% and reduces the MAE by more than half. 
These improvements underscore the significant impact of the introduced functional imaging constraint in improving the accuracy of the synthesized images. 
Additionally, LDM is inferior to our FICD, which may be due to potential information loss during its encoding process that maps images into low-dimensional latent space.  
Both VAE-based methods (\ie, VAE and VAEGAN) produce PET images that yield satisfactory quantitative results, surpassing both GAN-based methods (\ie, GAN and CycleGAN) and diffusion-based competing methods (\ie, DDPM and LDM). 
Among the GAN-based methods, CycleGAN achieves better results than GAN across all evaluation metrics.

\subsubsection{Qualitative Results}
Figure~\ref{qualitative} compares the axial slices of PET images synthesized by the FICD and six other methods, with the last two rows showing the ground-truth PET and input MRI, respectively. 
We further show the difference maps of each synthetic PET image in Fig.~\ref{error_map}, which are generated by calculating the difference between a synthetic PET and its ground truth. 
Visualizations of these images from the sagittal and coronal planes are available in Figs.~S1-S2 of \emph{Supplementary Materials}. 
From Fig.~\ref{qualitative}, we can observe that the PET images generated by FICD are distinguished by their superior quality and similarity to the ground truth. 
This is particularly evident in gray matter regions, including the \emph{temporal lobes}, \emph{precuneus}, and \emph{parietal cortex}.
These regions exhibit typical patterns of hypometabolism associated with Alzheimer's disease, and thus, accurate representation of these regions is essential for early detection and diagnosis~\cite{shivamurthy2015brain,dumba2019clinical,chetelat2020amyloid}.

\begin{table}[!t]
\setlength{\abovecaptionskip}{0pt}
\setlength{\belowcaptionskip}{-2pt}
\setlength{\abovedisplayskip}{0pt}
\setlength\belowdisplayskip{0pt}
\renewcommand{\arraystretch}{0.86}
\scriptsize
\centering
\caption{Results of different methods in cognitive function prediction for CLAS-SCD at 7-year follow-up and ADNI-SMC at 2-year follow-up, with the baseline imaging data as input.}
\setlength\tabcolsep{1pt}
\begin{tabular}{l|ccc|ccc|ccc}
\toprule
\multirow{2}{*}{~Method} &
\multicolumn{3}{c|}{MMSE~of~CLAS-SCD}&
\multicolumn{3}{c|}{MMSE~of~ADNI-SMC}&
\multicolumn{3}{c}{CDR~of~ADNI-SMC}\\
\cmidrule{2-10}
& CC$\uparrow$  & RMSE$\downarrow$  & MAE$\downarrow$ & CC$\uparrow$  & RMSE$\downarrow$  & MAE$\downarrow$& CC$\uparrow$  & RMSE$\downarrow$  & MAE$\downarrow$ \\
\midrule
~GAN &0.0527&3.4817&2.7573&0.2391&3.2200&2.9941&0.4483&1.5876&1.0944\\
~CycleGAN &0.1084&3.3857&2.7751&0.3548&2.1290&1.9563&0.4574&1.2650&0.9123\\
~VAE &0.1450&3.1875&2.4738&0.3200&\textbf{1.1921}&\textbf{1.6664}&0.3208&1.0339&0.8554\\
~VAEGAN &0.1724&\textbf{3.1571}&\textbf{2.4230}&0.2952&2.2013&1.6909&0.3175&1.2697&0.9756\\
~DDPM &0.1029&3.4389&2.7417&0.2423&4.3843&3.9687&0.3953&\textbf{1.0012}&\textbf{0.8014}\\
~LDM &0.1107&3.1838&2.4991&0.3204&4.6849&4.4609&0.2149&1.2404&0.872\\
~FICD~(Ours) &\textbf{0.1937}&3.3258&2.7092&\textbf{0.3863}&2.4238&2.0940&\textbf{0.5897}&1.2320&0.9852\\
\bottomrule
\end{tabular}
\label{image_input_regression}
\end{table}

The PET images synthesized by GAN and CycleGAN exhibit good similarity to the ground truth but with more noticeable noise. 
Besides, the images synthesized by VAE and VAEGAN generally show limited diversity, indicating a relatively uniform output regardless of variations in the input, compared with other methods. 
This uniformity can often be attributed to a weak association between the posterior distribution and the model input, a phenomenon frequently encountered in VAE~\cite{lucas2019don}. 
This may prevent the model from recognizing and encoding the unique characteristics of individual subjects. 
As a result, while the model may achieve low quantitative errors, it lacks the specificity required for detailed analysis of individual disease cases, compromising its utility in personalized diagnostics.
Besides, VAE-generated PET tends to be blurrier compared to those from other methods. 
Among various competing methods, LDM generates images of acceptable quality. 
However, it does not preserve image details as accurately as FICD.
The DDPM produces images with brightness around the brain's periphery but a diminished representation of the brain details.
The synthesized output predominantly maintains the structure of the input MRI, resulting in what might be considered a PET-style MRI scan.
This may be attributed to the fact that ground-truth PET data is not included in  optimization.

\begin{figure*}[!t]
\setlength{\abovecaptionskip}{-2pt}
\setlength{\belowcaptionskip}{0pt}
\setlength{\abovedisplayskip}{0pt}
\setlength\belowdisplayskip{0pt}
\centering
\includegraphics[width=0.98\textwidth]{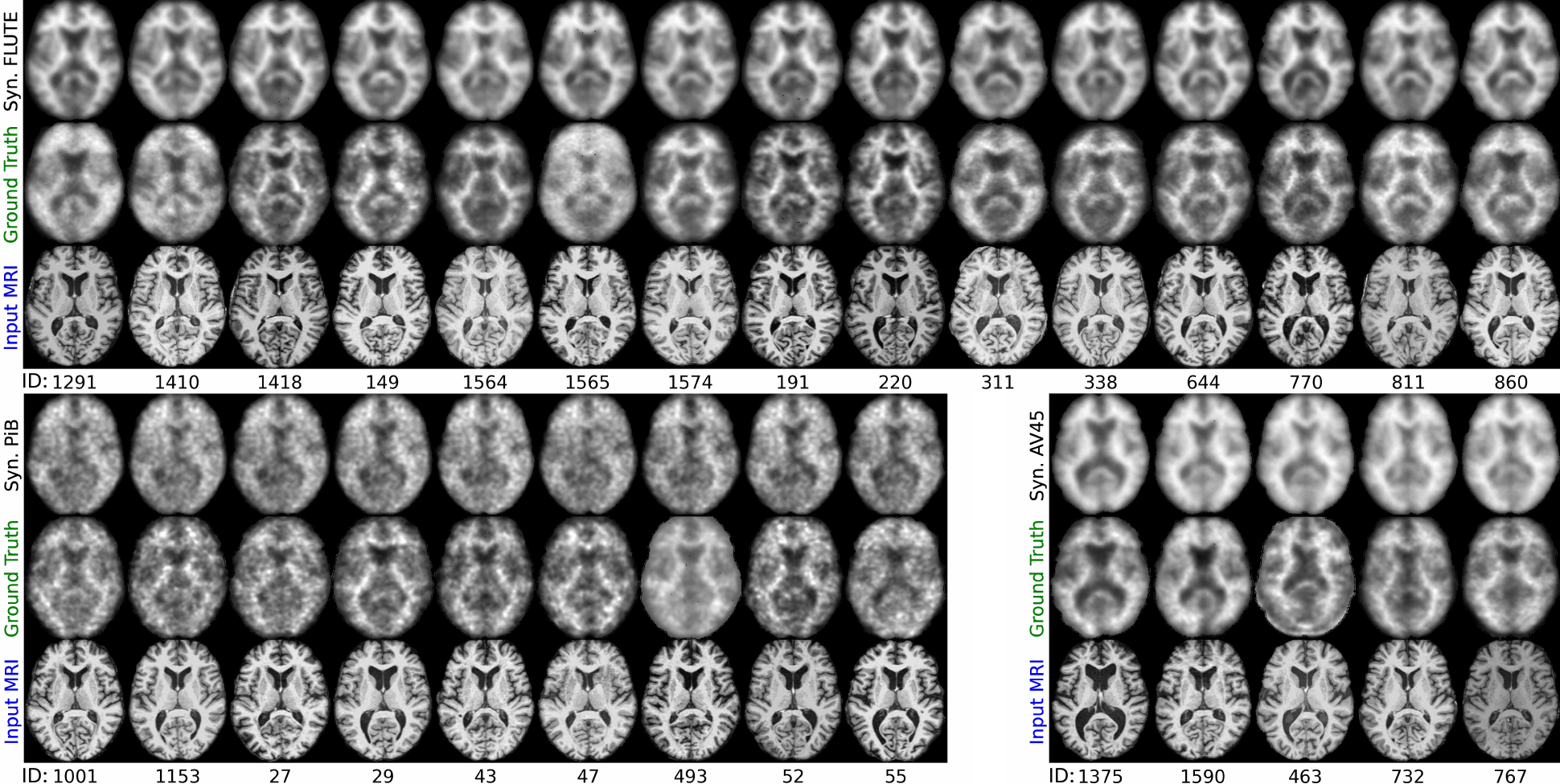}
\caption{Visualization of synthesized PET images using the FICD for three different amyloid PET tracers: FLUTE (top), PiB (bottom left), and AV45 (bottom right), compared with the ground-truth PET images and the corresponding input MRI scans from AIBL~\cite{ellis2009australian}.}
\label{aibl_show}
\end{figure*}

\begin{table}[!t]
\setlength{\abovecaptionskip}{0pt}
\setlength{\belowcaptionskip}{0pt}
\setlength{\abovedisplayskip}{0pt}
\setlength\belowdisplayskip{0pt}
\setlength{\belowcaptionskip}{0.86pt}
\setlength{\tabcolsep}{0.6pt}
\scriptsize
\centering
\caption{Results of fine-tuning the proposed FICD to synthesize PET images of different radiotracers from the AIBL cohort.}
\label{sites}
\begin{tabular}{l|cccc}
\toprule
Radiotracer & PSNR$\uparrow$ & SSIM$\uparrow$ & MAE$\downarrow$ & NMI$\uparrow$\\
\midrule
AV45 &$22.8995$$\pm$$2.7352$&$0.8527$$\pm$$0.0643$&$0.0319$$\pm$$0.0102$&$0.7335$$\pm$$0.0861$\\
PiB &$21.4160$$\pm$$0.9953$&$0.8093$$\pm$$0.0096$&$0.0366$$\pm$$0.0069$&$0.7174$$\pm$$0.0107$\\
FLUTE &$26.9723$$\pm$$1.6334$&$0.9089$$\pm$$0.0148$&$0.0194$$\pm$$0.0040$&$0.8717$$\pm$$0.0542$\\
\bottomrule
\end{tabular}
\label{finetune}
\end{table}

\subsection{Results of Disease Progression Forecasting}
The results of combining real and synthetic images for AD and CN subjects and transferring their categorical information to progressive SCD and stable SCD subjects are presented in Table~\ref{image_input_classification}.
From the table, we have a few observations.
First, the images synthesized by our proposed FICD demonstrate superior performance in classification tasks across both datasets, outperforming other methods across all evaluation metrics.
These results indicate that the images synthesized by FICD more effectively capture AD-specific features critical for analyzing AD-related diseases.

The GAN-based methods deliver generally higher results compared with VAE, although they achieve lower quantitative results.
These results align with observations noted in the qualitative results section that GAN-based results can also capture good brain details, although the GAN-generated images are prone to noise. 
In contrast, VAE performs less effectively in both tasks, as the synthesized images lack sufficient discriminative information for prediction.
On the other hand, VAEGAN and LDM produce comparable results, consistent with their quantitative performance.
DDPM, however, shows lower efficacy in these tasks.
Notably, the classification between pSMC and sSMC in ADNI-SMC yields higher results across all synthesizing methods than that of pSCD vs. sSCD classification on CLAS, which could be attributed to less domain gap between the test data and training data since they all come from ADNI.

\subsection{Results of Future Cognitive Function Prediction}

In this group of experiments, we aim to predict future brain cognitive function based on baseline MRI and PET, where missing PET images are imputed through a specific method.   
Table~\ref{image_input_regression} shows the prediction results of two cognitive scores and Fig.~\ref{fig_regression} shows the scatter plot of the results. 
From these results, we can observe that the cognitive scores predicted by FICD have the highest correlation with the ground truth, on two types of clinical scores (\ie, MMSE and CDR) and two sources of datasets (\ie, CLAS-SCD and ADNI-SMC). 
These results indicate that PET images synthesized by FICD produce the best results in predicting future brain cognitive function using baseline neuroimaging data.

\subsection{Results of Generalization Evaluation}
\subsubsection{Quantitative Results} 
Table~\ref{finetune} presents the quantitative results of applying FICD to amyloid PET in AIBL with three radiotracers. 
Note that FICD is initially trained on FDG-PET images from ADNI and fine-tuned for each radiotracer.  
Among them, FLUTE-PET achieves impressive results, closely paralleling those obtained with FDG-PET. 
The outcomes for PiB-PET and AV45-PET are relatively lower, which may be attributed to less training data and inconsistencies in image quality across different subjects scanned with different radiotracers. 

\begin{figure*}[!t]
\setlength{\abovecaptionskip}{-4pt}
\setlength{\belowcaptionskip}{0pt}
\setlength{\abovedisplayskip}{0pt}
\setlength\belowdisplayskip{0pt}
\centering
\includegraphics[width=0.98\textwidth]{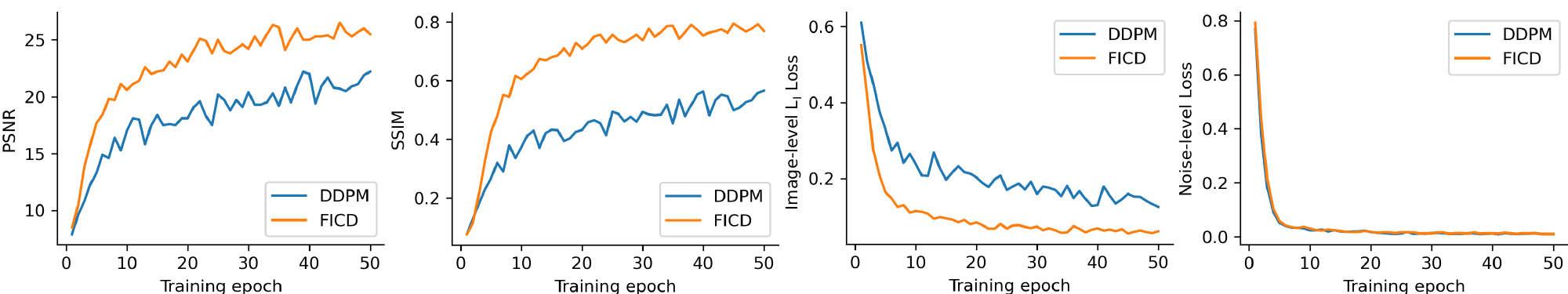}
\caption{Results of PSNR, SSIM, image-level $l_1$ loss, and noise-level loss achieved by DDPM and FICD in Task 1, with epoch number in the x-axis. 
}
\label{analysis_img_loss}
\end{figure*}

\begin{table*}[!t]
\setlength{\abovecaptionskip}{0pt}
\setlength{\belowcaptionskip}{0pt}
\setlength{\abovedisplayskip}{0pt}
\setlength\belowdisplayskip{0pt}
\renewcommand{\arraystretch}{0.86}
\scriptsize
\centering
\caption{Results of preclinical AD progression prediction on subjects from two preclinical AD datasets, with different imaging modalities at baseline as input.}
\setlength\tabcolsep{0.05pt}

\begin{tabular*}{1\textwidth}{@{\extracolsep{\fill}}l| c|  cc cc cc c|c cc cc cc}
\toprule
\multirow{2}{*}{Modality} & 
\multirow{2}{*}{\makecell[c]{Training\\Subject~\#}}&
\multicolumn{6}{c}{pSCD vs. sSCD classification on CLAS (no real PET available)}&&&
\multicolumn{6}{c}{pSMC vs. sSMC classification on ADNI} \\
\cmidrule{3-16}
&&AUC (\%) $\uparrow$&ACC (\%)$\uparrow$ &SEN (\%)$\uparrow$ &SPE (\%)$\uparrow$ &BAC (\%)$\uparrow$ &F1s (\%)$\uparrow$ &&&AUC (\%)$\uparrow$ &ACC (\%)$\uparrow$ &SEN (\%)$\uparrow$ &SPE (\%)$\uparrow$ &BAC (\%)$\uparrow$ &F1s (\%)$\uparrow$  \\ 
\midrule

MRI only
&795&56.42±4.04&51.20±4.27&50.83±6.67&51.37±3.14&51.10±4.90&40.00±5.25
&&&68.92±4.28&59.02±3.59&63.16±5.77&57.14±2.61&60.15±4.19&48.98±4.47\\ 

MRI+Real PET
&533
& -- & -- & -- & -- & -- & --
&&&68.70±5.75&60.98±4.45&66.32±7.14&58.57±3.23&62.44±5.18&51.43±5.45\\ 

FICD~(Ours) &795&\textbf{63.77}±2.72
&\textbf{58.67}±2.39
&\textbf{62.50}±3.73
&\textbf{56.86}±1.75
&\textbf{59.68}±2.74
&\textbf{49.18}±2.93
&&&\textbf{72.78}±2.83
&\textbf{63.61}±2.62
&\textbf{70.53}±4.21
&\textbf{60.48}±1.90
&\textbf{65.50}±3.06
&\textbf{54.69}±3.27\\
\bottomrule
\end{tabular*}
\label{Training_strategy}
\end{table*}

\subsubsection{Qualitative Results}
Figure~\ref{aibl_show} shows the qualitative result of PET with the three tracers. 
Aligning with previous quantitative results, FLUTE-PET scans are synthesized with very high fidelity. 
The PiB-PET scans are not as smooth as FLUTE-PET, which is caused by the radiotracer's characteristic that $^{11}$C for PiB-PET has a short radioactive decay half-life (20 minutes)~\cite{clark2011use}. 
This introduces a degree of unpredictability to the images. 
Despite this, synthesized PiB-PET and AV45-PET scans effectively capture detailed functional information, indicating that FICD has good generalizability. 
To be noted, while the synthesized images consistently maintain uniform quality, there is noticeable variability in the quality of ground-truth PET  from AIBL.
This is because AIBL differs significantly in the types of PET tracers used (\eg, PiB, FLUTE, and AV45).  
These factors introduce variability in ground-truth PET images due to different tracer kinetics and baseline metabolism patterns. 
This variability may influence the performance of our model, particularly in how well the model trained on one dataset (\eg, ADNI) can be applied to another (\eg, AIBL).  
To improve the model's generalizability across various datasets, domain adaptation techniques~\cite{kouw2019review} can be used to address differences in PET image characteristics.

\section{Discussion}
\label{S5}
\subsection{{Influence of Functional Imaging  Constraint}}%
We further investigate the influence of the proposed functional imaging constraint by comparing the training efficiency of DDPM and our FICD.
According to Eqs.~\eqref{noise_t-1}-\eqref{x_t-1}, the accurate estimate of $X^0$ (\ie~$\tilde X^0$) is essential as it influences the inference process to get $X^{t-1}$.
During each training epoch, we calculate  $\tilde X^0$ using Eq.~\eqref{x_0} and compare it with the ground truth.
The comparison is evaluated through PSNR, SSIM, and image-level $l_1$ loss defined in Eq.~\eqref{eq_immagLoss}. 
The noise-level loss in Eq.~\eqref{eq_noiseLoss} of both methods is also recorded. 
Since DDPM does not employ a functional imaging constraint in training, we compute the $l_1$ loss between its $\tilde X^0$ and the ground truth for comparing purposes only.

Figure~\ref{analysis_img_loss} depicts the trends in PSNR, SSIM, image-level $l_1$ loss, and noise-level loss throughout the training epochs for FICD and DDPM. 
From Fig.~\ref{analysis_img_loss}, we can see that even though the two methods have similar converging speeds in terms of noise-level loss, their corresponding $\tilde X^0$ have distinct results indicated by PSNR, SSIM, and image-level $l_1$ loss. 
The FICD shows rapid convergence in metrics evaluating $X^0$, whereas DDPM's convergence is significantly slower and yields less favorable outcomes. 
This indicates that in the absence of a functional imaging constraint, the denoising diffusion model struggles to produce high-quality images.

\begin{table}[!t]
\setlength{\abovecaptionskip}{0pt}
\setlength{\belowcaptionskip}{0pt}
\setlength{\abovedisplayskip}{0pt}
\setlength\belowdisplayskip{0pt}
\setlength{\tabcolsep}{0.4pt}
\renewcommand{\arraystretch}{0.9}
\scriptsize
\centering
\caption{Quantitative evaluation results of synthetic PET images generated by the proposed FICD using MRI as condition and its variant FICD-G using MRI gradient map as condition, with best results shown in bold. 
}
\label{sites}
\begin{tabular}{l|cccc}
\toprule
Method & PSNR$\uparrow$ & SSIM$\uparrow$ & MAE$\downarrow$ & NMI$\uparrow$\\ 
\midrule
FICD-G & $26.7001$$\pm$$0.8556$ & $0.9018$$\pm$$0.0222$& $0.0199$$\pm$$0.0021$& $0.8333$$\pm$$0.0304$\\
FICD & $\mathbf{27.7210}$$\pm$$1.1552$ & $\mathbf{0.9083}$$\pm$$0.0231$ & $\mathbf{0.0176}$$\pm$$0.0026$ & $\mathbf{0.8514}$$\pm$$0.0347$\\
\bottomrule
\end{tabular}
\label{gradient}
\end{table}

\subsection{Influence of PET Imputation}
We further investigate the influence of PET imputation on downstream tasks by examining Task 2 outcomes across three scenarios: 
1) MRI only: when the subjects of the entire cohort have only MRI data, 
2) MRI+Real PET: when two modalities are used but only a subset of the entire cohort has paired real MRI and PET images available,
and 3) FICD: when the entire cohort has paired multimodal images, with the missing PET synthesized by FICD.
The results presented in Table~\ref{Training_strategy} indicate that the highest accuracy is achieved when synthesized images are used to complete the modalities for all cohorts, significantly outperforming other scenarios. 
In cases where only real MRI and PET are utilized, the performance is not good, 
which may be attributed to the limited sample size (\ie, only 533 subjects). 
When only using MRI, there's a marginal increase in results, however, these outcomes don't reach the enhanced performance levels achieved by our approach with synthesized PET images.

\subsection{Influence of Input Condition}
\label{S5_condition}
 
We also investigate the influence of conditional inputs on the proposed FICD. 
As mentioned in Section~\ref{S3}, in addition to MRI, we can also apply the gradient of MRI as a condition to FICD, and we refer to the FICD conditioned on the MRI gradient map as \textbf{FICD-G}. 
Here, we compare the influence of these two types of conditions used in FICD by performing image synthesis (same in Task 1) and the downstream classification task (same in Task 2).
The quantitative evaluation results are reported in Table~\ref{gradient}, with visualization results reported in Fig.~S3 and downstream task results shown in Table~S1 of the \emph{Supplementary Materials}. 
All the experimental settings remain consistent across both methods, and evaluations are conducted with an MC sampling time of 5. 
From Table~\ref{gradient} and Table~S1, we can observe that utilizing the MRI scan directly as the condition yields marginally better results. 
Figure~S3 suggests that the results 
achieved by FICD-G (with MRI gradient as a condition) are also noteworthy and present a viable alternative to FICD.

\subsection{Comparison with Recent State-Of-The-Arts} 
In addition to the six competing methods in the main experiments, we further compare our FICD with three recent state-of-the-art (SOTA) approaches in image generation, including unified anatomy-aware cyclic adversarial network (\textbf{UCAN})~\cite{zhou2021synthesizing}, 
conditional flexible self-attention GAN (\textbf{CF-SAGAN})~\cite{wei2020predicting}, and
joint diffusion attention model (\textbf{JDAM})~\cite{xie2023synthesizing}. 
The experimental settings for these methods and their network architectures are introduced in Section~5 of the \emph{Supplementary Materials}.  
The quantitative evaluation results achieved by the proposed FICD and the three recent SOTA methods for FDG-PET generation on ADNI are reported in Table~\ref{tab_NewSOTA_quantitative}. 
The visualization of PET images synthesized by different methods 
is shown in Fig.~S6 of \emph{Supplementary Materials}.  
These results show that FICD outperforms the three methods in  quantitative evaluation and visual quality, providing superior fidelity in FDG-PET synthesis.

\begin{table}[!t]
\setlength{\abovecaptionskip}{0pt}
\setlength{\belowcaptionskip}{0pt}
\setlength{\abovedisplayskip}{0pt}
\setlength\belowdisplayskip{-0pt}
\setlength{\tabcolsep}{0.01pt}
\scriptsize
\renewcommand{\arraystretch}{0.86}
 \scriptsize
\centering
\caption{Quantitative evaluation results achieved by the proposed FICD and three recent SOTA methods for FDG-PET generation on ADNI.}
\label{sites}
\begin{tabular}{l|cccc}
\toprule
Method & PSNR$\uparrow$ & SSIM$\uparrow$ & MAE$\downarrow$ & NMI$\uparrow$\\ 
\midrule
UCAN  & $24.2043$$\pm$$1.0794$ & $0.8522$$\pm$$0.0182$ & $0.0257$$\pm$$0.0031$ & $0.7698$$\pm$$0.0248$\\
JDAM   & $22.8481$$\pm$$0.6720$ & $0.8380$$\pm$$0.0315$ & $0.0394$$\pm$$0.0034$ & $0.9797$$\pm$$0.0347$\\
CF-SAGAN & $27.1015$$\pm$$0.9857$ & $0.8775$$\pm$$0.0206$ & $0.0189$$\pm$$0.0023$ & $0.8393$$\pm$$0.0311$\\
FICD & $\mathbf{27.8847}$$\pm$$1.1676$ & $\mathbf{0.9124}$$\pm$$0.0239$ & $\mathbf{0.0173}$$\pm$$0.0026$ & $\mathbf{0.8603}$$\pm$$0.0355$\\
\bottomrule
\end{tabular}
\label{tab_NewSOTA_quantitative}
\end{table}

\subsection{Application of FIC to Another Diffusion Model} 
\label{S56_FICLDM}
The proposed functional imaging constraint (FIC) can be integrated to enhance other diffusion models, including the latent diffusion model (LDM)~\cite{rombach2022high}.   
In this work, we also develop a functional imaging constrained LDM method called \textbf{FIC-LDM}, with its architecture shown in Fig.~S7 in the \emph{Supplementary Materials}. 
Specifically, the FIC-LDM employs a feature encoder to extract low-dimensional latent features (with the latent map dimension of $10\times12\times10$), facilitating MRI-to-PET domain translation in the latent space.
The translated features are then decoded to synthesize the PET image. 
Table~\ref{tab_FIC_LDM} reports the results of FIC-LDM alongside its two variants: \textbf{FIC-LDM20} and \textbf{FIC-LDM40}, with the latent map dimension of $20\times24\times20$ and $40\times48\times40$, respectively. 
For comparative analysis, we also report the results of the original LDM (latent map dimension: $10\times12\times10$) and its two variants \textbf{LDM20} and \textbf{LDM40} (with latent map dimensions of $20\times24\times20$ and $40\times48\times40$, respectively).

Table~\ref{tab_FIC_LDM} shows that integrating FIC into LDM improves its performance on almost all the evaluation metrics across all latent map dimensions. 
As the dimension of the latent map increases, the positive impact of FIC on LDM's performance becomes increasingly significant. 
On the other hand, FIC-LDM and its two variants yield inferior results compared with FICD. 
With the H100 GPU cluster, the inference time for FICD is $30$ seconds, while this inference time is shortened to $1.47$ seconds using FIC-LDM40 and only $0.15$ seconds using FIC-LDM.   
Therefore, FIC-LDM and its variants can be considered as alternatives to FICD to reduce the inference time, but will sacrifice performance to some extent. 
In Fig.~S9 of \emph{Supplementary Materials}, we visualize the axial slices of PET images synthesized by the five methods. 
From Fig.~S9, we can observe that the PET images synthesized by FIC-LDM and its two variants contain more details compared with those produced by LDM and its variants. 
The difference maps also show smaller differences in FIC-LDM. 
The PET images synthesized with FICD demonstrate the highest fidelity among all results.

\begin{table}[!t]
\setlength{\abovecaptionskip}{0pt}
\setlength{\belowcaptionskip}{0pt}
\setlength{\abovedisplayskip}{0pt}
\setlength\belowdisplayskip{0pt}
\setlength{\tabcolsep}{0.01pt}
\renewcommand{\arraystretch}{0.86}
\scriptsize
\centering
\caption{Quantitative evaluation results achieved by FICD, FIC-LDM, LDM, and their variants for FDG-PET generation on ADNI.}
\label{sites}
\begin{tabular}{l|cccc}
\toprule
Method & PSNR$\uparrow$ & SSIM$\uparrow$ & MAE$\downarrow$ & NMI$\uparrow$\\ 
\midrule
LDM & $26.3088$$\pm$$0.9892$ & $0.8770$$\pm$$0.0215$ & $0.0206$$\pm$$0.0026$ & $0.8008$$\pm$$0.0287$\\
LDM20  & $25.3247$$\pm$$1.1241$ &$0.8813$$\pm$$0.0271$&$0.0239$$\pm$$0.0034$& $0.7861$$\pm$$0.0275$  \\ 
LDM40  & $24.0604$$\pm$$0.5649$ &$0.8853$$\pm$$0.0266$&$0.0276$$\pm$$0.0019$& $0.7786$$\pm$$0.0276$  \\ 
\midrule
FIC-LDM  & $26.3360$$\pm$$0.9624$ &$0.8792$$\pm$$0.0229$&$0.0207$$\pm$$0.0026$& $0.8035$$\pm$$0.0268$  \\ 
FIC-LDM20 & $26.9780$$\pm$$0.9757$ & $0.8979$$\pm$$0.0252$& $0.0192$$\pm$$0.0024$& $0.8247$$\pm$$0.0267$ \\
FIC-LDM40 & $26.7129$$\pm$$0.8895$ & $0.8982$$\pm$$0.0261$& $0.02021$$\pm$$0.0022$& $0.8382$$\pm$$0.0296$\\
\midrule
FICD & $\mathbf{27.8847}$$\pm$$1.1676$ & $\mathbf{0.9124}$$\pm$$0.0239$ & $\mathbf{0.0173}$$\pm$$0.0026$ & $\mathbf{0.8603}$$\pm$$0.0355$ \\
\bottomrule
\end{tabular}
\label{tab_FIC_LDM}
\end{table}

\begin{figure*}[!tbp]
\setlength{\belowdisplayskip}{-0pt}
\setlength{\abovedisplayskip}{-0pt}
\setlength{\abovecaptionskip}{-2pt}
\setlength{\belowcaptionskip}{-0pt}
\centering
\includegraphics[width=0.98\textwidth]{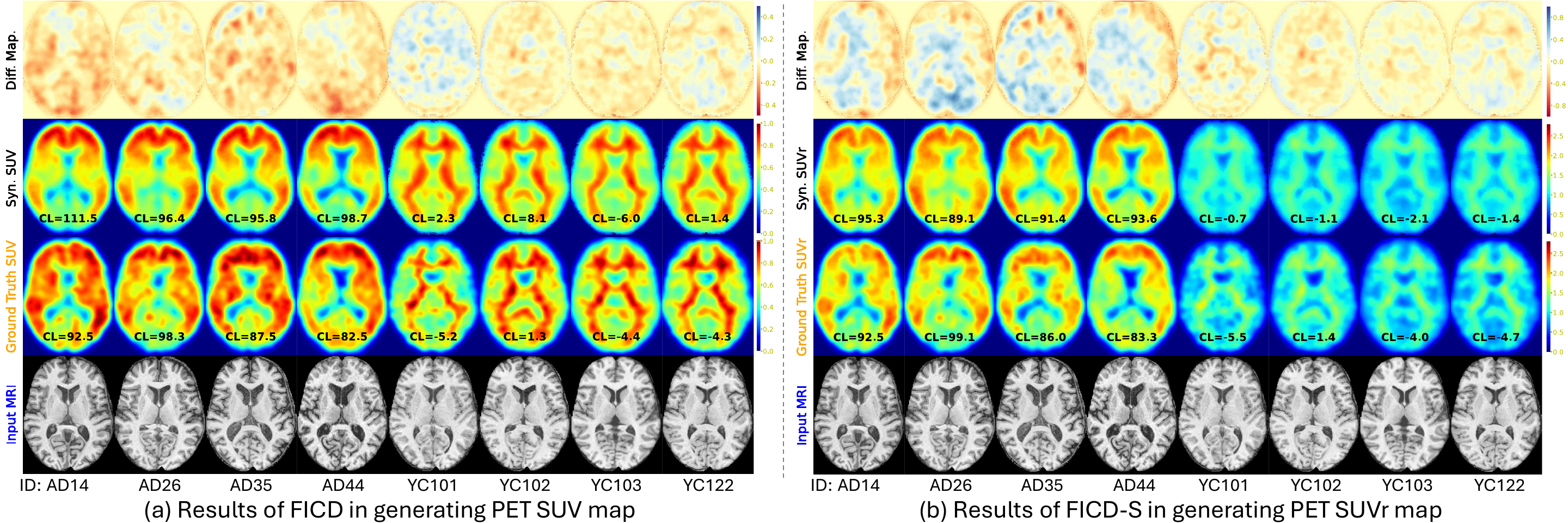}
\caption{(a) Synthesized (Syn.) SUV images and difference (Diff.) maps achieved by FICD, and (b) synthesized (Syn.) SUVr images and difference (Diff.) maps for eight test subjects in the Centiloid project achieved by FICD-S. The ground-truth SUV maps, SUVr maps, and input MRI are displayed at the bottom with the corresponding subject IDs. CL: Centroid scale.}
\label{suvr_result_show}
\end{figure*}

\subsection{Alternative Functional Constraint and Clinical Evaluation of Synthetic Images}
As a vital functional imaging tool, PET is particularly effective for assessing metabolic processes in the brain, which is crucial for diagnosing conditions like AD and MCI. 
A key metric used in PET to assess brain function is the standardized uptake value (SUV), which reflects the concentration of a radiotracer in a region normalized by the dose and the patient's body weight. 
The standardized uptake value ratio (SUVr) extends this by comparing radiotracer concentrations in target regions to reference areas (cerebellum in this case), allowing for the analysis of relative brain activity that provides quantifiable differences between disease stages. 
As shown in the Centiloid Project~\cite{klunk2015centiloid}, the average SUVr of PiB PET at the global cortical target (CTX) volume-of-interest (VOI) — which includes key brain regions typically burdened with amyloid deposits in AD — can serve as the standard quantitative amyloid imaging measure of young control (YC) and AD groups, by defining 0- and 100-anchor points of a Centiloid (CL) scale~\cite{pemberton2022quantification} for YC and AD groups, respectively. 
The CL scale is an unbounded 0 (the average gray matter signal from young healthy controls) to 100 (typical AD signal) scale that measures a single subject's amyloid load.  
On this basis, we develop a new functional imaging constraint based on SUVr maps to enhance the clinical relevance of synthesized images, leading to a more precise assessment of brain function. 
Specifically, instead of using the original PET image-based constraint in FICD, we can use normalized SUVr maps along with masked CTX VOI as the constraint, which is at the functional map level, and especially focuses on the selected VOIs of PET images. 
This method is denoted \textbf{FICD-S}, with the detailed architecture shown in Fig.~S8 of \emph{Supplementary Materials}.

In this experiment, data from 79 participants from the Centiloid Project~\cite{klunk2015centiloid} is utilized, comprising 34 AD subjects and 45 YC subjects. 
Both SUV maps of PiB-PET images (not raw PET data) and their corresponding T1-weighted MRI scans are provided, with detailed image processing procedures shown in Section~9 of \emph{Supplementary Materials}.  
We allocate approximately 90\% of the data for model fine-tuning and 10\% for test. 
Details for data partition can be found in Table~S8 of \emph{Supplementary Materials}. 
Two approaches are compared, including FICD utilizing SUV map-based constraint (since no raw PET is available) and FICD-S employing constraints based on standardized SUVr map and masked CTX VOI. 
To quantitatively evaluate synthetic images, four metrics are utilized: PSNR, SSIM, MAE, and NMI. 
To evaluate the clinical significance of synthetic PET images, we use the CL value calculated on generated SUVr maps as the evaluation metric in this experiment.

\begin{table}[!t]
\setlength{\abovecaptionskip}{0pt}
\setlength{\belowcaptionskip}{0pt}
\setlength{\abovedisplayskip}{0pt}
\setlength\belowdisplayskip{0pt}
\setlength{\tabcolsep}{0.8pt}
\renewcommand{\arraystretch}{0.86}
\scriptsize
\centering
\caption{Table~9 Quantitative evaluation results of PET SUV maps synthesized by FICD and SUVr maps generated by FICD-S for test subjects from the Centiloid Project~\cite{klunk2015centiloid} with PiB-PET data.}
\label{sites}
\begin{tabular}{l|cccc}
\toprule
Method & PSNR$\uparrow$ & SSIM$\uparrow$ & MAE$\downarrow$ & NMI$\uparrow$\\ 
\midrule

FICD  & $25.9833$$\pm$$1.3740$ & $0.7266$$\pm$$0.2309$ & $0.0294$$\pm$$0.0118$ & $0.8873$$\pm$$0.0358$\\
FICD-S & $26.1862$$\pm$$0.7923$ & $0.8551$$\pm$$0.0095$& $0.0553$$\pm$$0.0156$& $0.8567$$\pm$$0.0389$\\

\bottomrule
\end{tabular}
\label{suvr_synthesizing}
\end{table}

1) \textbf{Image Quality Evaluation Results}. 
The quantitative evaluation results for synthesizing SUV using FICD and SUVr using FICD-S are presented in Table~\ref{suvr_synthesizing}.
The results indicate that both types of constraints yield comparable outcomes. 
Additionally, Fig.~\ref{suvr_result_show} visually displays the synthetic SUV maps produced by FICD and the SUVr maps generated by FICD-S for eight test subjects, alongside the difference maps comparing synthetic images to their ground truth, as well as the ground-truth images and input MRI scans. 
From Fig.~\ref{suvr_result_show}~(b), we can observe distinct differences between the synthesized SUVr maps for AD and YC subjects. 
The synthesized AD SUVr map reveals substantial amyloid deposits in the frontal and temporal cortices, as well as the precuneus, which are regions commonly associated with a high amyloid burden in AD~\cite{klunk2015centiloid}.
In contrast, the synthesized YC SUVr maps show minimal amyloid deposition in these areas. 
This suggests that the SUVr maps synthesized by FICD-S are capable of accurately reflecting amyloid distribution patterns consistent with different disease stages, effectively distinguishing between AD and YC. 
In addition, it can be seen from Fig.~\ref{suvr_result_show}~(a) that the SUV maps synthesized by FICD  demonstrate high image fidelity compared with the ground truth and clearly distinguish between AD and YC subjects.

2) \textbf{Clinical Evaluation Results}.
Based on the SUVr maps generated by FICD-S, we achieve the CL values of $93.2\pm2.7$ for test AD group and $-1.3\pm0.5$ for test YC group. 
This is significant considering that the ground-truth CL values for the test AD and YC groups are $94.8 \pm 10.7$ and $-3.2 \pm 2.7$, respectively.  
The CL values of FICD are $100.9 \pm 5.8$ for test AD subjects and $1.5 \pm 5.0$ for  
 test YC subjects, with the ground truth of $93.3 \pm 8.1$ and $-3.2 \pm 2.6$ based on SUV maps, respectively. 
The ground truth is slightly different from FICD-S because in this case the SUV maps are smoothed before the calculation of SUVr images.  
According to various CL thresholds established in the literature and used for inclusion in current clinical trials~\cite{su2018utilizing,pemberton2022quantification}, CL$<$$10$ accurately reflects the absence of neuritic plaques (excluding AD), while CL$\geq$$50$ indicates a strong correlation with AD diagnosis. 
This suggests that our FICD-S and FICD yield discriminative CL values that can distinguish between AD and YC subjects and is therefore of high clinical significance. 
More discussions are given in \emph{Supplementary Materials}.

\subsection{Limitations and Future Work} 
Some limitations of the current study should be noted. 
\emph{On one hand}, in this work, the FICD is trained exclusively on cognitively normal subjects and is utilized to synthesize PET images across various disease categories. 
This may constrain the model's effectiveness, potentially overlooking subtle disease-specific brain pathology. 
As a future work, we could broaden its learning scope by including data with different brain states and enhance the training data with synthetic variations that simulate scanner differences and biological variability to increase the model's generalizability.  
On the other hand, current methods use Gaussian noise as input and do not take into account the specificity of PET data distribution. 
It is interesting to explicitly consider this important prior knowledge to further improve the learning  performance~\cite{everaert2023diffusion}, which will be another future work.

\section{Conclusion}
\label{S6}
This paper presents a functional imaging constrained diffusion (FICD) framework that synthesizes 3D brain PET images from MRI using a constrained diffusion model (CDM). FICD ensures stable training by progressively adding and then removing random noise to PET images through CDM, incorporating a functional imaging constraint for voxel-wise accuracy against ground-truth PET. Quantitative and qualitative assessments on a dataset of 293 subjects with paired T1-weighted MRI and FDG-PET show that FICD surpasses current state-of-the-art methods. Additional validation on three downstream tasks involving 1,262 subjects for brain state prediction and amyloid PET synthesis confirms FICD's effectiveness and generalizability.

\section*{Acknowledgments}

The authors would like to thank Dr. Heidi Roth and Dr. Weili Lin for their assistance with clinical evaluation of images produced in this project. 
This work utilized the high-performance computing resources provided by the Image Storage and Analysis Core of the Biomedical Research Imaging Center of 
UNC at Chapel Hill. 
A part of the data is from ADNI and AIBL. 
The ADNI and AIBL investigators provide data but are not involved in data processing, analysis, and writing. 
A list of ADNI investigators is accessible \href{https://adni.loni.usc.edu/wp-content/uploads/how\_to\_apply/ADNI\_Acknowledgement\_List.pdf}{online}.

\footnotesize
\bibliography{bibfile.bib}
\bibliographystyle{IEEEtran}
\end{document}


\title{Functional Imaging 
Constrained Diffusion 
for Brain PET Synthesis from Structural MRI\\
		-- \emph{Supplementary Materials}}

\author{Minhui~Yu, 
Mengqi~Wu,
Ling~Yue,
Andrea~Bozoki,
Mingxia~Liu,~\IEEEmembership{Senior Member,~IEEE}

\IEEEcompsocitemizethanks{
\IEEEcompsocthanksitem M.~Yu, M.~Wu and M.~Liu are with the Department of Radiology and BRIC, University of North Carolina at Chapel Hill, Chapel Hill, NC 27599 USA. M.~Yu and M.~Wu are also with the Joint Department of Biomedical Engineering, University of North Carolina at Chapel Hill and North Carolina State University, Chapel Hill, NC 27599, USA. 
L.~Yue is with the Department of Geriatric Psychiatry, Shanghai Mental Health Center, Shanghai Jiao Tong University School of Medicine, Shanghai 200240, China. 
A.~Bozoki is with the Department of Neurology, University of North Carolina at Chapel Hill, Chapel Hill, NC 27599, USA. 
Corresponding author: M.~Liu (email: mxliu@@med.unc.edu). 
\protect\\
}
}
	\markboth{}%
	{Functional Imaging 
Constrained Denoising Diffusion for Brain PET Synthesis from Structural MRI}
	
	\maketitle
In what follows, we first visually show the synthetic PET images and difference maps generated by the proposed FICD and six competing methods from coronal and sagittal views in Section~\ref{S1}. 
We then present the visualization and downstream prediction results of FICD using different input conditions in Section~\ref{S2}. 
After that, we investigate the model variability in inference with different Monte-Carol (MC) sampling times in Section~\ref{S3}. 
We then study the performance of the synthetic PET images against real ones in a downstream task in Section~\ref{S4} and visualize the outputs of FICD and recent state-of-the-art methods in Section~\ref{S5}. 
We further present the detailed architectures of two variants of FICD (\ie, FIC-LDM and FICD-S)  in Section~\ref{S6}, visualize the outputs of FIC-LDM, LDM, and their variants in Section~\ref{S7}, and introduce the computation costs of FICD and the competing methods in Section~\ref{S8}. 
Finally, we present the data processing procedure for images from the Centiloid Project~\cite{klunk2015centiloid} in Section~\ref{S9}, include more explanation of the proposed method in Section~\ref{S10}, and show detailed information on the studied subjects involved in the experiments in Section~\ref{S11}.

\section{Visualization of Synthetic PET in Sagittal and Coronal Planes}
\label{S1}
\begin{figure*}[htbp]
\setlength{\belowdisplayskip}{-1pt}
\setlength{\abovedisplayskip}{-1pt}
\setlength{\abovecaptionskip}{-1pt}
\setlength{\belowcaptionskip}{-1pt}
\center
 \includegraphics[width= 1\textwidth]{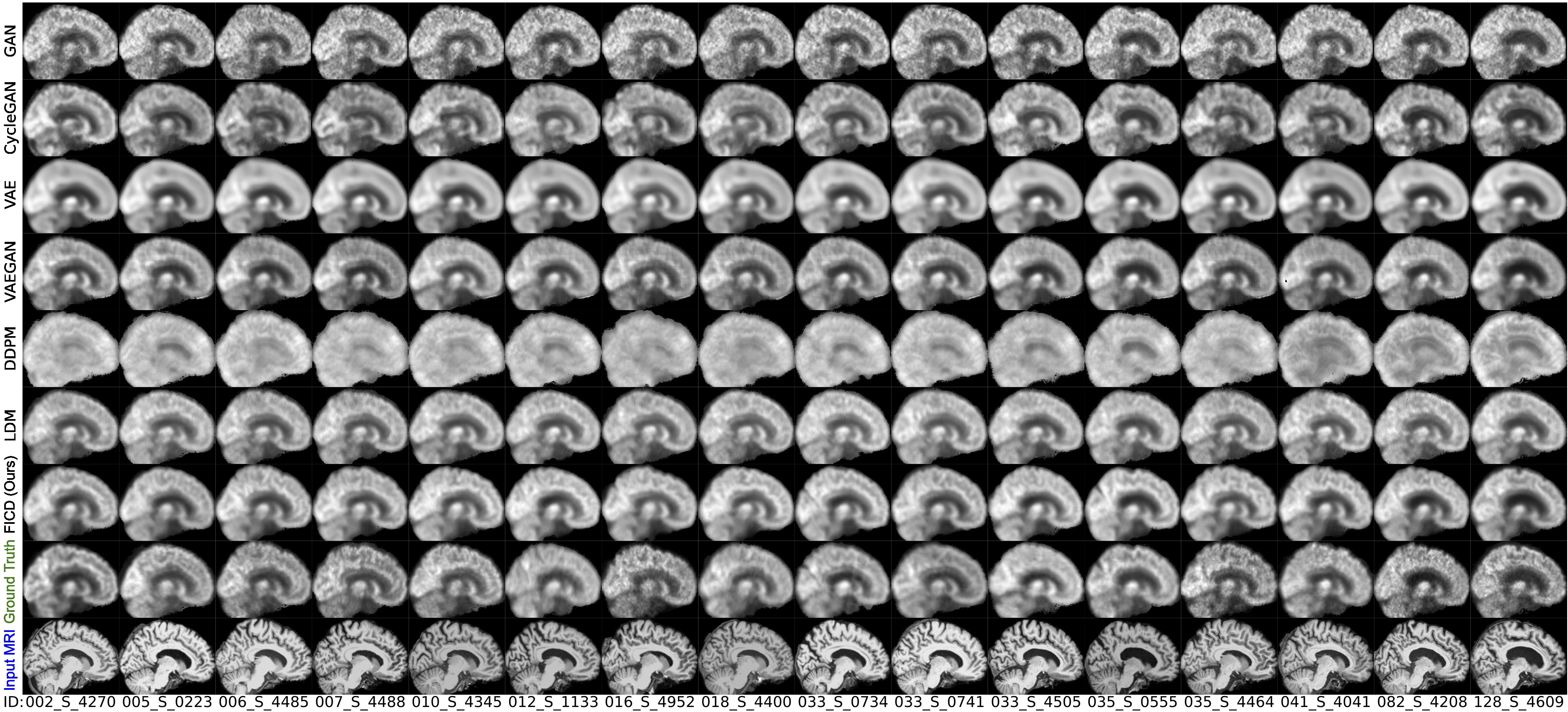}
  \includegraphics[width= 1\textwidth]{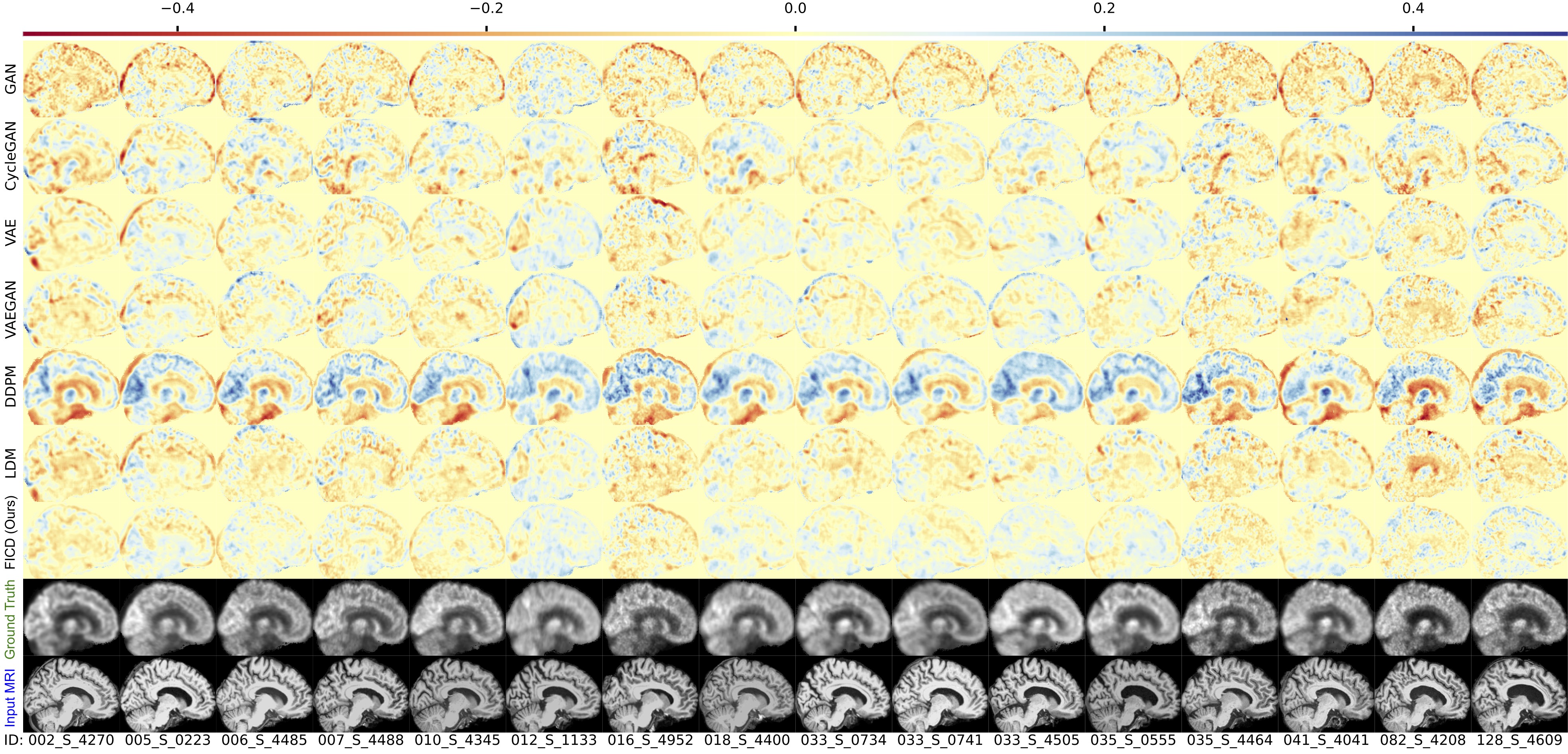}
 \caption{Sagittal plane of synthetic PET images generated by the proposed FICD and six competing methods (top), using MRI scans from test data group of ADNI CN subjects as inputs. 
Each difference map illustrates the difference between a synthetic PET image and its ground truth (bottom).}
 \label{fig_synPET_sagittal}
\end{figure*}

\begin{figure*}[!tbp]
\setlength{\belowdisplayskip}{-1pt}
\setlength{\abovedisplayskip}{-12pt}
\setlength{\abovecaptionskip}{-1pt}
\setlength{\belowcaptionskip}{-1pt}
\center
 \includegraphics[width= 1\textwidth]{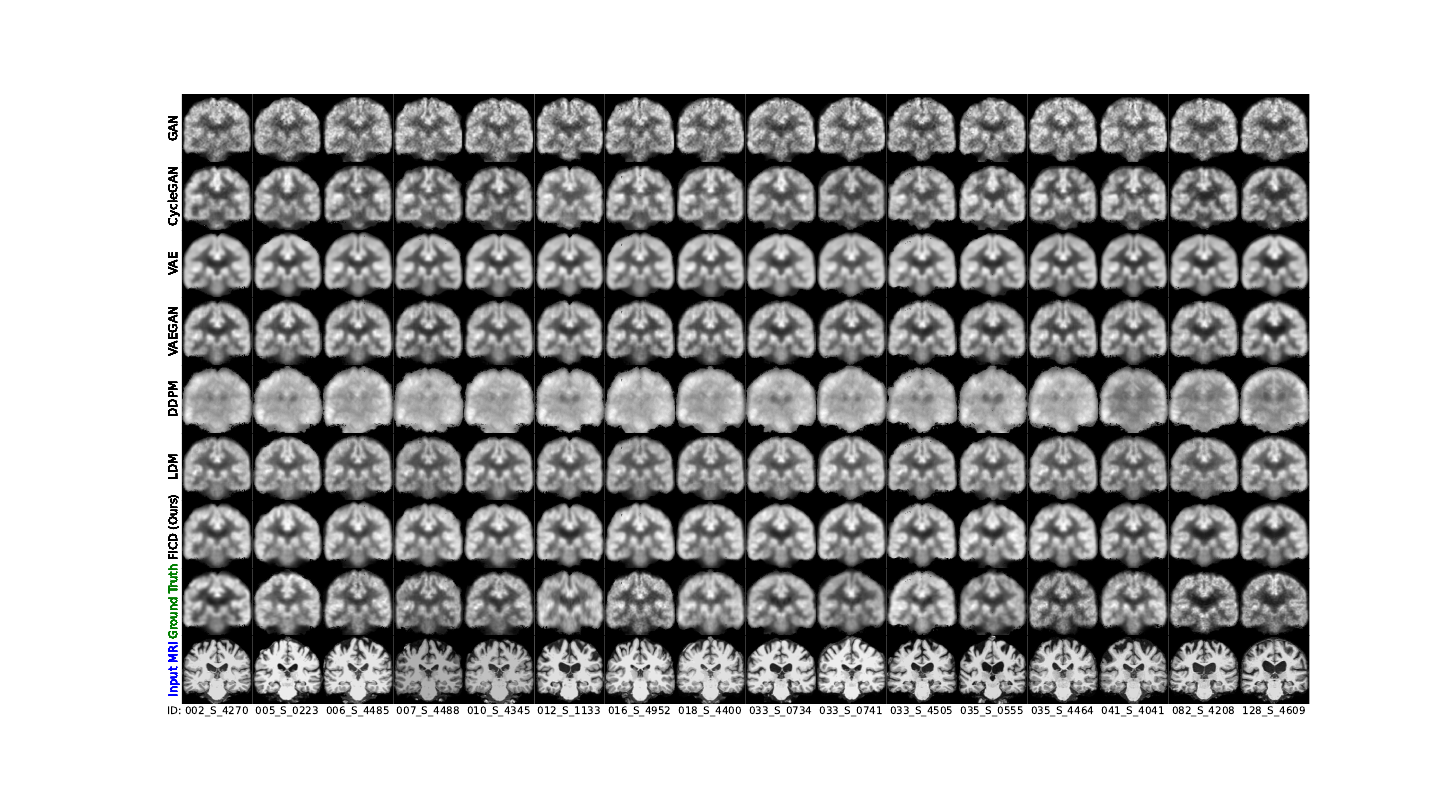}
  \includegraphics[width= 1\textwidth]{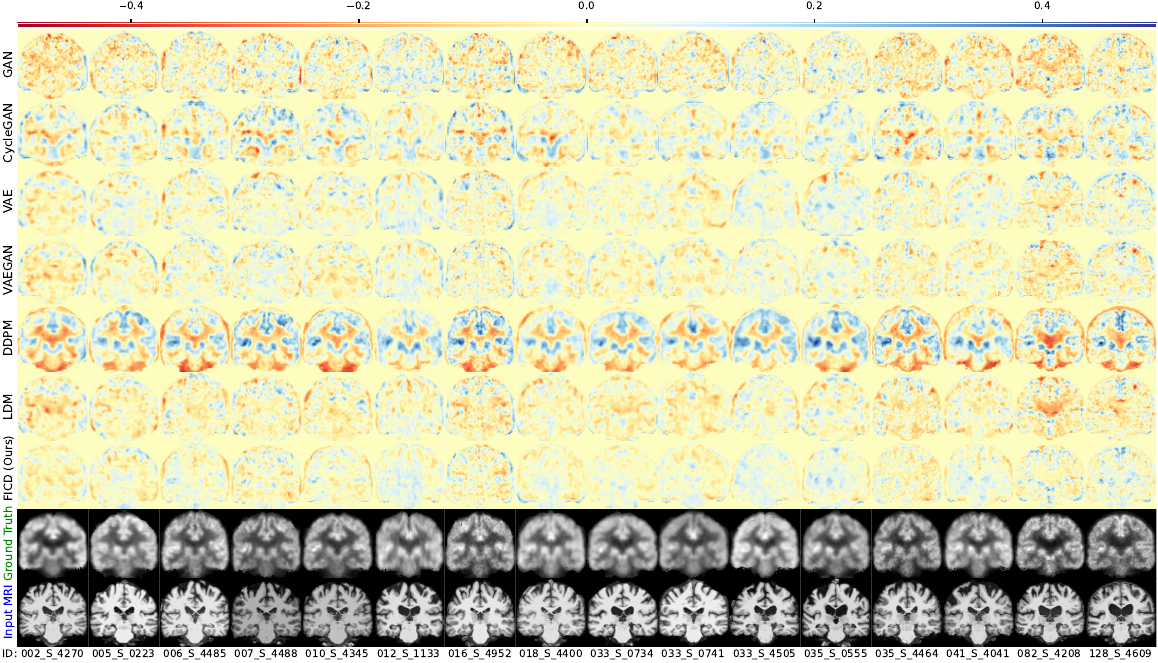}
 \caption{Coronal plane of synthetic PET images generated by the proposed FICD and six competing methods (top), using MRI scans from test data group of ADNI CN subjects as inputs. 
 Each difference map illustrates the difference between a synthetic PET image and its ground truth (bottom).}
 \label{fig_synPET_coronal}
\end{figure*}

We show the synthetic PET images and the corresponding difference maps in the sagittal view (Fig.~\ref{fig_synPET_sagittal}) and the coronal view (Fig.~\ref{fig_synPET_coronal}),
synthesized with T1-weighted MRI as input by our method and six competing methods. 
Note that each difference map is generated by calculating the difference between a synthetic PET and its ground truth.  
For comparison, their corresponding ground-truth PET and T1-weighted MRI inputs are also displayed.

\section{Visualization and Downstream Task Results with Different Input Condition}
\label{S2}
\begin{figure}[!tbp]
\setlength{\belowdisplayskip}{-0pt}
\setlength{\abovedisplayskip}{-0pt}
\setlength{\abovecaptionskip}{-4pt}
\setlength{\belowcaptionskip}{-0pt}
\centering
\includegraphics[width=0.49\textwidth]{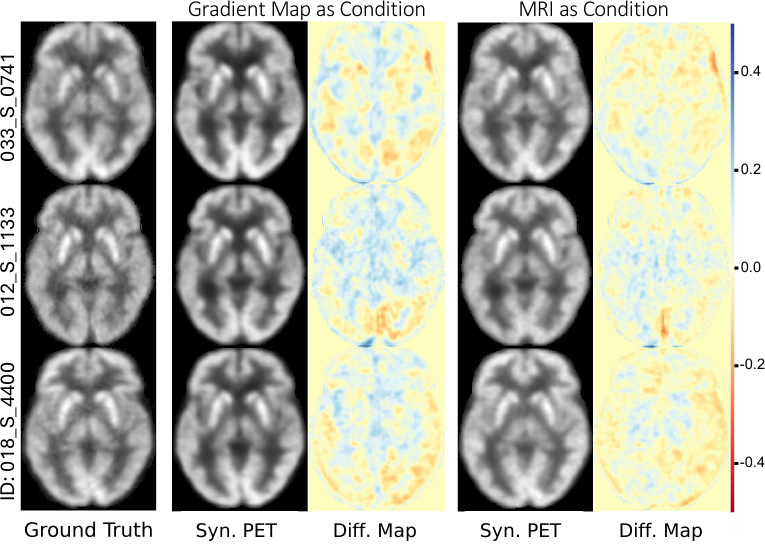}
\caption{Synthesized (Syn.) PET images and difference (Diff.) maps achieved by FICD using MRI gradient map and MRI as the condition, with ground truth from ADNI~\cite{jack2008alzheimer} shown on the left for comparison.}
\label{fig_gradientMRI}
\end{figure}

Aside from the quantitative results reported in Table~6 in the main text, we further compare the outputs of FICD using MRI gradient map versus using MRI as the condition, with results shown in Fig.~\ref{fig_gradientMRI}. 
It can be seen from Fig.~\ref{fig_gradientMRI} that both results provide clear brain functional information closely resembling the ground truth, while the gradient map-conditioned results show greater image contrast. 
Table~\ref{gradient_downstream} shows the results of FICD and its variant FICD-G in the downstream Task 2 (\ie, disease progression forecasting) on the ADNI~\cite{jack2008alzheimer} and the CLAS~\cite{xiao2016china} cohorts.

\begin{table*}[ht]
\setlength{\abovecaptionskip}{0pt}
\setlength{\belowcaptionskip}{0pt}
\setlength{\abovedisplayskip}{0pt}
\setlength\belowdisplayskip{0pt}
\renewcommand{\arraystretch}{0.9}
\scriptsize
\centering
\caption{Results (mean$\pm$standard deviation) of different methods in the tasks of CLAS-SCD progression prediction (\ie, pSCD vs. sSCD classification) and ADNI-SMC progression prediction (\ie, pSMC vs. sSMC classification) with the input of PET and MRI scans.}
\setlength\tabcolsep{2pt}
\begin{tabular*}{1\textwidth}{@{\extracolsep{\fill}}l|cc cc cc c|c cc cc cc}
\toprule
\multirow{2}{*}{~Method} &
\multicolumn{6}{c}{pSCD vs. sSCD classification on CLAS-SCD} &&
\multicolumn{6}{c}{pSMC vs. sSMC classification on ADNI-SMC} \\
\cmidrule{2-14}
&AUC (\%) $\uparrow$&ACC (\%)$\uparrow$ &SEN (\%)$\uparrow$ &SPE (\%)$\uparrow$ &BAC (\%)$\uparrow$ &F1s (\%)$\uparrow$ &&AUC (\%)$\uparrow$ &ACC (\%)$\uparrow$ &SEN (\%)$\uparrow$ &SPE (\%)$\uparrow$ &BAC (\%)$\uparrow$ &F1s (\%)$\uparrow$  \\ 

\midrule
~FICD-G
&58.97±5.92&57.07±4.33&60.00±6.77&55.69±3.19&57.84±4.98&47.21±5.33
&&70.93±7.15&61.64±3.82&67.37±6.14&59.05±2.78&63.21±4.46&52.24±4.76\\  
~FICD &\textbf{63.77}±2.72&\textbf{58.67}±2.39&\textbf{62.50}±3.73&\textbf{56.86}±1.75&\textbf{59.68}±2.74&\textbf{49.18}±2.93
&&\textbf{72.78}±2.83&\textbf{63.61}±2.62&\textbf{70.53}±4.21&\textbf{60.48}±1.90&\textbf{65.50}±3.06&\textbf{54.69}±3.27\\
\bottomrule
\end{tabular*}
\label{gradient_downstream}
\end{table*}

\section{Analysis of Output Variability in Inference}
\label{S3}
\begin{figure*}[ht]
\setlength{\abovecaptionskip}{-4pt}
\setlength{\belowcaptionskip}{-0pt}
\setlength{\abovedisplayskip}{0pt}
\setlength\belowdisplayskip{0pt}
\centering
\includegraphics[width=1\textwidth]{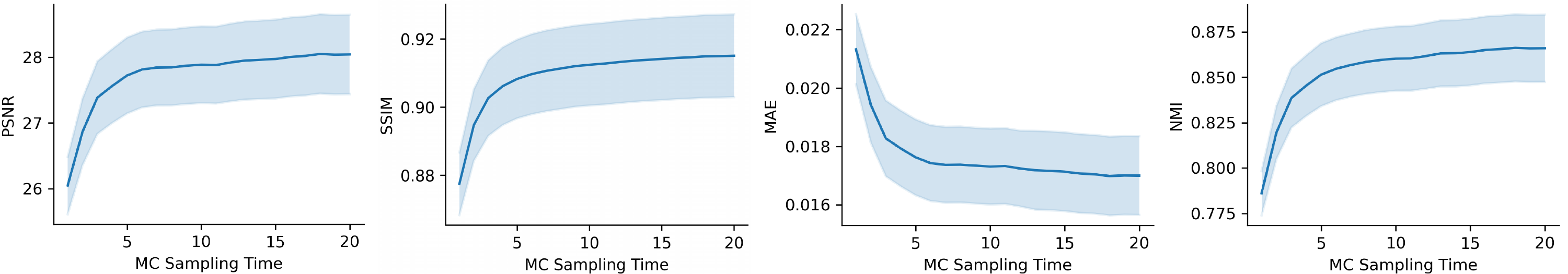}
\caption{Quantitative results achieved by FICD using different Monte-Carlo (MC) sampling times in Task 1.}
\label{montecarlo}
\end{figure*}

\begin{figure*}[!tbp]
\setlength{\belowdisplayskip}{-0pt}
\setlength{\abovedisplayskip}{-0pt}
\setlength{\abovecaptionskip}{-4pt}
\setlength{\belowcaptionskip}{-0pt}
\centering
\includegraphics[width=1\textwidth]{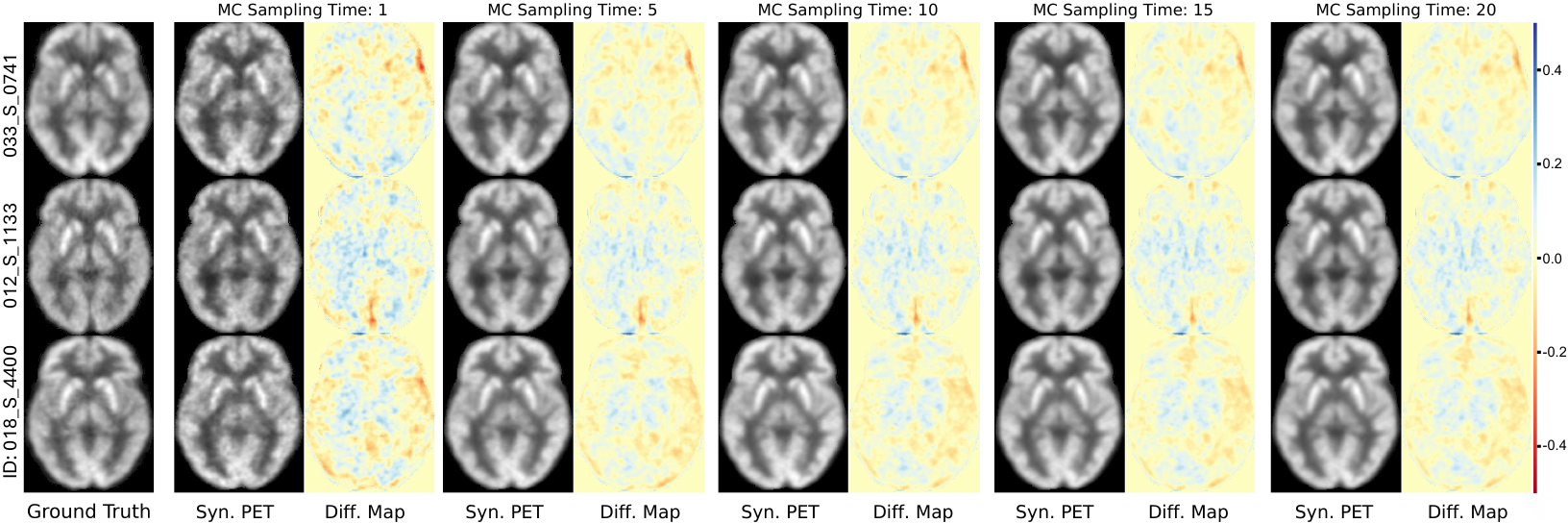}
\caption{Synthesized (Syn.) PET images and difference (Diff.) maps achieved by FICD using varying Monte-Carlo (MC) sampling times, with ground-truth PET images from ADNI~\cite{jack2008alzheimer} shown on the left for comparison.
}
\label{mc_short_error}
\end{figure*}

As described in the main text, the input for the inference phase is a random noise concatenated with MRI, and the random noise is updated in each timestep to generate corresponding PET, leading to a slight difference in synthesized images.
Following~\cite{lyu2022conversion}, we incorporate Monte-Carol (MC) sampling into our framework to mitigate this issue. 
We further investigate the influence of varying MC sampling times on the final synthesized PET in Task 1 (\ie, image synthesis). 
The quantitative outcomes of this analysis are presented in Fig.~\ref{montecarlo}, which shows a significant enhancement in terms of image quality as the MC sampling times are elevated from 1 to 5.
Beyond a sampling time of 5, the performance enhancements from additional sampling become marginal. 
Figure~\ref{mc_short_error} provides a visual comparison of synthesized images produced at MC sampling times from 1 to 20. 
From this figure, we can observe that with the increase in sampling times, the images gradually become smoother and the difference between a synthetic PET and its ground truth tends to be smaller, indicating an enhancement in image quality. 
When the MC sampling time is greater than 5, one can obtain a very stable output image with only a limited improvement in image quality. 
Considering the time-consuming nature of the inference phase, it becomes crucial to balance the benefits of increased sampling times against the associated computational costs. 

\section{Comparison of True and Synthetic PET on ADNI} 
\label{S4}

To validate the effectiveness of the synthesized PET images, we also compare the performance of PET images synthesized by FICD against real PET data.
Specifically, we replace varying portions of real PET — 20\%, 40\%, 60\%, 80\%, and 100\% — with synthesized images for downstream Task 2 (\ie, disease progression forecasting). 
The results of this comparison are presented in Table \ref{real_syn_comparison}. 
The table indicates that, while there is some fluctuation with varying portions of data substitution, the overall performance of FICD remains relatively consistent without significant changes. 
This implies that the synthetic PET images output by FICD are both reliable and clinically useful.

\begin{table*}[!tbp]
\setlength{\abovecaptionskip}{0pt}
\setlength{\belowcaptionskip}{0pt}
\setlength{\abovedisplayskip}{0pt}
\renewcommand{\arraystretch}{0.9}
\scriptsize
\centering
\caption{Results of preclinical AD progression prediction (\ie, pSMC vs. sSMC classification) on 533 subjects from ADNI, with MRI and FDG-PET data at baseline as input. 
`Ratio' is the percentage of synthetic PET generated by the proposed FICD, for instance, `20\%' means that 20\% synthetic PET and 80\% real PET are used.}

\begin{tabular*}{0.7\textwidth}{@{\extracolsep{\fill}}l| cc cc cc }
\toprule
\multirow{1}{*}{Ratio}  
&
AUC (\%)$\uparrow$ &ACC (\%)$\uparrow$ &SEN (\%)$\uparrow$ &SPE (\%)$\uparrow$ &BAC (\%)$\uparrow$ &F1s (\%)$\uparrow$  \\ 
\midrule

0\%&
68.70±5.75&60.98±4.45&66.32±7.14&58.57±3.23&62.44±5.18&51.43±5.45\\ 

20\%&
69.32±3.06&64.26±1.61&71.58±2.58&60.95±1.17&66.27±1.87&55.51±2.00\\ 

40\%&
69.27±4.66&60.98±2.62&66.32±4.21&58.57±1.90&62.44±3.06&51.43±3.27\\ 

60\%&69.67±4.56&60.98±3.34&66.32±5.37&58.57±2.43&62.44±3.90&51.43±4.16\\ 

80\%&67.59±5.99&60.98±3.93&66.32±6.32&58.57±2.86&62.44±4.59&51.43±4.90\\ 

100\%&
68.58±2.99&63.61±1.61&70.53±2.58&60.48±1.17&65.50±1.87&54.60±2.00\\ 

\bottomrule
\end{tabular*}
\label{real_syn_comparison}
\end{table*}

\begin{figure*}[htbp]
\setlength{\belowdisplayskip}{-1pt}
\setlength{\abovedisplayskip}{-1pt}
\setlength{\abovecaptionskip}{-1pt}
\setlength{\belowcaptionskip}{-1pt}
\center
\includegraphics[width= \textwidth]{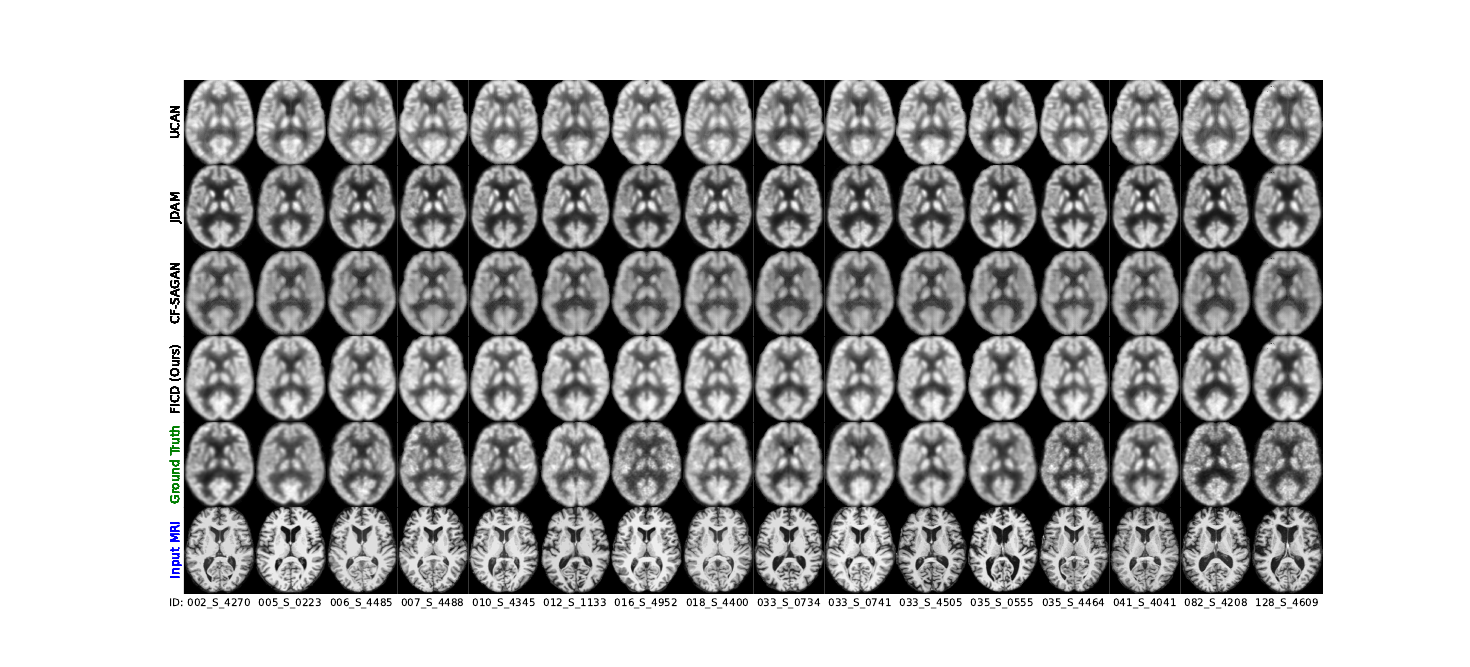}
\includegraphics[width= \textwidth]{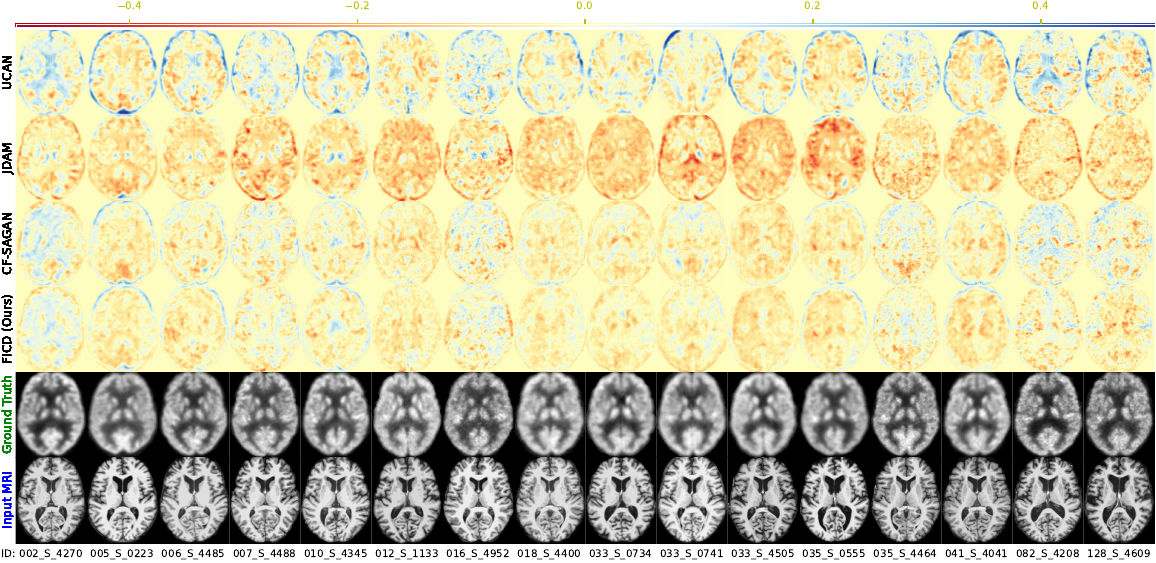}
\caption{Visualization of (top) PET images and synthesized by four methods (\ie, UCAN, JDAM, CF-SAGAN, and FICD) on cognitively normal subjects from the test set in ADNI, and (bottom) difference maps between synthetic and ground-truth PET images. The ground-truth PET images and input MRI are displayed at the bottom with corresponding subject IDs.}
 \label{fig_rencentSOTA}
\end{figure*}

\section{Visualization Comparison of FICD and Recent State-Of-The-Art Methods}
\label{S5}
In addition to the quantitative results provided in Table~8 in the main text, we also show the outputs of FICD and three recent state-of-the-art methods (\ie, \textbf{UCAN}~\cite{zhou2021synthesizing}, \textbf{JDAM}~\cite{xie2023synthesizing}, and \textbf{CF-SAGAN}~\cite{wei2020predicting}).  The PET images synthesized by the three methods and FICD and different maps on cognitively normal subjects from the test set in ADNI are shown in Fig.~\ref{fig_rencentSOTA}. 
For clarity, we introduce the detailed architectures of the three recent methods in the following. 

\begin{itemize}
    \item \textbf{UCAN}: 
This method utilizes a cyclic adversarial network consisting of a single generator and discriminator. 
The generator is a 3D U-Net comprising encoders with convolutional layers containing 64, 128, and 128 channels, each followed by instance normalization and a Squeeze-and-Excitation layer, along with residual bottleneck layers and upsampling with skip connections, ending with a Tanh-activated output layer. 
The discriminator uses a PatchGAN structure with multiple convolutional layers to classify patches as real or fake. 
Each input image is divided into 4 patches, each with dimensions of 96×96×96.

\item \textbf{JDAM}: 
This method utilizes a 2D-based approach where each image is divided into 90 slices along the axial plane. 
JDAM involves two key phases: a diffusion phase and a sampling phase. 
In the diffusion phase, PET data undergoes gradual conversion into Gaussian noise, while the MRI data is used as a condition and kept unchanged.
During this process, a joint probability distribution (JPD) PET is learned. 
The sampling phase involves using a predictor-corrector approach. 
The predictor carries out a reverse diffusion, while the corrector applies Langevin dynamics for sampling. 
The architecture of the model includes six residual blocks for downsampling, with channels of sizes 128, 256, 256, 256, 256, and 256, followed by six residual blocks for upsampling.

\item \textbf{CF-SAGAN}: 
This method involves a two-step training process. 
First, a sketcher is trained for 50 epochs to synthesize PET from MRI using a GAN-based UNet. The generator includes an encoder containing residual blocks with 8, 16, 32, 64, and 128 channels, and a symmetric decoder, and the discriminator has four convolutional layers. The training process is optimized using an $l_1$-based generator loss and a discriminator loss. 
In the second step, the generated outputs of the sketcher are used as input to train a refiner for an additional 30 epochs, utilizing the same model architecture as the sketcher. The refiner training is optimized with a weighted $l_1$-based generator loss, where the weights are derived from the brain mask and a segmentation map of white matter. 
\end{itemize}

For a fair comparison, we typically use the default setting of the three competing methods and make a concerted effort to ensure that the network architecture and hyperparameters are comparable to the proposed FICD.

\section{Details of Proposed FIC-LDM and FICD-S}
\label{S6}
\begin{figure*}[!t]
\setlength{\abovecaptionskip}{0pt}
\setlength{\belowcaptionskip}{0pt}
\setlength{\abovedisplayskip}{0pt}
\setlength\belowdisplayskip{0pt}
\setlength{\abovecaptionskip}{1pt}
\centering
\includegraphics[width=0.96\textwidth]{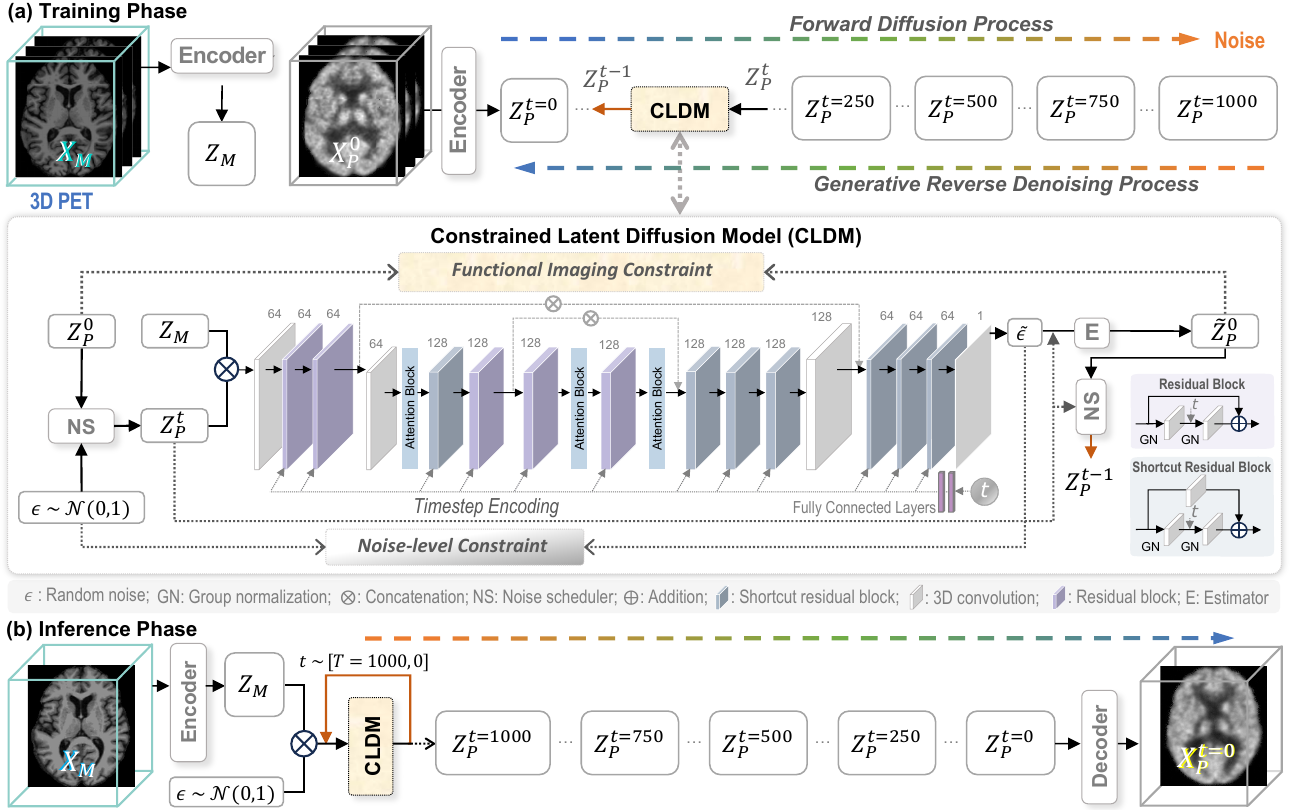}
\caption{Illustration of the proposed functional imaging constrained latent diffusion model (FIC-LDM) for 3D brain PET image synthesis with paired structural MRI as the condition. 
(a) The training phase consists of a \emph{forward diffusion process} that incrementally adds noise to an input PET image, and a \emph{generative reverse denoising process} that gradually removes the noise. 
The latent features $Z_M$ of an MRI and $Z^0_P$ of a PET scan are the inputs of the constrained latent diffusion model (CLDM), which is the latent version of our CDM in FICD.  
(b) During inference, an MRI latent feature map and a pure noise of the same size are input to the CLDM, which progressively removes the noise to generate the estimated PET latent feature map. 
This map is then decoded by a pre-trained decoder to generate the synthesized PET.}
\label{framework_FICLDM}
\end{figure*}

\begin{figure*}[!t]
\setlength{\abovecaptionskip}{0pt}
\setlength{\belowcaptionskip}{0pt}
\setlength{\abovedisplayskip}{0pt}
\setlength\belowdisplayskip{0pt}
\setlength{\abovecaptionskip}{1pt}
\centering
\includegraphics[width=0.96\textwidth]{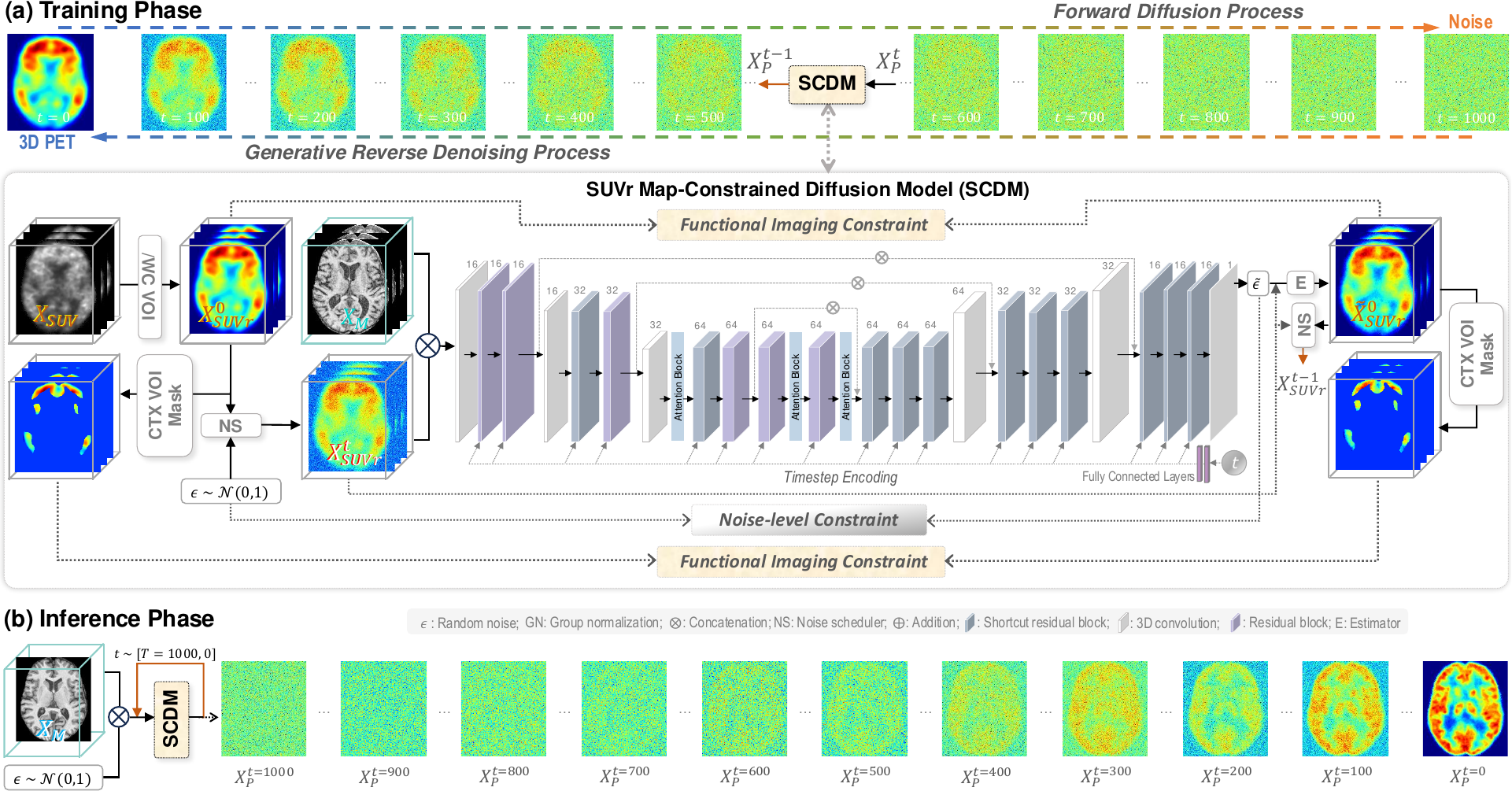}
\caption{Illustration of the proposed standardized uptake value (SUV) map-constrained diffusion model (FICD-S) for synthesizing 3D brain PET images conditioned on paired structural MRI.  
(a) The training phase includes a \emph{forward diffusion process} that incrementally adds noise to an input PET image, and a \emph{generative reverse denoising process} that gradually removes the noise. 
The SUV maps are normalized by the mean value of the whole cerebellum (WC) to derive SUV ratio (SUVr) maps. 
The SUVr maps and SUVr maps masked by global cortical target (CTX) volume-of-interest (VOI) are used as ground truth. 
(b) During inference, an MRI and a pure noise are input to the SUVr map-constrained diffusion model (SCDM), which progressively removes the noise to generate a synthetic SUVr.
}
\label{framework_FICDS}
\end{figure*}

As mentioned in the main text, in addition to FICD, we also design a functional imaging constrained latent diffusion model (called \textbf{FIC-LDM}) and a SUVr map-constrained method (called \textbf{FICD-S}). 
For clarity, we illustrate the detailed architectures of FIC-LDM and FICD-S in Fig.~\ref{framework_FICLDM} and Fig.~\ref{framework_FICDS}. 
In the following, we briefly introduce the details of these two methods.

\begin{itemize}
    \item \textbf{FIC-LDM}: 
As shown in Fig.~\ref{framework_FICLDM}, the training phase of FIC-LDM contains a \emph{forward diffusion process} and a \emph{generative reverse denoising process}. 
The input training PET image $X^0_P$ and the condition $X_M$ (\ie, MR images) are both encoded using the same encoder as the LDM, which is pre-trained alongside a decoder as part of an autoencoder by learning the reconstruction of PET using the training images. 
The corresponding latent features $Z_M$ and $Z^0_P$ are the inputs of the constrained latent diffusion model (CLDM), which is the latent version of our CDM in FICD. 
In CLDM, at timestep $t$, a noise scheduler $\mathbf{NS}$ introduces random noise $\epsilon\sim N(0,1)$ into the $Z^0_P$ to create the noisy image $Z^t_P$. 
Then the added noise $\epsilon$ is predicted by a U-Net-based neural network, and the predicted noise $\tilde \epsilon$ is used to estimate the denoised PET in the previous timestep $Z^{t-1}_P$ and in the final timestep $\tilde Z_P^0$. 
The predicted noise in CLDM is optimized by a noise-level constraint, and the estimated $\tilde Z^0_P$ is optimized by the proposed functional imaging constraint.
In the inference phase, an MRI latent feature map and a pure noise of the same size are input to the CLDM, which progressively removes the noise to generate the estimated PET latent features. 
The estimated PET latent features are then decoded by the pre-trained decoder to generate the synthesized PET.

\item \textbf{FICD-S}: 
As shown in Fig.~\ref{framework_FICDS}, the training phrase of FICD-S includes a \emph{forward diffusion process} and a \emph{generative reverse denoising process}. 
In the standardized uptake value ratio (SUVr) map-constrained diffusion model (SCDM), the SUV map $X_{SUV}$ is firstly divided by the mean intensity of the whole cerebellum to calculate the SUVr map $X^0_{SUVr}$. 
At timestep $t$, a noise scheduler $\mathbf{NS}$ introduces random noise $\epsilon\sim N(0,1)$ into the $X^0_{SUVr}$ to create the noisy image $X^t_{SUVr}$. 
The added noise $\epsilon$ is then predicted using a U-Net-based neural network, which takes the concatenated inputs of $X^t_{SUVr}$ and condition $X_M$ (\ie, MR images).
The predicted noise $\tilde \epsilon$ is used to estimate the denoised SUVr at the previous timestep $X^{t-1}_{SUVr}$ and in the final timestep $\tilde X_{SUVr}^0$. 
A global cortical target (CTX) volume-of-interest (VOI) mask is applied to $\tilde X_{SUVr}^0$ to obtain a weighted output.
The predicted noise in SCDM is optimized through a noise-level constraint.
Additionally, both the estimated $\tilde X^0_{SUVr}$ and the weighted estimate of $\tilde X^0_{SUVr}$ are optimized by the proposed functional imaging constraint.
During the inference phase, $X_M$ and a pure noise of the same size are concatenated and input to the SCDM, which progressively removes the noise to generate the estimated SUVr.

\end{itemize}

\section{Visualization Results of FIC-LDM, LDM, and Their Variants}
\label{S7}
\begin{figure*}[htbp]
\setlength{\belowdisplayskip}{-1pt}
\setlength{\abovedisplayskip}{-1pt}
\setlength{\abovecaptionskip}{-3pt}
\setlength{\belowcaptionskip}{-1pt}
\center
 \includegraphics[width= 0.96\textwidth]{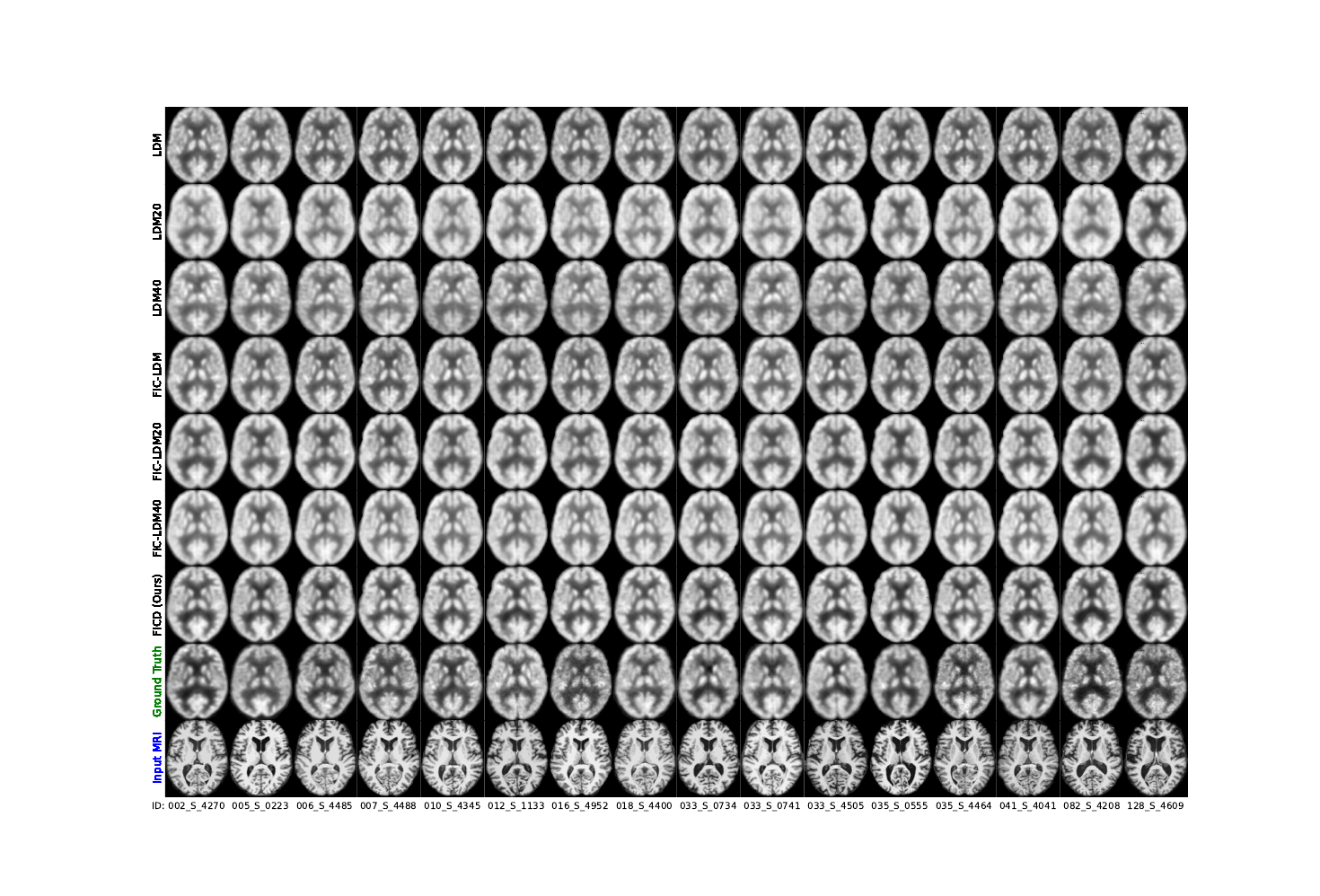}
  \includegraphics[width= 0.96\textwidth]{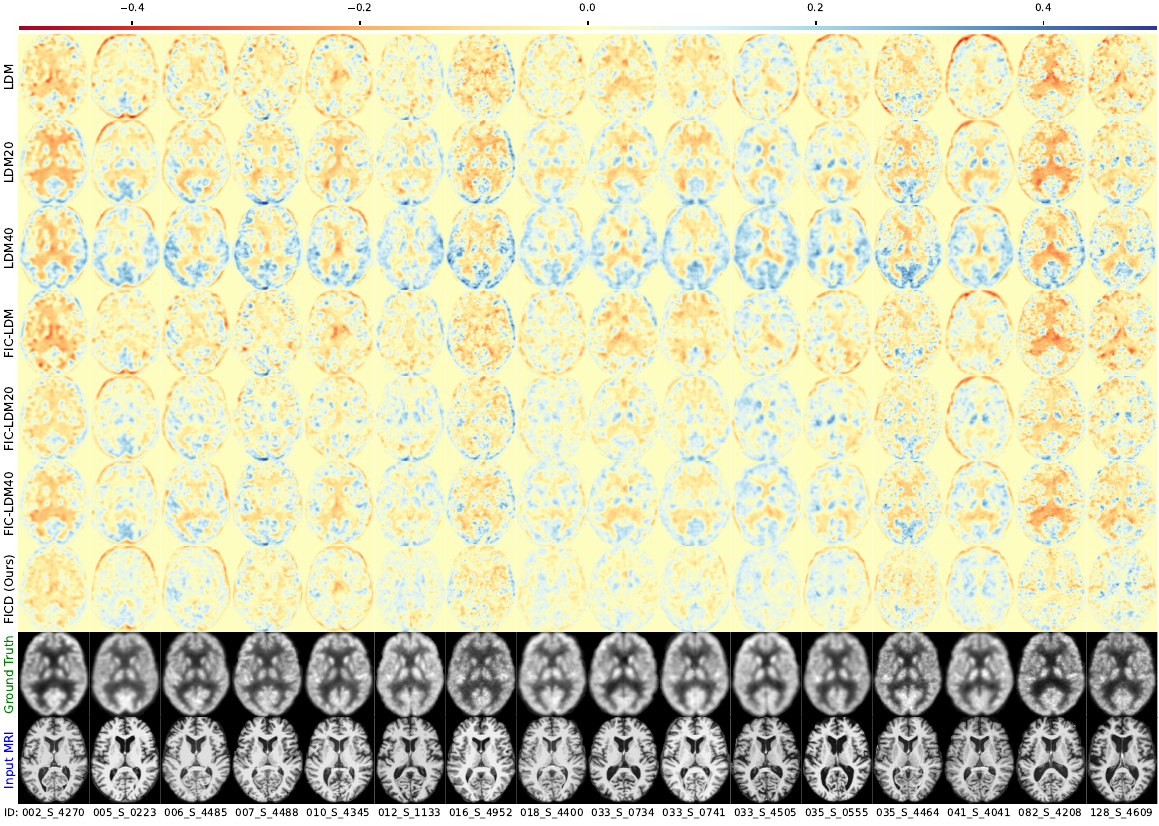}
\caption{Visualization of (top) PET images synthesized by seven methods (\ie, LDM, LDM20, LDM40, FIC-LDM, FIC-LDM20, FIC-LDM40, and FICD) on cognitively normal subjects from the test set in ADNI~\cite{jack2008alzheimer}, and (bottom) difference maps between synthetic and ground-truth PET images. The ground-truth PET images and input MRI are displayed at the bottom with the corresponding subject IDs.} 
 \label{fig_fic_ldm}
\end{figure*}

In addition to the quantitative results provided in Table~8 in the main text, we further visualize the outputs of the functional imaging constrained latent diffusion model (\textbf{FIC-LDM}) and its two variants (\ie, \textbf{FIC-LDM20} and \textbf{FIC-LDM40}), the latent diffusion model (LDM) and its two variants (\ie, \textbf{LDM20} and \textbf{LDM40}), as well as FICD. 
Specifically, both FIC-LDM and LDM feature a latent map dimension of $10 \times 12 \times 10$. FIC-LDM20 and LDM20 have a latent map dimension of $20 \times 24 \times 20$. Additionally, FIC-LDM40 and LDM40 possess a latent map dimension of $40 \times 48 \times 40$.
The PET images synthesized by the seven methods and different maps on cognitively normal subjects from the test set in ADNI are shown in Fig.~\ref{fig_fic_ldm}.

\begin{table*}[!t]
\setlength{\abovecaptionskip}{0pt} 
\setlength{\belowcaptionskip}{0pt}  
\setlength\abovedisplayskip{0pt}
\setlength\belowdisplayskip{0pt}
\centering
\caption{Computational cost comparison across all methods using an RTX 3090 GPU with
24 GB of memory. M: Million; S: Second.}
\footnotesize
\setlength{\tabcolsep}{2pt}
\begin{tabular}{l|p{3cm}p{3cm}}
\toprule
     ~~Method &  Parameters (M) & Inference Time (S)\\ 
\midrule
     ~~GAN 
     & 3.66 & 11.03\\ 
     ~~CycleGAN 
     & 1.81 & 13.50\\ 
     ~~VAE 
     & 36.64 & 0.14\\ 
     ~~VAEGAN 
     & 9.29 & 0.14\\ 
     ~~DDPM 
     & 3.04 & 597.00\\ 
     ~~UCAN & 0.95 & 4.34 \\ 
     ~~JDAM 
     & 52.81 & 6,509.34\\ 
     ~~CF-SAGAN 
     & 2.83 & 0.06\\ 
     ~~FIC-LDM, LDM 
     & 16.43 & 13.00\\ 
     ~~FIC-LDM20, LDM20 
     & 10.60 & 23.00\\ 
     ~~FIC-LDM40, LDM40 
     & 34.09 & 54.00\\ 
     ~~FICD, FICD-S 
     & 3.04 & 634.00\\ 
     \bottomrule
\end{tabular}
\label{table_timeCost}
\end{table*}

\section{Computational Cost Comparison}
\label{S8}
Given that all methods in this study are based on deep learning, we evaluated their computational costs by assessing the number of trainable parameters and the inference time for a batch size of one, using an RTX 3090 GPU with 24 GB of memory. 
As presented in Table~\ref{table_timeCost}, the proposed FICD model uses fewer trainable parameters than most competing approaches. 
It should also be noted that DDPM-based methods (\ie, DDPM, FICD, and FICD-S) are associated with relatively higher time costs for inference. 
To address this, our FIC-LDM and its variants provide an alternative to FICD, significantly decreasing inference time, though with some trade-off in overall performance (see Table~8 in the main text). 
Additionally, we implement the proposed methods on a GPU cluster with four H100 GPUs (each with 80 GB of memory) and find the inference time of FICD and FICD-S for generating a synthesized 3D image is approximately 30 seconds,  
while this inference time is shortened to $1.47$ seconds using FIC-LDM40 and only $0.15$ seconds using FIC-LDM. 
This significantly improves the inference speed and makes FICD and its variants well-adaptable in practical applications.

\section{Data Processing for Images from Centiloid Project}
\label{S9}
As mentioned in Section 5.6 of the main text, we design a standardized uptake value (SUV) map-constrained method, called \textbf{FICD-S}, and evaluate the clinical utility of synthetic PET images generated by FICD-S and FICD on the Centiloid Project~\cite{klunk2015centiloid}. 
This project is designed to standardize quantitative amyloid imaging measures in multi-tracer amyloid PET data. 
There are 34 Alzheimer's disease (AD) patients and 45 young control (YC) subjects in this cohort, with each subject equipped with paired SUV maps of PiB-PET images (not raw PET data) and T1-weighted MRI scans. 
The data preprocessing follows the pipeline provided on the website of the Centiloid Project, with an additional image smoothing step to reduce noise. 
After processing, all resultant global cortical target (CTX) SUVr values have less than 5\% difference from the values published in the Centiloid paper, and the CL values of each group have less than 2\% difference from the provided values. 
This ensures that the data we process meets the quality control requirements of the Centiloid Project (\href{https://www.gaain.org/centiloid-project}{https://www.gaain.org/centiloid-project}). 
The SUVr image is gained from using the whole cerebellum as the reference volume-of-interest (VOI), by dividing the SUV map by the mean intensity of the cerebellum.  
Since the intensity of SUVr of the subjects at the same disease stage has a similar range, \ie, around $[0, 3.2]$ for AD subjects and $[0, 2]$ for YC subjects, we normalize the data range to roughly $[-1,1]$ by a fixed linear mapping. 
This would enable us to recover the original intensity of SUVr images by inverse calculation to the synthesized output images. 
In this experiment, since we only have an SUV map for each subject (no raw PiB-PET data), we use the SUV map-based constraint to replace the original PET image-based constraint in FICD. 
The FICD-S method will directly output SUVr maps based on T1-weighted MRIs and noise. 

Once the SUVr map is obtained, the mean SUVr value at CTX VOI is calculated for each subject, and these values are averaged within each group. 
In FICD-S, the calculated group mean for the YC subjects is $SUVr_{YC}=1.008$ and the mean for AD subjects is $SUVr_{AD}=1.996$. 
In FICD, the calculated group mean for the YC subjects is $SUVr_{YC}=1.026$ and the mean for AD subjects is $SUVr_{AD}=2.008$. 
These values are then set to 0 and 100 CL, respectively, and the CL value for each individual subject is calculated using the following expression~\cite{klunk2015centiloid}:
\begin{equation}
CL = 100 * \frac{SUVr_{IND}-SUVr_{YC}}{SUVr_{AD}-SUVr_{YC}} \nonumber
\label{CL}
\end{equation}
where $SUVr_{IND}$ is an individual's SUVr value.

\begin{figure}[!t]
\setlength{\abovecaptionskip}{0pt}
\setlength{\belowcaptionskip}{0pt}
\setlength{\abovedisplayskip}{0pt}
\setlength\belowdisplayskip{0pt}
\setlength{\abovecaptionskip}{1pt}
\centering
\includegraphics[width=0.6\textwidth]{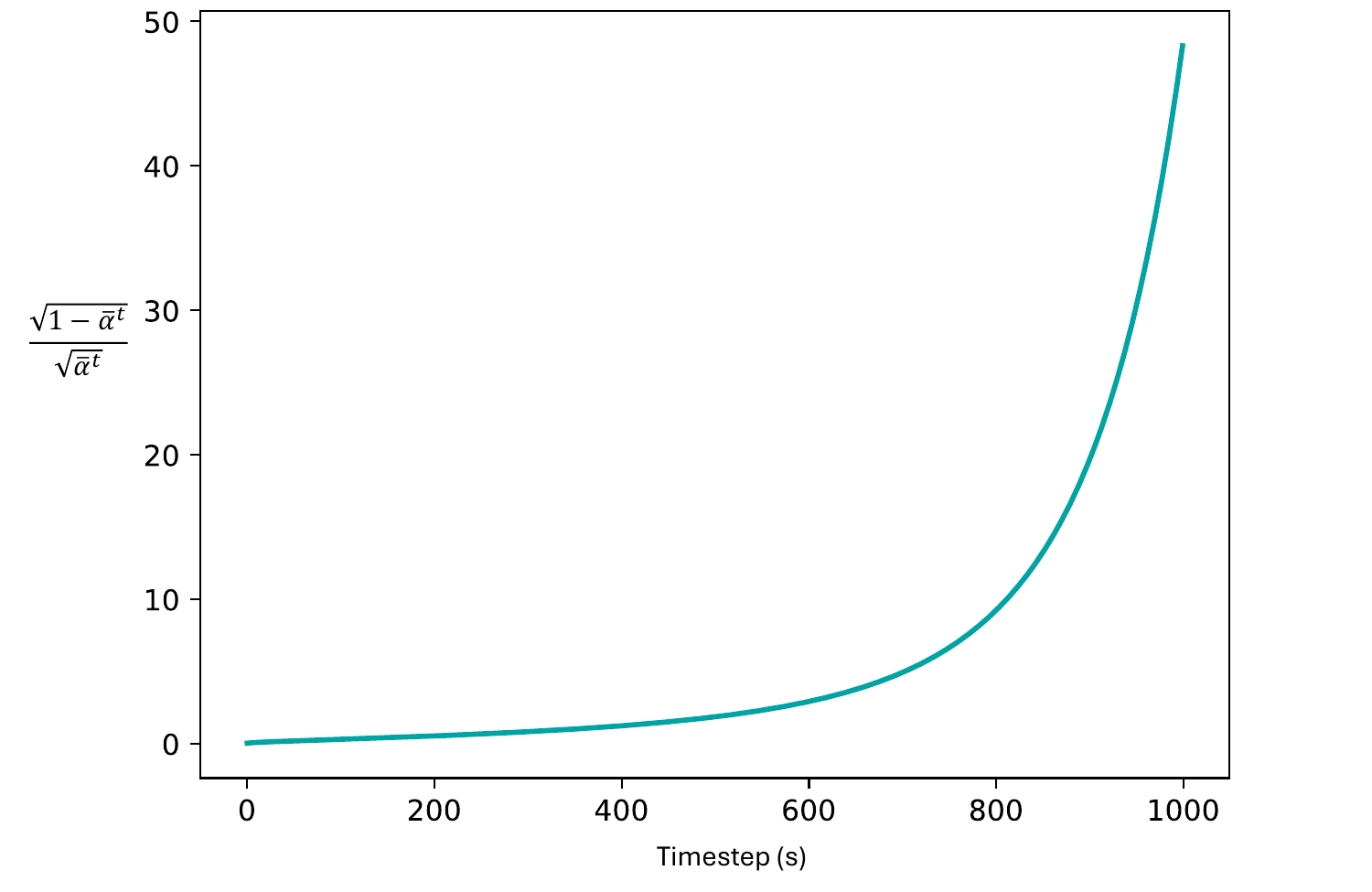}
\caption{The trend line of $\frac{\sqrt{1-\bar{\alpha}^t}}{\sqrt{\bar{\alpha}^t}}$ in FICD over timesteps.}
\label{alpha}
\end{figure}

\section{More Explanation of Proposed FIC}
\label{S10}
In addition to the conventional noise-level loss $L_N$ used in diffusion models in Eq.~(4), we develop a new image-level constraint FIC, \ie, $L_I$ in Eq.~(6) in the main text. 
By rewriting Eq.~(3), we have the new expression of $X_P^0$ as follows 
\begin{equation}
\small
X^0_P = \frac{(X^t_P - \epsilon \sqrt{1 - \bar{\alpha}^t})}{\sqrt{\bar{\alpha}^t}}. \nonumber
\label{x_0_rewrite}
\end{equation}
By combining Eq.~(5), we have the following
\begin{equation}
\small
X^0_P - \tilde X_P^0 = \frac{\sqrt{1-\bar{\alpha}^t}}{\sqrt{\bar{\alpha}^t}}(\tilde \epsilon - \epsilon),  \nonumber
\label{x_0_rewrite}
\end{equation}
where $\epsilon$ is the true noise and $\tilde{\epsilon}$ is the predicted noise. 
Since $\bar{\alpha}^t := \prod_{s=1}^{t} \alpha^s$
is a time-dependent hyperparameter, the value of $\frac{\sqrt{1-\bar{\alpha}^t}}{\sqrt{\bar{\alpha}^t}}$ varies with the timestep, following the trend shown in Fig.~\ref{alpha}. 
Thus the difference between $X_P^0$ and $\tilde X_P^0$ can be treated as a weighted noise loss, and the weight increases with each timestep.

In fact, employing a constant training weight for the noise loss $L_N$ in diffusion models would introduce bias in training, whereas varying the weights across timesteps can improve the quality of the generated images~\cite{yu2023debias}. 
With the proposed FIC, we assign higher weight as the timestep increases, thereby compelling the noise error $\tilde \epsilon - \epsilon$ to decrease more significantly at larger timestep. 
This helps the training of diffusion models by avoiding generating low-quality images with relatively large noise scales at various timesteps. 
On the other hand, the motivation behind introducing the functional imaging constraint is to enhance the fidelity of synthesized PET images, \emph{ensuring voxel-wise alignment with their ground truth}. 
While the traditional noise-level constraint $L_N$ effectively minimizes the error in noise prediction at each denoising step, it indirectly influences the image quality by controlling the denoising process. 
In contrast, FIC directly targets the alignment and accuracy of the final image output, focusing specifically on the functional information conveyed by voxel intensities that are critical to PET scans. 
The empirical results in Section 5.1 of the main text verify the effectiveness of FIC by comparing our FICD with the traditional diffusion model (\ie, DDPM) in terms of image fidelity.

{\color{black}
\section{Detailed information on studied subjects}
\label{S11}
We display the detailed subject data information, including the data source, training/test data partition, subject category, subject number, the radiotracer for PET images, and two types of clinical scores used in each task in Table~\ref{Demography}.
}

\begin{table}[t]
\setlength{\abovecaptionskip}{0pt}
\setlength{\belowcaptionskip}{0pt}
\setlength{\abovedisplayskip}{0pt}
\setlength\belowdisplayskip{0pt}
\renewcommand{\arraystretch}{0.9}
\setlength{\tabcolsep}{0.2pt}
\scriptsize
\centering
\caption{Clinical information of subjects used in this work. 
Mini-mental state examination (MMSE) and clinical dementia rating (CDR) scores are presented as mean ± standard deviation. 
Four types of radiotracers are involved in PET imaging: $^{18}$F-fluorodeoxyglucose (FDG), Pittsburgh Compound-B (PiB), $^{18}$F-flutemetamol (FLUTE), and Florbetapir (AV45).}%
\setlength\tabcolsep{8pt}
\begin{tabular}{l|ccccccc}
\toprule
~Task & Cohort&Group&Class&Subject~\#&PET Radiotracer&MMSE&CDR\\
\midrule
  ~\multirow{2}{*}{Task~1}&\multirow{2}{*}{ADNI }
  ~&Training&CN&263&FDG& --&--\\
  ~&&Test&CN&30&FDG&--&--\\
\midrule
    ~\multirow{8}{*}{Task~2 \& Task~3}&\multirow{2}{*}{ADNI}
    ~&Training&AD&359&FDG
    &23.31±2.05&4.29±1.81\\
  ~&&Training&CN&436&FDG
  &29.08±1.12&0.04±0.19\\
\cmidrule(lr){2-8}
  ~&\multirow{2}{*}{CLAS-SCD}
  ~&Test&pSCD&24&FDG
  &26.75±2.55&--\\
  ~&&Test&sSCD&51&FDG
  &27.73±2.15&--\\
\cmidrule(lr){2-8}
  ~&\multirow{2}{*}{ADNI-SMC}&Test&pSMC&19&FDG&28.95±1.36&1.45±1.14\\
  ~&&Test&sSMC&42&FDG&29.07±0.99&0.67±0.24\\
\midrule
  ~\multirow{6}{*}{Task~4}&\multirow{2}{*}{AIBL}
    ~&Fine-tuning&CN&107&PiB&--&--\\
  ~&&Test&CN&12&PiB&--&--\\
\cmidrule(lr){2-8}
  ~&\multirow{2}{*}{AIBL}&Fine-tuning&CN&128&FLUTE&--&--\\
  ~&&Test&CN&15&FLUTE&--&--\\
\cmidrule(lr){2-8}
  ~&\multirow{2}{*}{AIBL}&Fine-tuning&CN&62&AV45&--&--\\
  ~&&Test&CN&7&AV45&--&--\\
\bottomrule
\end{tabular}
\label{Demography}
\end{table}

To facilitate reproducible research, 
in Table~\ref{traing_test}, we list all the subject IDs from Alzheimer's Disease Neuroimaging Initiative (\textbf{ADNI}) dataset~\cite{jack2008alzheimer} that we use for training and test the synthetic models in Task 1. 
In Table~\ref{scd_Subject}, we list all subjects from the Chinese Longitudinal Aging Study (\textbf{CLAS})~\cite{xiao2016china} and significant memory concerns (SMC) from the ADNI dataset~\cite{jack2008alzheimer} that are used in Tasks 2-3. 
Table~\ref{aibl_sbj} shows the list of the subjects from the Australian Imaging, Biomarkers and Lifestyle (\textbf{AIBL}) database~\cite{ellis2009australian} that are used for fine-tuning and test in Task 4. 
In Table~\ref{centiloid_sbj}, we list all data from the Centiloid Project~\cite{klunk2015centiloid} that are used for fine-tuning and test in the clinical evaluation task (see Section 5.6 in the main text).

\begin{table*}[!tbp]
\setlength{\belowdisplayskip}{-1pt}
\setlength{\abovedisplayskip}{-1pt}
\setlength{\abovecaptionskip}{-1pt}
\setlength{\belowcaptionskip}{-1pt}
\renewcommand{\arraystretch}{0.9}
\setlength{\tabcolsep}{0.3pt}
\scriptsize
\centering
\caption{
Subject IDs of data from Alzheimer's Disease Neuroimaging Initiative ({ADNI}) dataset that are used for training and test in Task 1.}%
\begin{tabular*}{1\textwidth}{@{\extracolsep{\fill}}l|cc cc cc cc cc c}
\toprule
~Group &
\multicolumn{11}{c}{Subject ID} \\
\midrule
~\multirow{24}{*}{Training}
& 002\_S\_2010
& 002\_S\_4213
& 002\_S\_4225
& 002\_S\_4262
& 003\_S\_4119
& 003\_S\_4288
& 003\_S\_4350
& 003\_S\_4441
& 003\_S\_4555
& 003\_S\_4644
& 003\_S\_4872\\
& 003\_S\_4900
& 005\_S\_0610
& 006\_S\_0484
& 006\_S\_0498
& 006\_S\_0731
& 006\_S\_4150
& 006\_S\_4357
& 006\_S\_4449
& 007\_S\_4387
& 007\_S\_4516
& 007\_S\_4620\\
& 007\_S\_4637
& 009\_S\_0751
& 009\_S\_0842
& 009\_S\_0862
& 009\_S\_4337
& 009\_S\_4388
& 009\_S\_4612
& 010\_S\_0067
& 010\_S\_0419
& 010\_S\_0420
& 010\_S\_0472\\
& 010\_S\_4442
& 011\_S\_0002
& 011\_S\_0005
& 011\_S\_0008
& 011\_S\_0021
& 011\_S\_0023
& 011\_S\_4075
& 011\_S\_4105
& 011\_S\_4120
& 011\_S\_4222
& 011\_S\_4278\\
& 012\_S\_0637
& 012\_S\_4026
& 012\_S\_4545
& 012\_S\_4643
& 013\_S\_0502
& 013\_S\_0575
& 013\_S\_4579
& 013\_S\_4580
& 014\_S\_4080
& 014\_S\_4093
& 014\_S\_4401\\
& 014\_S\_4576
& 014\_S\_4577
& 016\_S\_0359
& 016\_S\_4097
& 016\_S\_4121
& 016\_S\_4638
& 016\_S\_4688
& 016\_S\_4951
& 018\_S\_0043
& 018\_S\_0055
& 018\_S\_4313\\
& 018\_S\_4349
& 018\_S\_4399
& 019\_S\_4367
& 019\_S\_4835
& 020\_S\_0097
& 020\_S\_0883
& 021\_S\_0647
& 021\_S\_4254
& 021\_S\_4276
& 021\_S\_4335
& 021\_S\_4421\\
& 022\_S\_0014
& 022\_S\_0066
& 022\_S\_0096
& 022\_S\_0130
& 022\_S\_4173
& 022\_S\_4196
& 022\_S\_4266
& 022\_S\_4291
& 022\_S\_4320
& 023\_S\_4164
& 023\_S\_4448\\
& 024\_S\_0985
& 024\_S\_1063
& 024\_S\_4084
& 024\_S\_4158
& 027\_S\_0074
& 027\_S\_0120
& 029\_S\_0843
& 029\_S\_0845
& 029\_S\_0866
& 029\_S\_4279
& 029\_S\_4290\\
& 029\_S\_4384
& 029\_S\_4385
& 029\_S\_4585
& 029\_S\_4652
& 031\_S\_0618
& 031\_S\_4032
& 031\_S\_4218
& 031\_S\_4474
& 031\_S\_4496
& 032\_S\_0095
& 032\_S\_4277
\\
& 032\_S\_4348
& 032\_S\_4429
& 032\_S\_4921
& 033\_S\_4176
& 033\_S\_4177
& 033\_S\_4179
& 033\_S\_4508
& 035\_S\_0048
& 035\_S\_4082
& 036\_S\_0576
& 036\_S\_0672\\
& 036\_S\_0813
& 036\_S\_4389
& 036\_S\_4491
& 036\_S\_4878
& 037\_S\_0327
& 037\_S\_0454
& 037\_S\_0467
& 037\_S\_4028
& 037\_S\_4071
& 037\_S\_4308
& 037\_S\_4410\\
& 041\_S\_0262
& 041\_S\_0898
& 041\_S\_1002
& 041\_S\_4014
& 041\_S\_4037
& 041\_S\_4060
& 041\_S\_4200
& 041\_S\_4427
& 053\_S\_4578
& 057\_S\_0779
& 057\_S\_0818\\
& 057\_S\_0934
& 062\_S\_1099
& 068\_S\_4340
& 068\_S\_4424
& 070\_S\_4856
& 070\_S\_5040
& 072\_S\_0315
& 072\_S\_4391
& 073\_S\_0311
& 073\_S\_0312
& 073\_S\_0386\\
& 073\_S\_4155
& 073\_S\_4382
& 073\_S\_4393
& 073\_S\_4552
& 073\_S\_4559
& 073\_S\_4739
& 073\_S\_4762
& 073\_S\_4795
& 073\_S\_5023
& 082\_S\_0363
& 082\_S\_4090\\
& 082\_S\_4224
& 082\_S\_4339
& 082\_S\_4428
& 094\_S\_0489
& 094\_S\_0526
& 094\_S\_4234
& 094\_S\_4459
& 094\_S\_4503
& 094\_S\_4560
& 094\_S\_4649
& 098\_S\_0171\\
& 098\_S\_4003
& 098\_S\_4018
& 098\_S\_4050
& 098\_S\_4275
& 098\_S\_4506
& 099\_S\_0090
& 099\_S\_0352
& 099\_S\_0533
& 099\_S\_0534
& 099\_S\_4076
& 099\_S\_4086\\
& 099\_S\_4104
& 100\_S\_0047
& 100\_S\_4469
& 109\_S\_0967
& 109\_S\_4499
& 114\_S\_0173
& 114\_S\_0416
& 116\_S\_0360
& 116\_S\_0648
& 116\_S\_0657
& 116\_S\_4010\\
& 116\_S\_4043
& 116\_S\_4092
& 116\_S\_4453
& 116\_S\_4483
& 116\_S\_4855
& 123\_S\_4362
& 126\_S\_0506
& 126\_S\_0680
& 127\_S\_0259
& 127\_S\_2234
& 127\_S\_4148\\
& 127\_S\_4198
& 127\_S\_4604
& 127\_S\_4645
& 127\_S\_4843
& 128\_S\_0230
& 128\_S\_0245
& 128\_S\_0272
& 128\_S\_0500
& 128\_S\_0522
& 128\_S\_0863
& 128\_S\_2123
\\
& 128\_S\_4586
& 128\_S\_4599
& 128\_S\_4607
& 128\_S\_4832
& 129\_S\_0778
& 129\_S\_4369
& 129\_S\_4371
& 129\_S\_4396
& 129\_S\_4422
& 130\_S\_0232
& 130\_S\_1200\\
& 130\_S\_4343
& 130\_S\_4352
& 131\_S\_0123
& 131\_S\_0319
& 135\_S\_4446
& 135\_S\_4566
& 135\_S\_4598
& 136\_S\_4269
& 136\_S\_4433
& 137\_S\_0283
& 137\_S\_0301\\
& 137\_S\_0459
& 137\_S\_0686
& 137\_S\_0972
& 137\_S\_4466
& 137\_S\_4482
& 137\_S\_4520
& 137\_S\_4587
& 137\_S\_4632
& 153\_S\_4125
& 153\_S\_4139
& 153\_S\_4151\\
& 153\_S\_4372
& 941\_S\_1195
& 941\_S\_1197
& 941\_S\_1202
& 941\_S\_1203
& 941\_S\_4066
& 941\_S\_4100
& 941\_S\_4255
& 941\_S\_4365
& 941\_S\_4376\\

\midrule
~\multirow{3}{*}{Test}
& 002\_S\_4270
& 003\_S\_4081
& 005\_S\_0223
& 006\_S\_4485
& 007\_S\_4488
& 010\_S\_4345
& 011\_S\_0016
& 012\_S\_1133
& 013\_S\_4616
& 016\_S\_4952
& 018\_S\_4400\\
& 021\_S\_4558
& 031\_S\_4021
& 032\_S\_4386
& 033\_S\_0734
& 033\_S\_0741
& 033\_S\_4505
& 035\_S\_0555
& 035\_S\_4464
& 036\_S\_1023
& 041\_S\_4041
& 062\_S\_0768\\
& 068\_S\_4174
& 072\_S\_4103
& 082\_S\_4208
& 094\_S\_2201
& 109\_S\_1013
& 128\_S\_4609
& 941\_S\_1194
& 941\_S\_4292\\
\bottomrule
\end{tabular*}
\label{traing_test}
\end{table*}

\begin{table*}[!tbp]
\setlength{\belowdisplayskip}{-1pt}
\setlength{\abovedisplayskip}{-1pt}
\setlength{\abovecaptionskip}{-1pt}
\setlength{\belowcaptionskip}{-1pt}
\renewcommand{\arraystretch}{0.9}
\setlength{\tabcolsep}{0.3pt}
\scriptsize
\centering
\caption{Subject IDs of the two cohorts that are used in the two downstream tasks (\ie, Task 2 and Task 3) to evaluate the synthesized PET, including progressive subject cognitive decline (pSCD) and stable SCD (sSCD) subjects from CLAS based on a 7-year follow-up, and progressive significant memory concerns (pSMC) and stable SMC (sSMC) subjects from SMC cohort in ADNI, based on a 2-year follow-up.}%

\begin{tabular*}{1\textwidth}{@{\extracolsep{\fill}}l|c|cc cc cc cc c}
\toprule
~Cohort & Class &
\multicolumn{9}{c}{Subject ID} \\
\midrule
~\multirow{9}{*}{CLAS-SCD}&\multirow{3}{*}{~pSCD }&SMHCA0008 & SMHCA0017 & SMHCA0087 & SMHCA0144 & SMHCA0204 & SMHCA0363
& SMHCA0405 & SMHCA0514 & SMHCA0543\\
~&& SMHCA0638 & SMHCAB009 & SMHCAB023 
&SMHCAB033 & SMHCAB126 & SMHCAB127 & SMHCAB132 & SMHCAB138 & SMHCAB257\\
~&& SMHCAH045 & SMHCAH051 & SMHCAH129 & SMHCAH182 & SMHCAH311 & SMHCAH337 \\

\cmidrule{2-11}

~&\multirow{6}{*}{~sSCD }& SMHCA0004 & SMHCA0014 & SMHCA0015 & SMHCA0020 & SMHCA0028 & SMHCA0031&SMHCA0093 & SMHCA0133 &SMHCA0143\\
~&& SMHCA0147 & SMHCA0185 & SMHCA0186 &SMHCA0211 & SMHCA0230 & SMHCA0250 & SMHCA0262 &SMHCA0396 & SMHCA0402\\
~&&SMHCA0406 & SMHCA0425 & SMHCA0429 & SMHCA0480 & SMHCA0541 & SMHCA0559 &SMHCA0584 & SMHCA0600 & SMHCA0619 \\
~&& SMHCA0624 & SMHCA0643 & SMHCAB029 &SMHCAB080 & SMHCAB114 & SMHCAB128 & SMHCAB178 & SMHCAB204 & SMHCAB232 \\
~&&SMHCAH006 & SMHCAH028&SMHCAH036 & SMHCAH069 & SMHCAH073 & SMHCAH113 &SMHCAH118 & SMHCAH143 & SMHCAH171\\
~&& SMHCAH245&SMHCAH285 & SMHCAH294 &SMHCAH296 & SMHCAH297 & SMHCAH347 \\

\midrule

~\multirow{9}{*}{~ADNI-SMC }& \multirow{3}{*}{~pSMC }& 007\_S\_5265 & 021\_S\_5237 & 027\_S\_5277 & 029\_S\_5166 & 029\_S\_5219 & 032\_S\_5263 & 035\_S\_4785 & 041\_S\_5253 &051\_S\_5285 \\
~&& 051\_S\_5294 & 072\_S\_5207 & 082\_S\_5282 & 100\_S\_5096 & 114\_S\_5234 & 126\_S\_5214 &127\_S\_5132 & 130\_S\_5258&135\_S\_5273 \\
~&& 137\_S\_4862 \\

\cmidrule{2-11}

~&\multirow{5}{*}{~sSMC }& 002\_S\_5178 & 002\_S\_5230 & 003\_S\_5130 & 003\_S\_5154 & 003\_S\_5209 & 012\_S\_5157 & 012\_S\_5195 & 013\_S\_5171 & 020\_S\_5140\\
~& & 021\_S\_5177 & 021\_S\_5194 & 024\_S\_5290 & 027\_S\_5083 & 027\_S\_5093 & 027\_S\_5109 & 027\_S\_5118 & 027\_S\_5169 & 027\_S\_5170 \\
~&& 027\_S\_5288 & 029\_S\_5158 & 032\_S\_5289 & 033\_S\_5198 & 033\_S\_5259 & 037\_S\_5126 & 037\_S\_5222 & 041\_S\_5078 & 041\_S\_5097 \\
~&& 041\_S\_5100 & 041\_S\_5141 & 053\_S\_5272 & 053\_S\_5296 & 057\_S\_5292 & 082\_S\_5278 &  100\_S\_5091 & 126\_S\_5243 & 127\_S\_5185 \\
~&& 127\_S\_5200 & 127\_S\_5228 & 127\_S\_5266 & 130\_S\_5175 & 135\_S\_5113 & 941\_S\_5193 \\
\bottomrule
\end{tabular*}
\label{scd_Subject}
\end{table*}

\begin{table*}[!tbp]
\setlength{\belowdisplayskip}{-1pt}
\setlength{\abovedisplayskip}{-1pt}
\setlength{\abovecaptionskip}{-1pt}
\setlength{\belowcaptionskip}{-1pt}
\renewcommand{\arraystretch}{0.9}
\setlength{\tabcolsep}{0.3pt}
\scriptsize
\centering
\caption{
Subject IDs of data from the Australian Imaging, Biomarkers and Lifestyle (AIBL) database~\cite{ellis2009australian} that are used for fine-tuning and test in Task 4.}
\begin{tabular*}{1\textwidth}
{@{\extracolsep{\fill}}l|ccc ccc ccc ccc ccc ccc cc cc cc c}
\toprule
~Group &
\multicolumn{25}{c}{Subject ID} \\
\midrule
~\multirow{6}{*}{FLUTE-PET for Fine-tuning }
& 24
& 51
& 78
& 144
& 147
& 148
& 149
& 155
& 161
& 167
& 168
& 185
& 190
& 191
& 197
& 198
& 220
& 228
& 250
& 307
& 311
& 326
& 329
& 338
& 415\\
& 420
& 428
& 464
& 484
& 496
& 500
& 533
& 534
& 618
& 626
& 632
& 636
& 637
& 638
& 639
& 644
& 668
& 669
& 674
& 691
& 692
& 693
& 733
& 764
& 770\\
& 774
& 791
& 811
& 817
& 818
& 835
& 845
& 859
& 875
& 884
& 907
& 915
& 985
& 988
& 1215
& 1225
& 1228
& 1230
& 1234
& 1236
& 1237
& 1249
& 1251
& 1257
& 1258\\
& 1265
& 1278
& 1291
& 1295
& 1302
& 1303
& 1309
& 1311
& 1312
& 1332
& 1334
& 1335
& 1337
& 1339
& 1341
& 1344
& 1356
& 1361
& 1373
& 1378
& 1386
& 1392
& 1396
& 1405
& 1410\\
& 1413
& 1416
& 1417
& 1418
& 1419
& 1421
& 1423
& 1431
& 1432
& 1461
& 1483
& 1494
& 1503
& 1517
& 1520
& 1531
& 1541
& 1553
& 1563
& 1565
& 1566
& 1567
& 1569
& 1574
& 1585\\
& 1586
& 1587
& 1598
\\
\midrule
~\multirow{1}{*}{FLUTE-PET for Test}
& 234
& 313
& 709
& 786
& 860
& 1098
& 1218
& 1285
& 1330
& 1343
& 1360
& 1370
& 1412
& 1422
& 1564
\\
\midrule
~\multirow{3}{*}{AV45-PET for Fine-tuning}
& 2
& 127
& 146
& 154
& 158
& 179
& 290
& 318
& 319
& 330
& 519
& 535
& 536
& 565
& 612
& 670
& 704
& 708
& 711
& 725
& 727
& 760
& 804
& 805
& 812\\
& 837
& 841
& 852
& 858
& 863
& 864
& 973
& 1024
& 1027
& 1184
& 1186
& 1187
& 1193
& 1194
& 1198
& 1214
& 1224
& 1255
& 1281
& 1283
& 1286
& 1290
& 1301
& 1304
& 1411\\
& 1439
& 1441
& 1443
& 1444
& 1451
& 1454
& 1455
& 1501
& 1578
& 1588
& 1595
& 1609\\

\midrule
~\multirow{1}{*}{AV45-PET for Test}
& 463
& 732
& 767
& 880
& 1192
& 1375
& 1590\\
\midrule

~\multirow{5}{*}{PiB-PET for Fine-tuning}
& 3
& 4
& 14
& 17
& 18
& 20
& 21
& 22
& 23
& 25
& 26
& 27
& 28
& 29
& 33
& 36
& 38
& 39
& 40
& 42
& 43
& 44
& 46
& 47
& 50\\
& 52
& 53
& 55
& 57
& 60
& 61
& 62
& 64
& 68
& 73
& 75
& 80
& 86
& 88
& 98
& 105
& 117
& 118
& 121
& 123
& 125
& 134
& 138
& 156
& 181\\
& 183
& 206
& 218
& 229
& 241
& 253
& 254
& 269
& 275
& 284
& 287
& 299
& 314
& 332
& 335
& 349
& 355
& 364
& 367
& 406
& 411
& 432
& 434
& 442
& 445\\
& 493
& 498
& 516
& 518
& 523
& 556
& 570
& 573
& 655
& 661
& 697
& 698
& 707
& 737
& 740
& 778
& 808
& 868
& 891
& 914
& 938
& 1000
& 1001
& 1109
& 1146
\\
& 1147
& 1174
& 1241
& 1322
& 1355
& 1384
& 1424
\\
\midrule
~\multirow{1}{*}{PiB-PET for Test}
& 16
& 31
& 59
& 79
& 270
& 316
& 382
& 407
& 529
& 541
& 867
& 1153
\\
 \bottomrule
\end{tabular*}
\label{aibl_sbj}
\end{table*}

\begin{table*}[!tbp]
\setlength{\belowdisplayskip}{-1pt}
\setlength{\abovedisplayskip}{-1pt}
\setlength{\abovecaptionskip}{-1pt}
\setlength{\belowcaptionskip}{-1pt}
\renewcommand{\arraystretch}{0.9}
\setlength{\tabcolsep}{0.3pt}
\scriptsize
\centering
\caption{
Subject IDs of data from the Centiloid Project~\cite{klunk2015centiloid} that are used for fine-tuning and test in the clinical evaluation task.}%
\begin{tabular*}{1\textwidth}
{@{\extracolsep{\fill}}l|ccc ccc ccc ccc ccc ccc cc cc cc c}
\toprule
~Group &
\multicolumn{20}{c}{Subject ID} \\
\midrule
~\multirow{5}{*}{PiB-PET for Fine-tuning } &
AD01 &
AD02 &
AD03 &
AD04 &
AD05 &
AD06 &
AD07 &
AD08 &
AD09 &
AD10 &
AD11 &
AD12 &
AD13 &
AD15 &
AD16 &
AD17 &
AD18 &
AD20 &
AD21 &
AD22 \\&
AD23 &
AD24 &
AD25 &
AD27 &
AD28 &
AD29 &
AD30 &
AD31 &
AD32 &
AD33 &
AD34 &
AD36 &
AD37 &
AD38 &
AD39 &
AD40 &
AD41 &
AD42 &
AD43 &
AD45 \\&
YC104 &
YC105 &
YC106 &
YC107 &
YC108 &
YC109 &
YC110 &
YC111 &
YC112 &
YC113 &
YC114 &
YC115 &
YC116 &
YC117 &
YC118 &
YC119 &
YC120 &
YC121 &
YC123 &
YC124 \\&
YC125 &
YC126 &
YC127 &
YC128 &
YC129 &
YC130 &
YC131 &
YC132 &
YC133 &
YC134 &

\\
\midrule
~\multirow{1}{*}{PiB-PET for Test}
& AD14
& AD19
& AD26
& AD35
& AD44
& YC101
& YC102
& YC103
& YC122
\\
 \bottomrule
\end{tabular*}
\label{centiloid_sbj}
\end{table*}


\footnotesize
\bibliographystyle{IEEEtran}
\bibliography{bibfile}